\documentclass[10pt,twocolumn]{article} 

\usepackage[english]{babel} 

\usepackage{microtype} 

\usepackage{amsmath,amsfonts,amsthm} 

\usepackage[svgnames]{xcolor} 

\usepackage[hang, small, labelfont=bf, up, textfont=it]{caption} 

\usepackage{booktabs} 

\usepackage{lastpage} 

\usepackage{graphicx} 

\usepackage{enumitem} 
\setlist{noitemsep} 

\usepackage{sectsty} 
\allsectionsfont{\usefont{OT1}{phv}{b}{n}} 

\usepackage{geometry} 

\usepackage[T1]{fontenc} 
\usepackage[utf8]{inputenc} 

\usepackage{XCharter} 

\usepackage{fancyhdr} 
\fancypagestyle{firstpage}{ 
	\fancyhf{}
}

\newcommand{\authorstyle}[1]{{\large\usefont{OT1}{phv}{b}{n}\color{DarkRed}#1}} 

\usepackage{titling} 

\newcommand{\HorRule}{\color{DarkGoldenrod}\rule{\linewidth}{1pt}} 

\pretitle{
	\vspace{-30pt} 
	\HorRule\vspace{10pt} 
	\fontsize{32}{36}\usefont{OT1}{phv}{b}{n}\selectfont 
	\color{DarkRed} 
}

\posttitle{\par\vskip 15pt} 

\preauthor{} 

\postauthor{ 
	\vspace{10pt} 
	\par\HorRule 
	\vspace{20pt} 
}

\usepackage{lettrine} 
\usepackage{fix-cm}	

\newcommand{\initial}[1]{ 
	\lettrine[lines=3,findent=4pt,nindent=0pt]{
		\color{DarkGoldenrod}
		{#1}
	}{}%
}

\usepackage{xstring} 

\newcommand{\lettrineabstract}[1]{
	\StrLeft{#1}{1}[\firstletter] 
	\initial{\firstletter}\textbf{\StrGobbleLeft{#1}{1}} 
}

\usepackage[backend=bibtex,style=authoryear,natbib=true]{biblatex} 

\usepackage[autostyle=true]{csquotes} 

\usepackage{hyperref}
\usepackage{amssymb}
\usepackage{amsmath}
\usepackage{unitsdef}
\usepackage{url}

\title{Diverse local epidemics reveal the distinct effects of population density, demographics, climate, depletion of susceptibles, and intervention in the first wave of COVID-19 in the United States} 

\author{
	\authorstyle{Niayesh Afshordi\textsuperscript{1,2,3,*},  Benjamin Holder\textsuperscript{4,1}, Mohammad Bahrami \textsuperscript{5}, and Daniel Lichtblau\textsuperscript{5}} 
	\newline\newline 
	$^{1}$Department of Physics and Astronomy, University of Waterloo, Waterloo, ON, N2L 3G1, Canada\\
$^{2}$Waterloo Centre for Astrophysics, University of Waterloo, Waterloo, ON, N2L 3G1, Canada\\
$^{3}$Perimeter Institute of Theoretical Physics, 31 Caroline St. N., Waterloo, ON, N2L 2Y5, Canada\\
$^4$ Department of Physics, Grand Valley State University, Grand Rapids, Michigan 49401, United States \\
$^5$ Wolfram Research Inc, 100 Trade Center Drive, Champaign, IL 61820-7237 ,USA\\
$^*$ {\it corresponding author}: \href{emailto:nafshordi@pitp.ca}{nafshordi@pitp.ca}
}

\date{\today} 

\begin{document}

\maketitle 

\thispagestyle{firstpage} 


\lettrineabstract{The SARS-CoV-2 pandemic has caused significant mortality and morbidity worldwide, sparing almost no community.  As the disease will likely remain a threat for years to come, an understanding of the precise influences of human demographics and settlement, as well as the dynamic factors of climate, susceptible depletion, and intervention, on the spread of localized epidemics will be vital for mounting an effective response. We consider the entire set of local epidemics in the United States; a broad selection of demographic, population density, and climate factors; and local mobility data, tracking social distancing interventions, to determine the key factors driving the spread and containment of the virus.  Assuming first a linear model for the rate of exponential growth (or decay) in cases/mortality, we find that population-weighted density, humidity, and median age dominate the dynamics of growth and decline, once interventions are 
accounted for. A focus on distinct metropolitan areas suggests that some locales benefited from the timing of a nearly simultaneous nationwide shutdown, and/or the regional climate conditions in mid-March; while others suffered significant outbreaks prior to intervention. Using a first-principles model of the infection spread, we then develop predictions for the impact of the relaxation of social distancing and local climate conditions.   A few regions, where a significant fraction of the population was infected, show evidence that the epidemic has partially resolved via depletion of the susceptible population (i.e., ``herd immunity''), while most regions in the United States remain overwhelmingly susceptible. These results will be important for optimal management of intervention strategies, which can be facilitated using our online dashboard.}

\section*{Introduction}

The new human coronavirus SARS-CoV-2 emerged in Wuhan Province, China in December 2019 \citep{chen2020,li2020}, reaching 10,000 confirmed cases and 200 deaths due to the disease (known as COVID-19) by the end of January this year. Although travel from China was halted by late-January, dozens of known introductions of the virus to North America occurred prior to that \citep{holshue2020,kucharski2020}, and dozens more known cases were imported to the US and Canada during February from Europe, the Middle East, and elsewhere.  Community transmission of unknown origin was first detected in California on February 26, followed quickly by Washington State \citep{chu_englund2020}, Illinois and Florida, but only on March 7 in New York City.  Retrospective genomic analyses have demonstrated that case-tracing and self-quarantine efforts were effective in preventing most known imported cases from propagating \citep{ladner2020, gonzalezreiche2020, worobey2020}, but that the eventual outbreaks on the West Coast \citep{worobey2020, chu_englund2020,deng2020} and New York \citep{gonzalezreiche2020} were likely seeded by unknown imports in mid-February.  By early March, cross-country spread was primarily due to interstate travel rather than international imports \citep{fauver_grubaugh2020}.  

In mid-March 2020, nearly every region of the country saw a period of uniform exponential growth in daily confirmed cases --- signifying robust community transmission --- followed by a plateau in late March, likely due to social mobility reduction.  The same qualitative dynamics were seen in COVID-19 mortality counts, delayed by approximately one week.  Although the qualitative picture was similar across locales, the quantitative aspects of localized epidemics --- including initial rate of growth, infections/deaths per capita, duration of plateau, and rapidity of resolution ---were quite diverse across the country. Understanding the origins of this diversity will be key to predicting how the relaxation of social distancing, annual changes in weather, and static local demographic/population characteristics will affect the resolution of the first wave of cases, and will drive coming waves, prior to the availability of a vaccine.

The exponential growth rate of a spreading epidemic is dependent on the biological features of the virus-host ecosystem --- including the incubation time, susceptibility of target cells to infection, and persistence of the virus particle outside of the host ---  but, through its dependence on the transmission rate between hosts, it is also a function of external factors such as population density, air humidity, and the fraction of hosts that are susceptible.  Initial studies have shown that SARS-CoV-2 has a larger rate of exponential growth (or, alternatively, a lower doubling time of cases\footnote{The doubling time is $\ln 2$ divided by the exponential growth rate.}) than many other circulating human viruses \citep{park2020}.  For comparison, the pandemic influenza of 2009, which also met a largely immunologically-naive population, had a doubling time of 5--\unit{10}{d} \citep{yu2012, storms2013}, while that of SARS-CoV-2 has been estimated at 2--\unit{5}{d} \citep{sanche2020, oliveiros2020} (growth rates of $\unit{\sim0.10}{{\rm d}^{-1}}$ vs.\ $\unit{\sim0.25}{{\rm d}^{-1}}$). It is not yet understood which factors contribute to this high level of infectiousness.  

While the dynamics of an epidemic (e.g., cases over time) must be described by numerical solutions to nonlinear models, the exponential growth rate, $\lambda$, usually has a simpler dependence on external factors. Unlike case or mortality incidence numbers, the growth rate does not scale with population size. It is a directly measurable quantity from the available incidence data, unlike, e.g., the reproduction number, which requires knowledge of the serial interval distribution \citep{wallinga_lipsitch2007, roberts_heesterbeek2007, dushoff_park2020}, something that is difficult to determine empirically \citep{champredon2015, nishiura2010}. Yet, the growth rate contains the same threshold as the reproduction number ($\lambda = 0$ vs.\ $R_0$ = 1), between a spreading epidemic (or an unstable uninfected equilibrium) and a contracting one (or an equilibrium that is resistant to flare-ups).  Thus, the growth rate is an informative direct measure on that space of underlying parameters.  

\begin{figure*}
\centering
    \includegraphics[width=\linewidth]{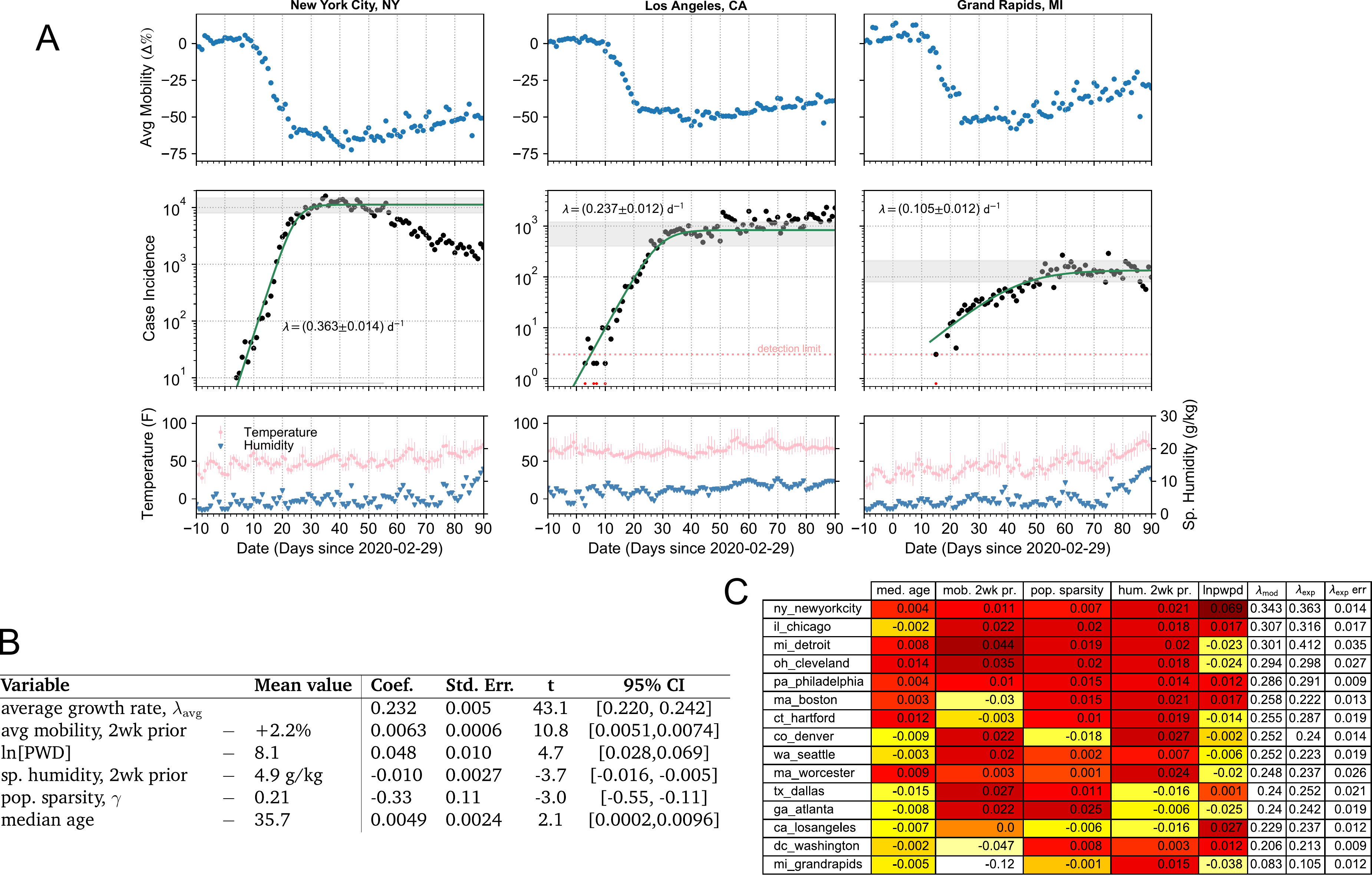}
    \caption{Mobility and COVID-19 incidence data examples, and the results of linear regression to extracted initial exponential growth rates, $\lambda_{\rm exp}$,  in the top 100 metropolitan regions. (A) Three example cities with different initial growth rates. Data for Google mobility (blue points), daily reported cases (black points), and weather (red and blue points, bottom) are shown with a logistic fit to cases (green line). Data at or below detection limit were excluded from fits (dates marked by red points). Thin grey bars at base of cases graphs indicate region considered ``flat'', with right end indicating the last point used for logistic fitting; averaging over ``flat'' values generates the thick grey bars to guide the eye.  [See Supp.\ Mat. for additional information and for complete data sets for all metropolitan regions.] (B) Weighted linear regression results in fit to $\lambda_{\rm exp}$ for all metropolitan regions. (C) Effect of each variable on growth rate (i.e., $\Delta \lambda$ values) for those regions with well-estimated case and death rates; white/yellow indicates a negative effect on $\lambda$, red indicates positive.}
    \label{fig:metroregions}
%
\vspace{10pt}
%
%
\includegraphics[width=0.49\linewidth]{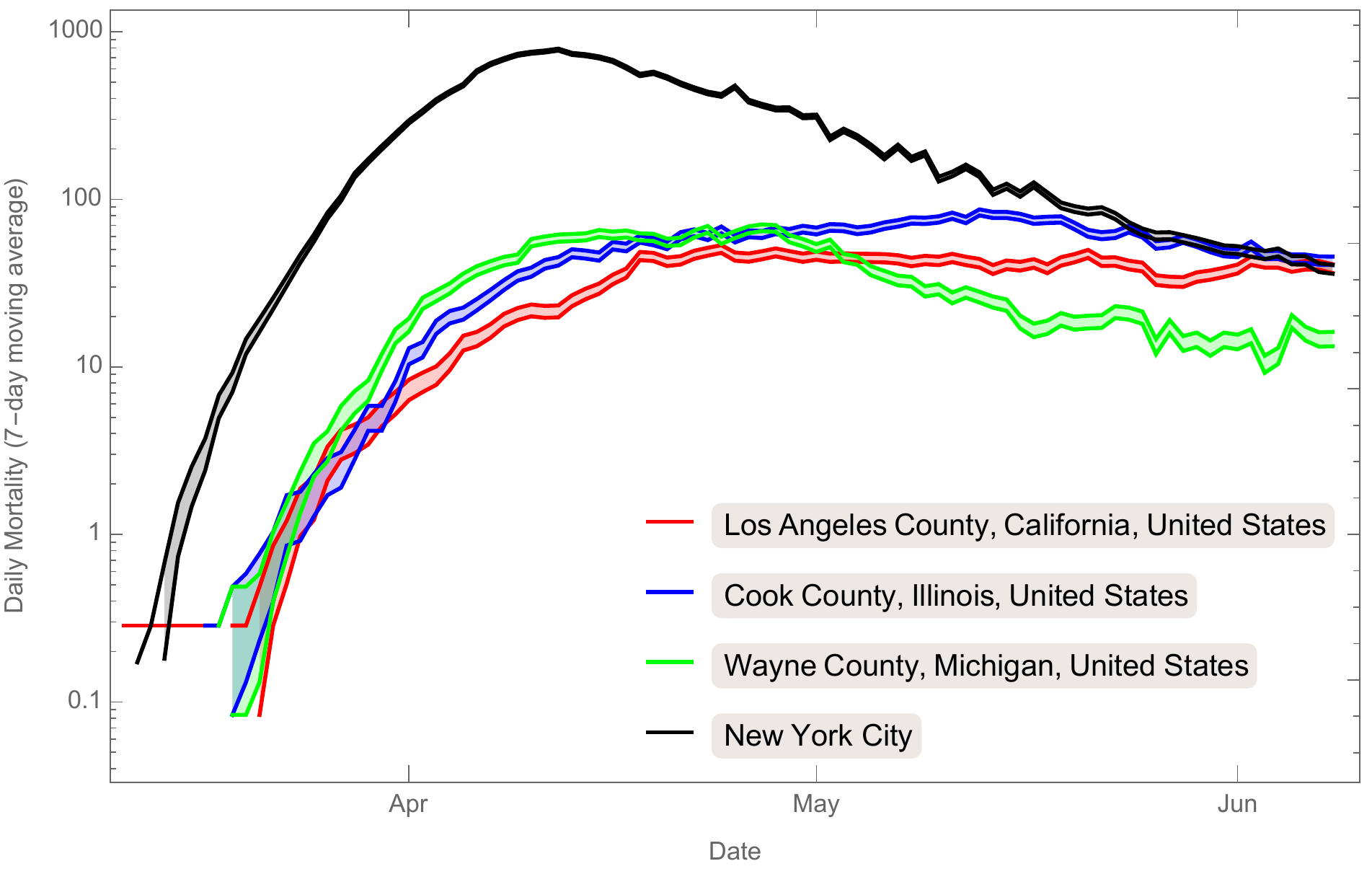}
\includegraphics[width=0.49\linewidth]{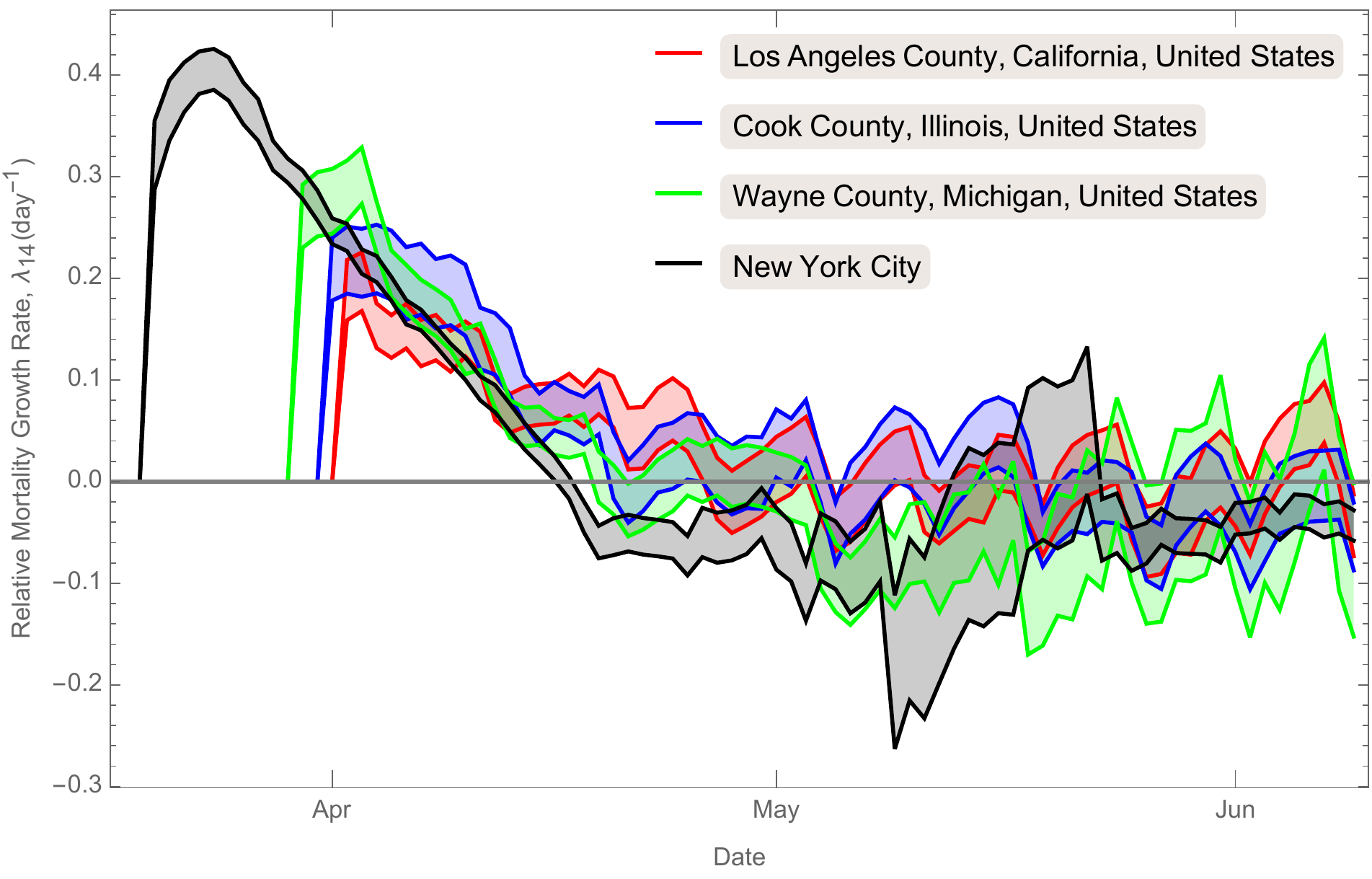}
\caption{COVID-19 mortality incidence (7-day rolling average, left) and exponential growth rate ($\lambda_{14}$, determined by regression of the logged mortality data over 14-day windows, right) for the four US counties with  >2400 confirmed COVID-19 reported deaths (as of 8th June, 2020).  }\label{US4_ID7}
\end{figure*}

In this work, we leverage the enormous data set of epidemics across the United States to evaluate the impact of demographics, population density and structure, weather, and non-pharmaceutical interventions (i.e., mobility restrictions) on the exponential rate of growth of COVID-19.  Following a brief analysis of the initial spread in metropolitan regions, we expand the meaning of the exponential rate to encompass all aspects of a local epidemic --- including growth, plateau and decline --- and use it as a tracer of the dynamics, where its time dependence and geographic variation are dictated solely by these external variables and per capita cumulative mortality. Finally, we use the results of that linear analysis to calibrate a new nonlinear model --- a renewal equation that utilizes the excursion probability of a random walk to determine the incubation period --- from which we develop local predictions about the impact of social mobility relaxation, the level of herd immunity, and the potential of rebound epidemics in the Summer and Fall. 

\section*{Results}

\subsection*{Initial growth of cases in metropolitan regions is exponential with rate depending on mobility, population, demographics, and humidity}

As an initial look at COVID-19's arrival in the United States, we considered the $\sim$100 most populous  metropolitan regions --- using maps of population density to select compact sets of counties representing each region (see Supplementary Material) --- and estimated the initial exponential growth rate of cases in each region. We performed a linear regression to a large set of demographic (sex, age, race) and population variables, along with weather and social mobility \citep{google2020} preceding the period of growth (Figure \ref{fig:metroregions}).   In the best fit model ($R^2=0.75$, ${\rm BIC} = -183$), the baseline value of the initial growth rate was $\lambda = \unit{0.21}{{\rm d}^{-1}}$ (doubling time of \unit{3.3}{d}), with average mobility two weeks prior to growth being the most significant factor (Figure \ref{fig:metroregions}B).  Of all variables considered, only four others were significant: population density (including both {\em population-weighted density} (PWD) --- also called the ``lived population density'' because it estimates the density for the average individual \citep{craig1985}--- and {\em population sparsity}, $\gamma$, a measure of the difference between PWD and standard population density, see Methods), $p<0.001$ and $p=0.006$;  specific humidity two weeks prior to growth, $p=0.001$; and median age, $p=0.04$.

While mobility reduction certainly caused the ``flattening'' of case incidence in every region by late-March, our results show (Figure \ref{fig:metroregions}C) that it likely played a key role in reducing the  {\em rate} of growth in Boston, Washington, DC, and Los Angeles, but was too late, with respect to the sudden appearance of the epidemic, to have such an effect in, e.g., Detroit and Cleveland. In the most extreme example, Grand Rapids, MI, seems to have benefited from a late arriving epidemic, such that its growth (with a long doubling time of \unit{7}{d}) occurred almost entirely post-lockdown.  

Specific humidity, a measure of absolute humidity, has been previously shown to be inversely correlated with respiratory virus transmission \citep{lowen2007, shaman2009, shaman2011, kudo2019}. Here, we found it to be a significant factor, but weaker than population density and mobility (Figure \ref{fig:metroregions}C). It could be argued that Dallas, Los Angeles, and Atlanta saw a small benefit from higher humidity at the time of the epidemic's arrival, while the dry late-winter conditions in the Midwest and Northeast were more favorable to rapid transmission of SARS-CoV-2.

\subsection*{\label{sec:I7}Exponential growth rate of mortality as a dynamical, pan-epidemic, measure}

In the remainder of this report, we consider the exponential rate of growth (or decay) in local confirmed deaths due to COVID-19. The statistics of mortality is poorer compared to reported cases, but it is much less dependent on unknown factors such as the criteria for testing, local policies, test kit availability, and asymptomatic individuals \citep{pearce2020}.  Although there is clear evidence that a large fraction of COVID-19 mortality is missed in the official counts \citep[e.g.,][]{leon2020, Modi2020.04.15.20067074}, mortality is likely less susceptible to rapid changes in reporting, and, as long as the number of reported deaths is a monotonic function of the actual number of deaths (e.g., a constant fraction, say $50\%$), the sign of the exponential growth rate will be unchanged, which is the crucial measure of the success in pandemic management.  

To minimize the impact of weekly changes, such as weekend reporting lulls, data dumps, and mobility changes from working days to weekends, we calculate the regression of $\ln\left[{\rm Mortality}\right]$ over a 14-day interval, and assign this value, $\lambda_{14}(t)$, and its standard error to the last day of the interval.  Since only the data for distinct 2-week periods are independent, we multiply the regression errors by $\sqrt{14}$ to account for correlations between the daily estimates.  Together with a ``rolling average'' of the mortality, this time-dependent measure of the exponential growth rate provides, at any day, the most up-to-date information on the progression of the epidemic (Figure \ref{US4_ID7}). 

In the following section, we consider a linear fit to $\lambda_{14}$, to determine the statistically-significant external (non-biological) factors influencing the dynamics of local exponential growth and decline of the epidemic.  We then develop a first-principles model for $\lambda_{14}$ that allows for extrapolation of these dependencies to predict the impact of future changes in social mobility and climate.

\subsection*{Epidemic mortality data explained by mobility, population, demographics, depletion of susceptible population and weather, throughout the first wave of COVID-19}

\begin{table*}
\begin{tabular}{l|lll}
\hline
 \text{Joint Fit to All potential drivers} & \text{Estimate} & \text{Std Err} & \text{t-Statistic} 
   \\
   \hline
Baseline Mortality Growth Rate $\lambda_{14}$ & 0.195 & 0.011 & 17.2 \\
 \text{COVID Death Fraction} & -59.4 & 6.1 &-9.7 \\
 \text{Social Mobility (2wks prior)} & 0.00238 & 0.00028 & 8.5  \\
$ \ln$(Population Weighted Density)-8.24 & 0.0412 & 0.0058 & 7.1  \\
\text{Social Mobility (4wks prior)}& 0.00122 & 0.00019 & 6.6\\
Population Sparsity-0.188 & -0.249 & 0.063 & -3.9  \\
 $\log$(\text{Annual Death})-4.04 & -0.0301 & 0.0091 & -3.3 \\
 \text{Median Age}-37.47 & 0.0038 & 0.0012 & 3.0 \\
 \text{People per Household}-2.76 & 0.023 & 0.014 & 1.6  \\
 \text{Specific Humidity (2wks prior)}-5.92 g/kg & -0.0033 & 0.0031 & -1.1   \\
 \text{Temperature (2wks prior)}-13.11 C & -0.00083 & 0.0013 & -0.6 \\
 \text{Temperature (4wks prior)}-11.60 C & -0.00060 & 0.0014 & -0.4  \\
 \text{Specific Humidity (4wks prior)}-5.53 g/kg & 0.00058 & 0.0032 & 0.2 \\
\end{tabular}

\begin{tabular}{l|lll}
\hline
 \text{Joint Fit to statistically significant drivers} & \text{Estimate} & \text{Std Err} & \text{t-Statistic} 
   \\
\hline
 Baseline Mortality Growth Rate $\lambda_{14}$ & 0.198 & 0.011 & 18.7\\
 \text{COVID Death Fraction} & -56.7 & 5.9 & -9.7 \\
 \text{Social Mobility (2wks prior)}&  0.00236 & 0.00027 & 8.8 \\
 \text{Social Mobility (4wks prior)} & 0.00131 & 0.00017 & 7.6  \\
 $ \ln$(Population Weighted Density)-8.24 & 0.0413 & 0.0058 & 7.2  \\
Population Sparsity-0.188 & -0.260 & 0.061 & -4.3 \\
 \text{Specific Humidity (2wks prior)}-5.92 g/kg & -0.0047 & 0.0011 & -4.1  \\
 $\log$(\text{Annual Death})-4.04 & -0.0324 & 0.0088 & -3.7  \\
 \text{Median Age}-37.48 & 0.0040 & 0.0012 & 3.3 \\
\end{tabular}
\caption{Joint Linear Fit to $\lambda_{14}(t)$ data (Top). Any dependence with t-Statistic below $2.5\sigma$ is considered not statistically significant. Joint Linear Fit to $\lambda_{14}(t)$, including only statistically significant dependencies (Bottom). For all coefficients, the population-weighted baseline is subtracted from the linear variable. \label{tab:linear}}

\vspace{12pt}

\begin{tabular}{l|l|l }
\hline
Parameter & Best-Fit $\pm$ Std Err & Description\\
\hline
$\tau = \tau_0 ({\rm Median~Age}/26.2~ {\rm years})^{C_A}$ & &Time from exposure to contagiousness\\
$\tau_0$(day) & $160 \pm 58$ & Normalization \\
$C_{A}$ & $-2.26 \pm 0.95$ & Age dependence \\
\hline
$d^{-1}$(day) & $17.6 \pm 2.2$ & Time from exposure to quarantine/recovery  \\
\hline
$C_D$ & $3460 \pm 610$ & Conversion constant, $f_D \rightarrow f_I$ \\
\hline
$\beta$: Equation (\ref{eq:beta})&  &  Rate constant for infection\\
$\ln\left[k\beta_0\tau_0^{-2}({\rm m}^2/{\rm day}^3)\right]$ & $0.37 \pm 1.25 $
 &  Normalization \\
$100 C_{\cal M}$ & $8.08 \pm 1.76$ & Dependence on Social Mobility  \\
$C_{\cal H}$ & $ -0.154 \pm 0.055$ & Dependence on specific humidity \\
$C_\gamma$ & $-5.52 \pm 2.35$ & Dependence on population sparsity \\
 $C_{A_D}$ & $-1.05 \pm 0.25$ & Dependence on total annual deaths \\
\end{tabular}
\caption{Best-fit parameters for the nonlinear model using parametrization defined in the text.} \label{tab:nonlinear}
\end{table*}

We considered a spatio-temporal dataset containing 3933 estimates of the exponential growth measure, $\lambda_{14}$, covering the three month period of 8 March 2020 -- 8 June 2020 in the 187 US counties for which information on COVID-19 mortality and all potential driving factors, below, were available (the main barrier was social mobility information, which limited us to a set of counties that included 69\% of US mortality). A joint, simultaneous, linear fit of these data to 12 potential driving factors (Table \ref{tab:linear}) revealed only 7 factors with {\it independent} statistical significance. Re-fitting only to these variables returned the optimal fit for the considered factors (${\rm BIC}=-5951$; $R^2 = 0.674$). 

We found, not surprisingly, that higher population density, median age, and social mobility correlated with positive exponential growth, while population sparsity, specific humidity, and susceptible depletion correlated with exponentially declining mortality. Notably the coefficients for each of these quantities was in the 95\% confidence intervals of those found in the analysis of metropolitan regions (and vice versa). Possibly the most surprising dependency was the negative correlation, at $\simeq -3.7\sigma$ between $\lambda_{14}$ and the {\it total} number of annual deaths in the county. In fact, this correlation was marginally more significant than a correlation with $\log$(population), which was $-3.3\sigma$. One possible interpretation of this negative correlation is that the number of annual death is a proxy for the number of potential outbreak clusters. The larger the number of clusters, the longer it might take for the epidemic to spread across their network, which would (at least initially) slow down the onset of the epidemic. 

\subsection*{\label{sec:nonlinear} Nonlinear model}
To obtain more predictive results, we developed a mechanistic nonlinear model for infection (see Supplementary Material for details). We followed the standard analogy to chemical reaction kinetics (infection rate is proportional to the product of susceptible and infectious densities), but defined the generation interval (approximately the incubation period) through the excursion probability in a 1D random walk, modulated by an exponential rate of exit from the infected class. This approach resulted in a {\em renewal equation} \citep{heesterbeek_dietz1996, champredon2015, champredon2018}, with a distribution of generation intervals that is more realistic than that of standard SIR/SEIR models, and which could be solved formally (in terms of the Lambert W function) for the growth rate in terms of the infection parameters:
	\begin{equation} \label{eq:nonlinmodel}
	\lambda = \frac{1}{2\tau} \left[ {\rm W} \left(\sqrt{\frac{\beta  \mathcal{S} \tau}{2}}\right) \right]^2 - d
	\end{equation} 
The model has four key dependencies, which we describe here, along with our assumptions about their own dependence on population, demographic, and climate variables. As mortality (on which our estimate of growth rate is based) lags infection (on which the renewal equation is based), we imposed a fixed time shift of $\Delta t$ for time-dependent variables:
	\begin{enumerate}
	\item We assumed that the susceptible population, which feeds new infections and drives the growth, is actually a sub-population of the community, consisting of highly-mobile and frequently interacting individuals, and that most deaths occurred in separate sub-population of largely immobile non-interacting individuals. Under these assumptions, we found (see Supp.\ Mat.) that the susceptible density, $S(t)$, could be estimated from the cumulative per capita death fraction, $f_D$, as:
	    \begin{equation*}
	   S(t-\Delta t) = S(0) \exp \left[ - C_D \, f_D(t) \right] \quad \left(f_D= D_{\rm tot}/N\right)\,,
	    \end{equation*}
	where $D_{\rm tot}$ is the cumulative mortality count, $N$ is the initial population, and the initial density is $S(0)=k\,{\rm PWD}$.
	\item We assumed that the logarithm of the ``rate constant'' for infection, $\beta$, depended linearly on social mobility, $m$, specific humidity, $h$, population sparsity, $\gamma$, and total annual death, $A_D$, as:
		\begin{equation}
		\begin{split}
		&\ln\left[\beta\left(\mathcal{M}, \mathcal{H}, \gamma, A_D \right)\right] = \ln\left[\beta_0 \right] \\
		&\quad + C_{\mathcal{M}} \left(\mathcal{M} - \bar{\mathcal{M}}\right) + C_{\mathcal{H}} \left(\mathcal{H} - \bar{\mathcal{H}} \right) \\
		&\quad + C_{\gamma} \left(\gamma - \bar{\gamma}\right) + C_{A_D} \left(A_D - \bar{A}_D\right)
		\end{split}
		\label{eq:beta}
		\end{equation}
	where a barred variable represents the (population-weighted) average value over all US counties, and where the mobility and humidity factors were time-shifted with respect to the growth rate estimation window: $\mathcal{M} = m\left(t - \Delta t\right)$ and $\mathcal{H}  = h\left(t - \Delta t\right)$.
	\item The characteristic time scale to infectiousness, $\tau$, is intrinsic to the biology and therefore we assumed it would depend only on the median age of the population, $A$.  We assumed a power law dependence:
		\begin{equation}
		\tau = \tau_0 \left(\frac{A}{A_0}\right)^{C_A}
		\end{equation}
	where we fixed the pivot age, $A_0$, to minimize the error in $\tau_0$.
	\item The exponential rate of exit from the infected class, $d$, was assumed constant, since we found no significant dependence for it on other factors in our analysis of US mortality. From the properties of the Lambert W function, when the infection rate or susceptibility density approach zero --- through mobility restrictions or susceptible depletion --- the growth rate will tend to $\lambda \approx - d$, its minimum value.
\end{enumerate}
\begin{figure*}
    \centering
    \includegraphics[width=0.48\linewidth]{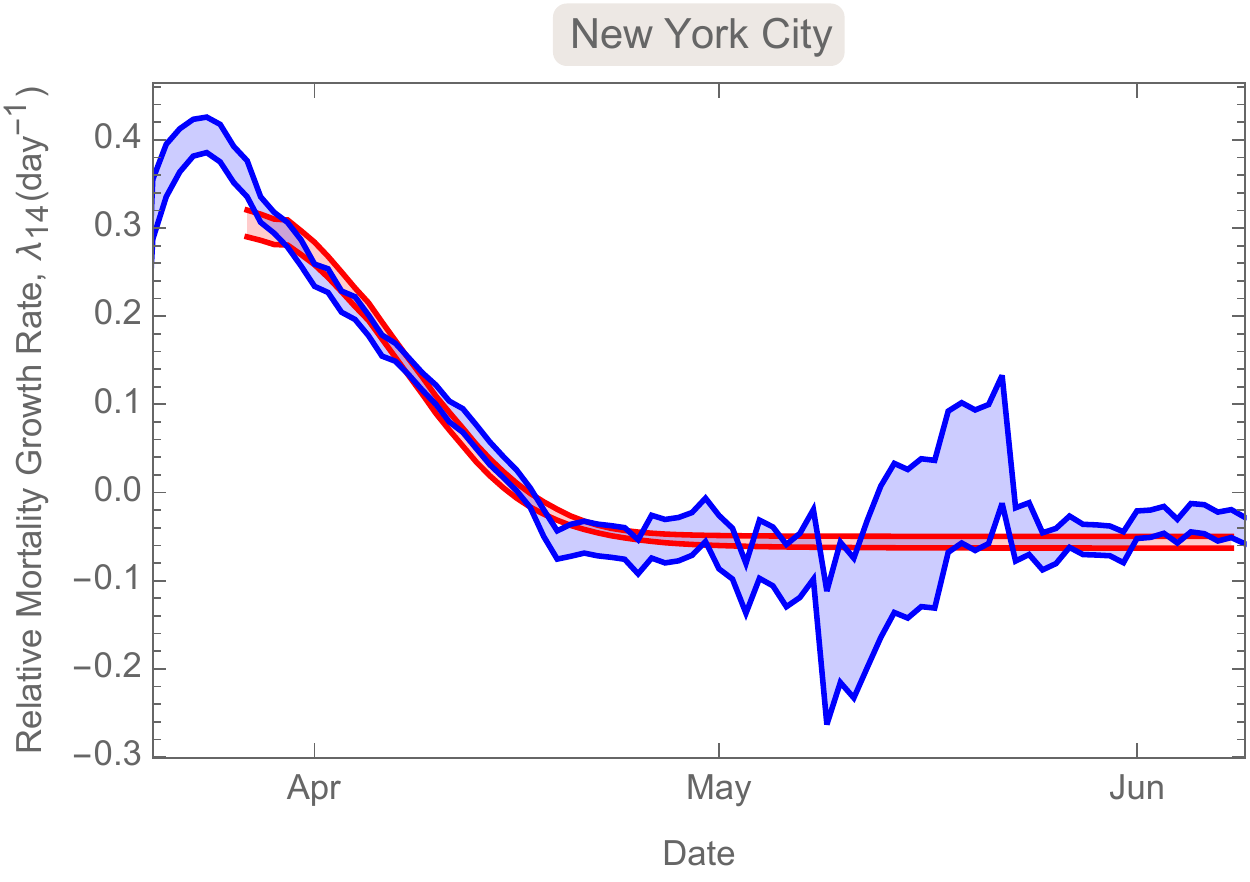} 
    \includegraphics[width=0.48\linewidth]{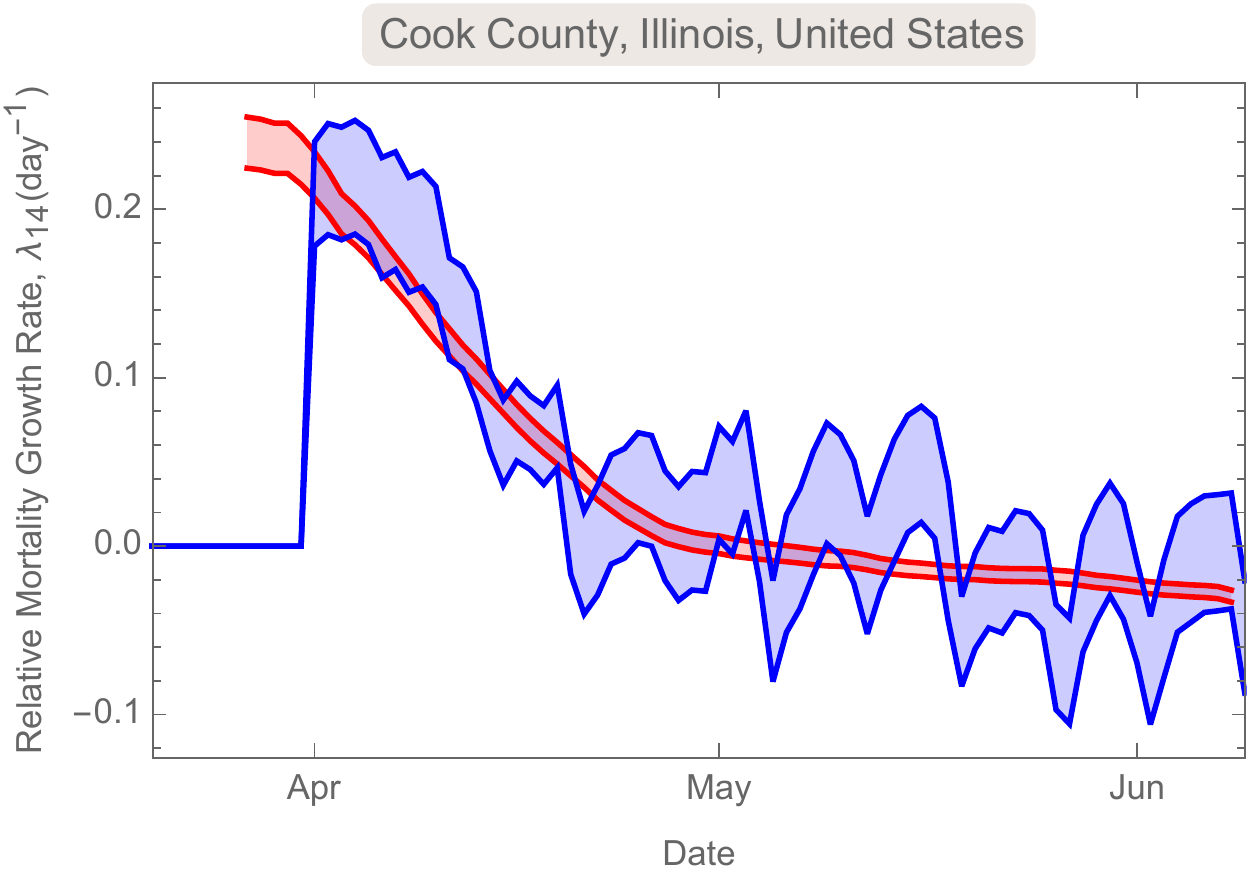}
     \includegraphics[width=0.48\linewidth]{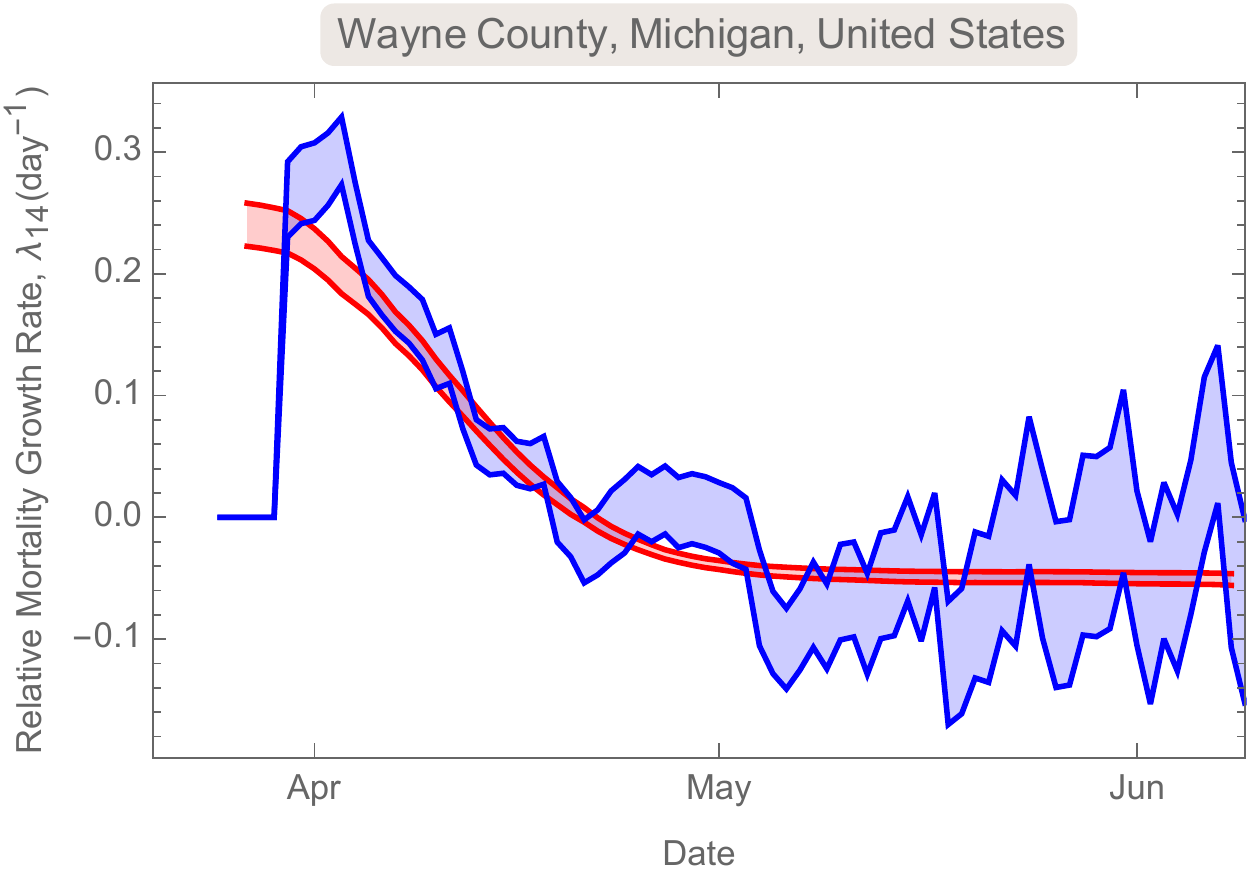}
      \includegraphics[width=0.48\linewidth]{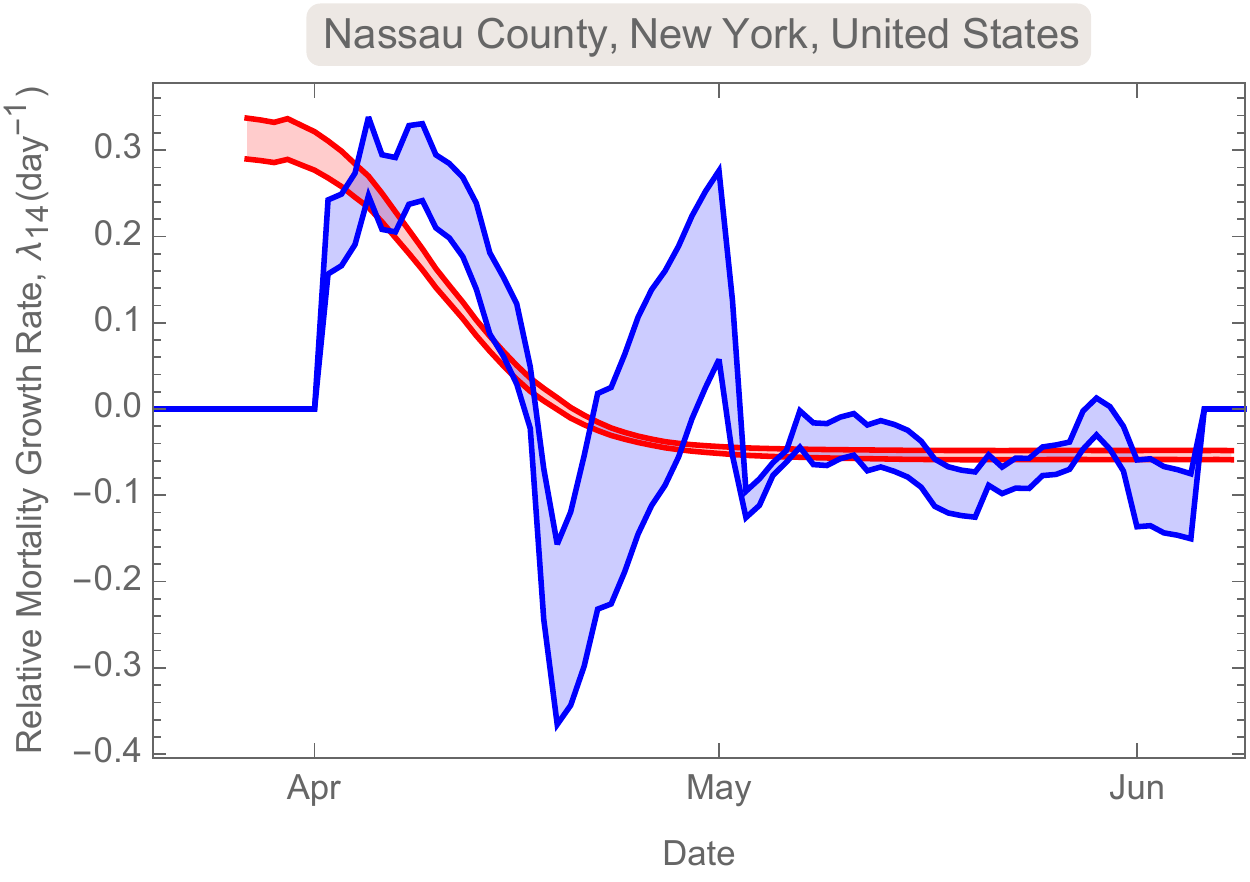}
      \includegraphics[width=0.48\linewidth]{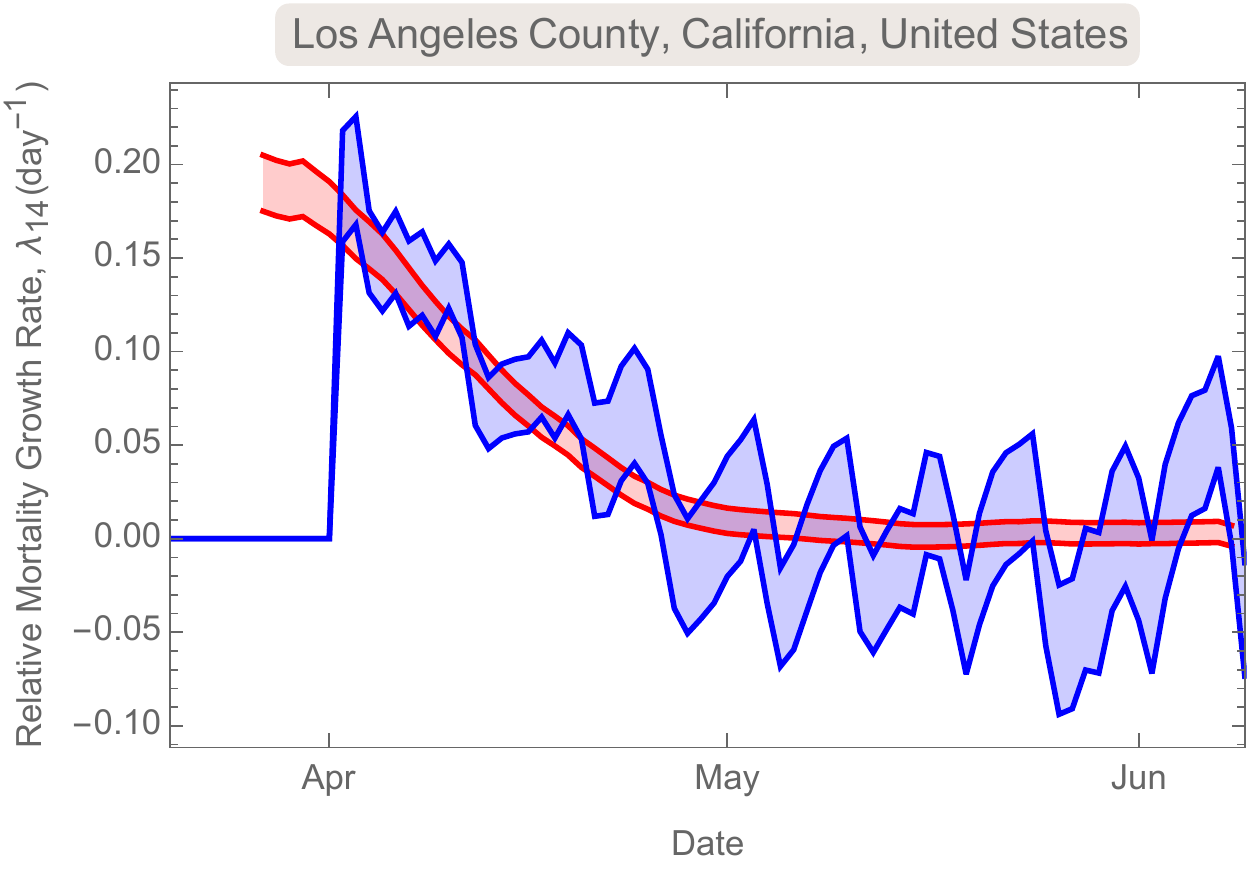}
       \includegraphics[width=0.48\linewidth]{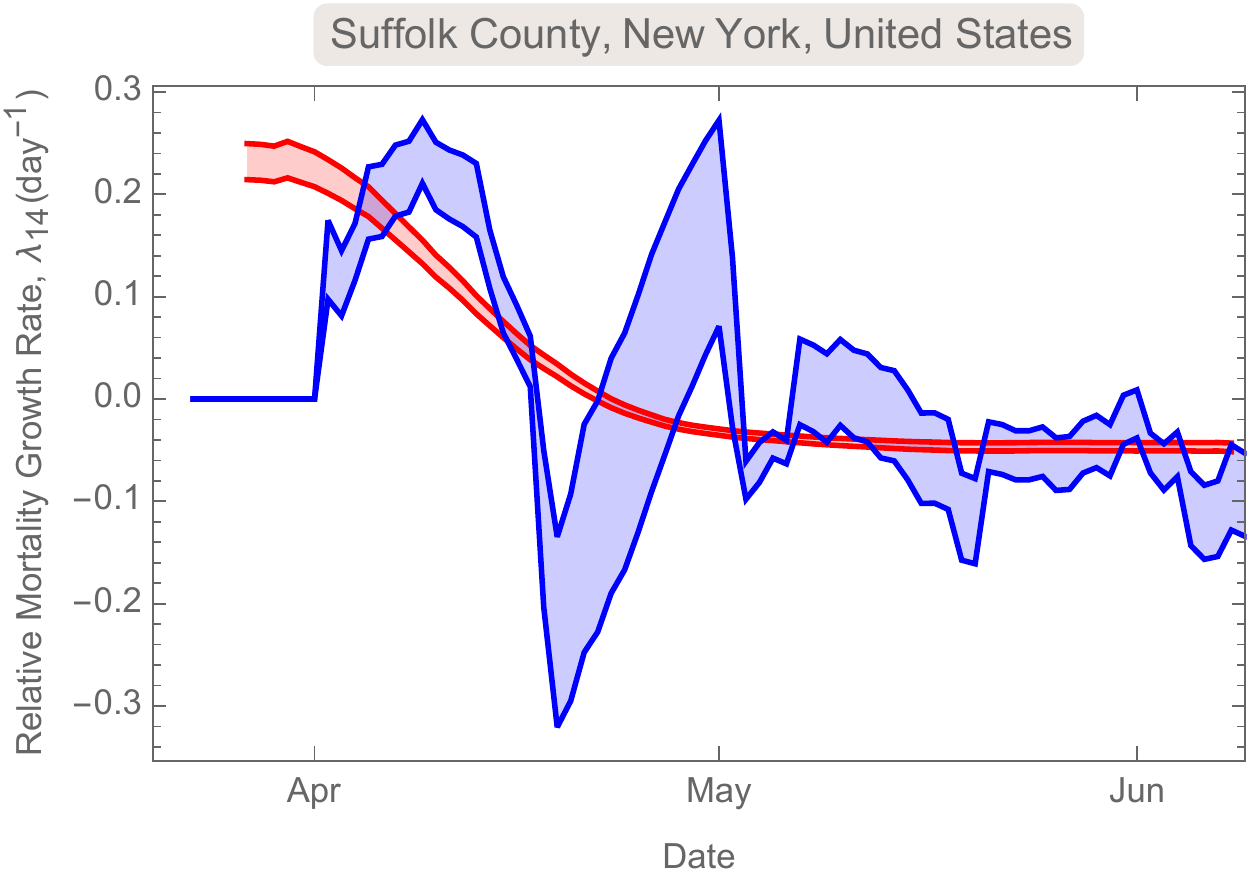}
    \caption{Nonlinear model prediction (Eqn.\ \ref{eq:nonlinmodel}, red) for the actual (blue) mortality growth rate, in the six counties with highest reported death. Bands show 1-$\sigma$ confidence region for both the model mean and the $\lambda_{14}$ value.}
    \label{fig:I7_postdiction}
\end{figure*}
With these parameterizations, we performed a nonlinear regression to $\lambda_{14}(t)$ using the entire set of US county mortality incidence time series (Table \ref{tab:nonlinear}). Compared to the linear model of the previous section (Table \ref{tab:linear}b), the fit improved by 7.6$\sigma$ (${\rm BIC} = -6008$ ;  $R^2=0.724$), despite both having 9 free parameters.  Through the estimated parameter values, the model makes predictions for an individual's probability of becoming infectious, and the distributions of incubation period and generation interval, all as a function of the median age of the population (see Supplementary Material).

\begin{figure*}
    \centering
        \includegraphics[width=0.45\linewidth]{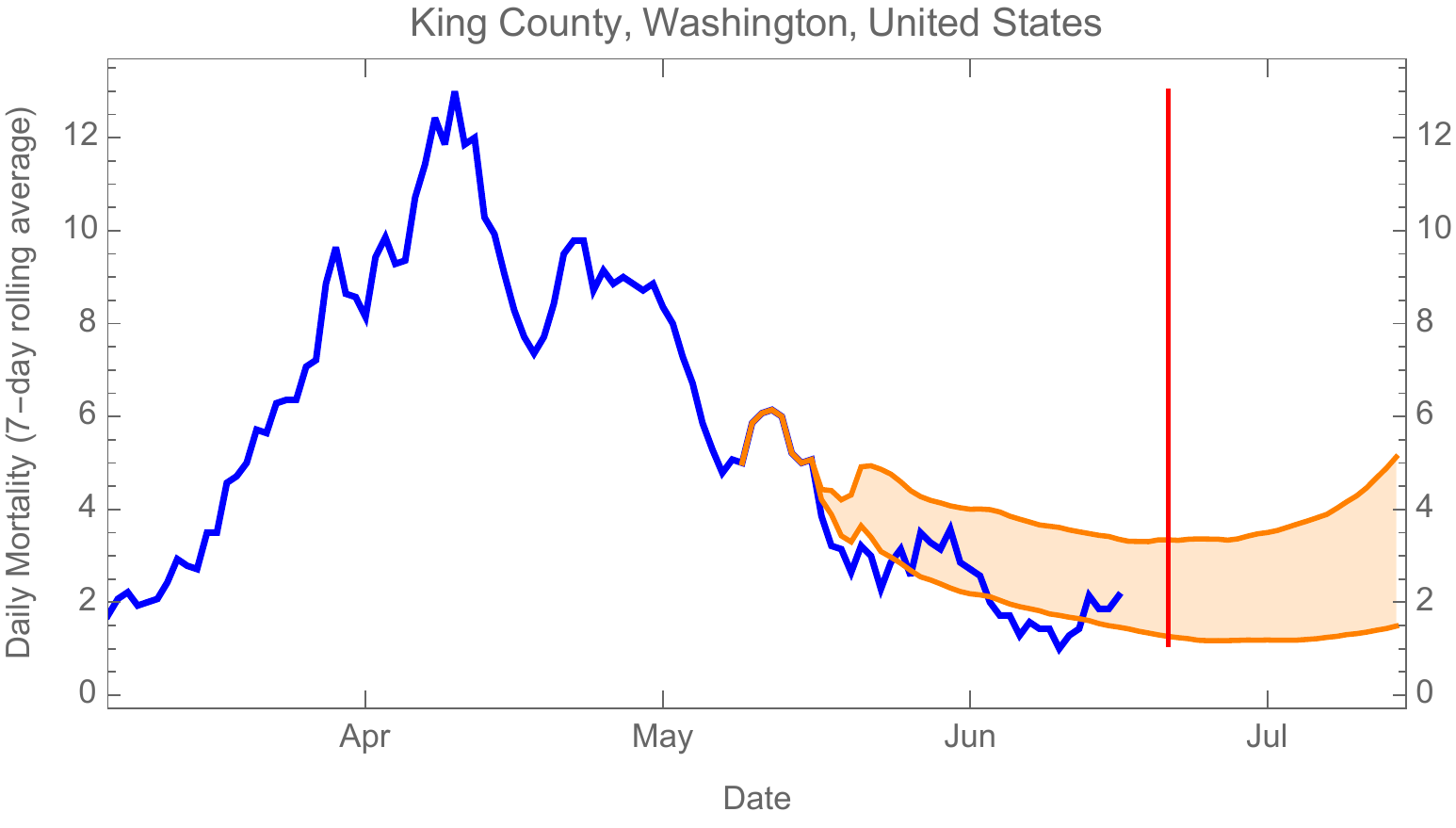}
    \includegraphics[width=0.45\linewidth]{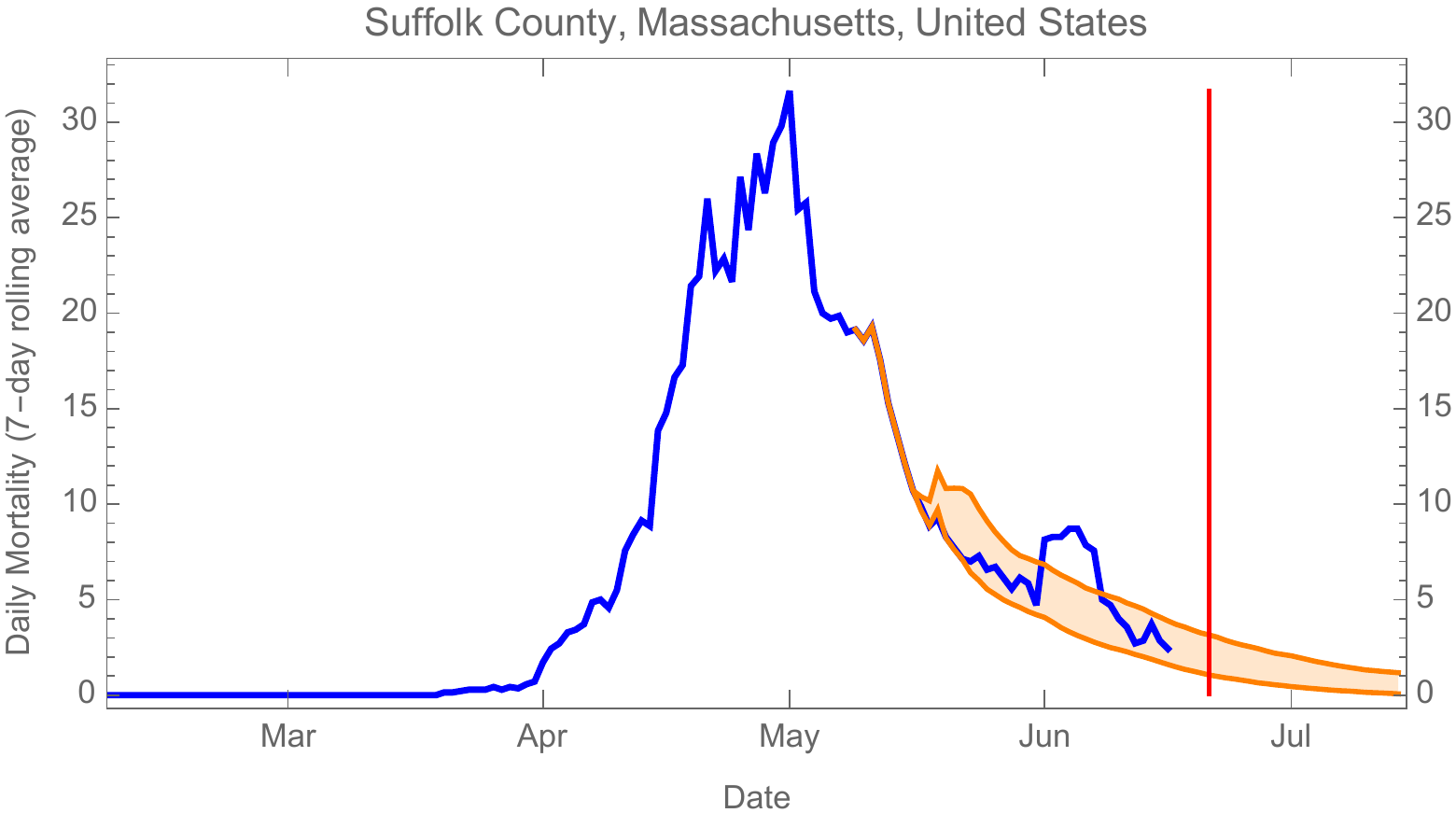}
        \includegraphics[width=0.45\linewidth]{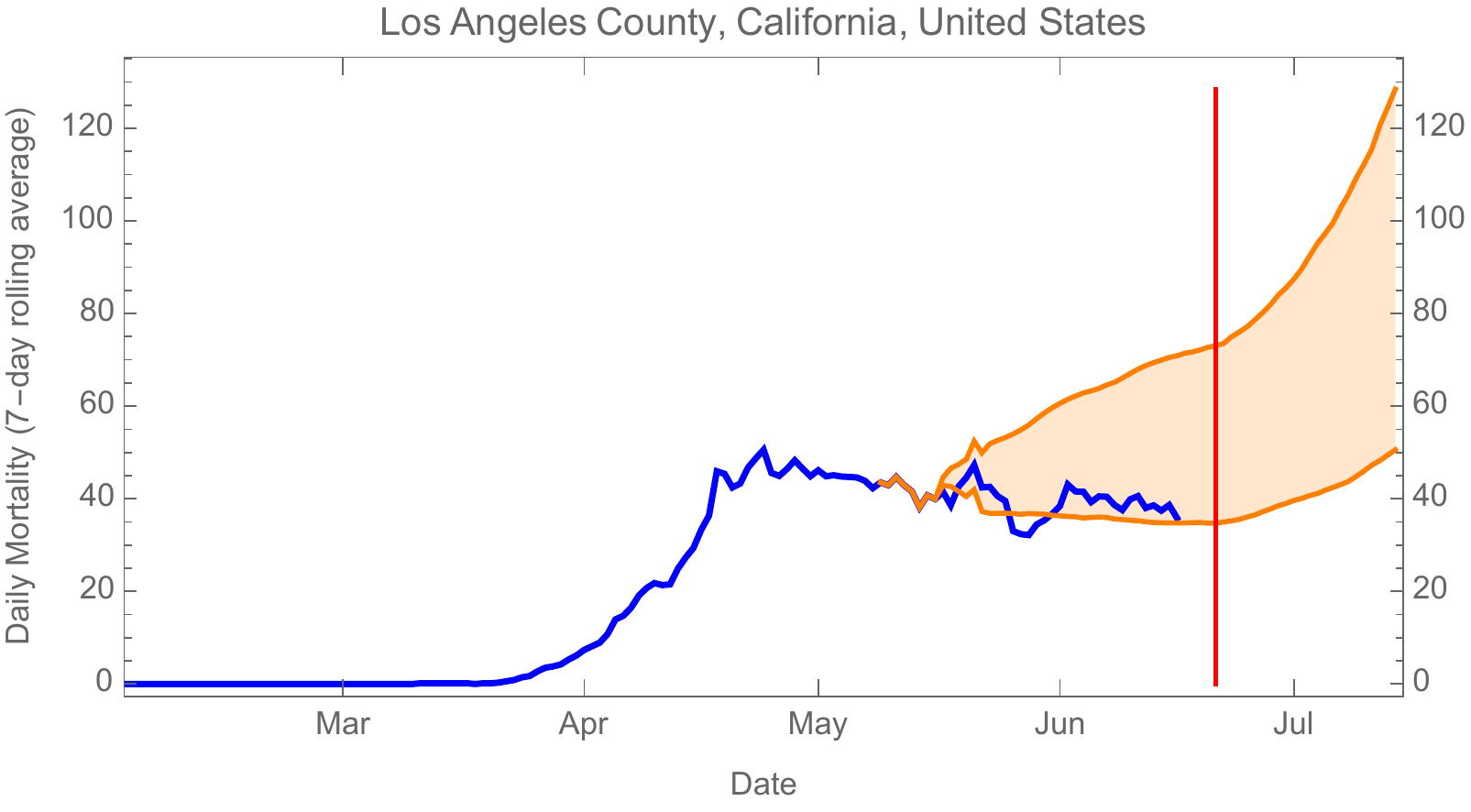}
        \includegraphics[width=0.45\linewidth]{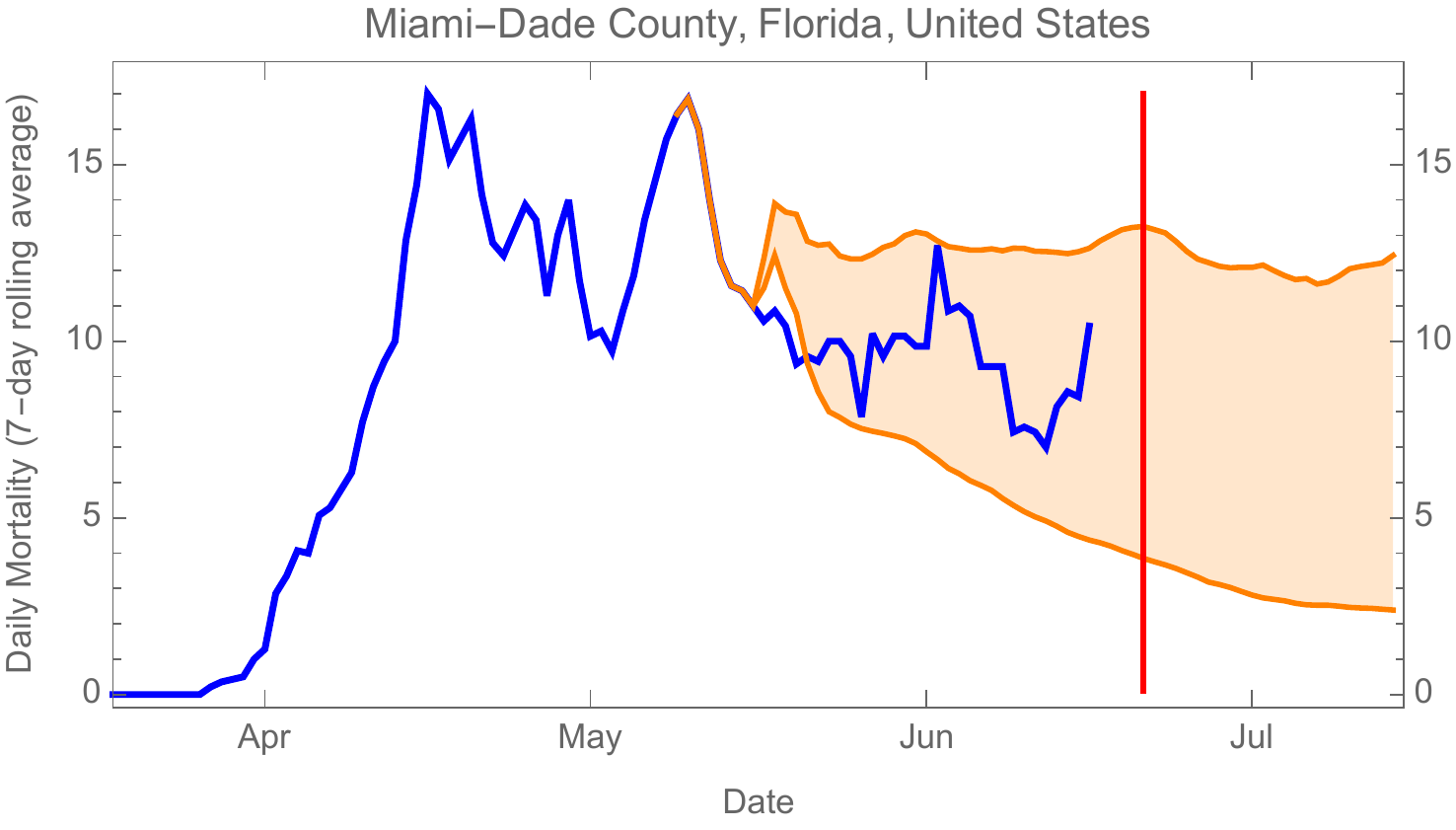}
    \caption{\label{fig:Mortality_prediction} Forecasts of COVID-19 mortality (orange) --- based on the best-fit nonlinear model to data prior to May 16th, 2020 --- versus actual reported mortality (blue) for 4 large US counties. The 68\% confidence range (orange regions) were determined from 100 random 60-day long simulations (see Supplementary Methods). The vertical red lines indicate June 21st. Forecasts for most US counties can be found at our online dashboard: \href{https://wolfr.am/COVID19Dash}{https://wolfr.am/COVID19Dash}  }
\end{figure*}
The model was very well fit to the mortality growth rate measurements for counties with a high mortality (Figure \ref{fig:I7_postdiction}). More quantitatively, the scatter of measured growth rates around the best-fit model predictions was (on average) only 13\% larger than the measurement errors, independent of the population of the county \footnote{See Supplementary Material for more detailed discussion of Error Diagnostics.}.  

Importantly, when the model was calibrated on only a subset of the data --- e.g., all but the final month for which mobility data is available --- its 68\% confidence prediction for the remaining data was accurate (Figure \ref{fig:Mortality_prediction}) given the known mobility and weather data for that final month.  This suggests that the model, once calibrated on the first wave of COVID-19 infections, can make reliable predictions about the ongoing epidemic, and future waves, in the United States.

\subsection*{Predictions for relaxed mobility restrictions, the onset of summer, and the potential second wave}

\begin{figure*}
    \centering
    \includegraphics[width=0.45\linewidth]{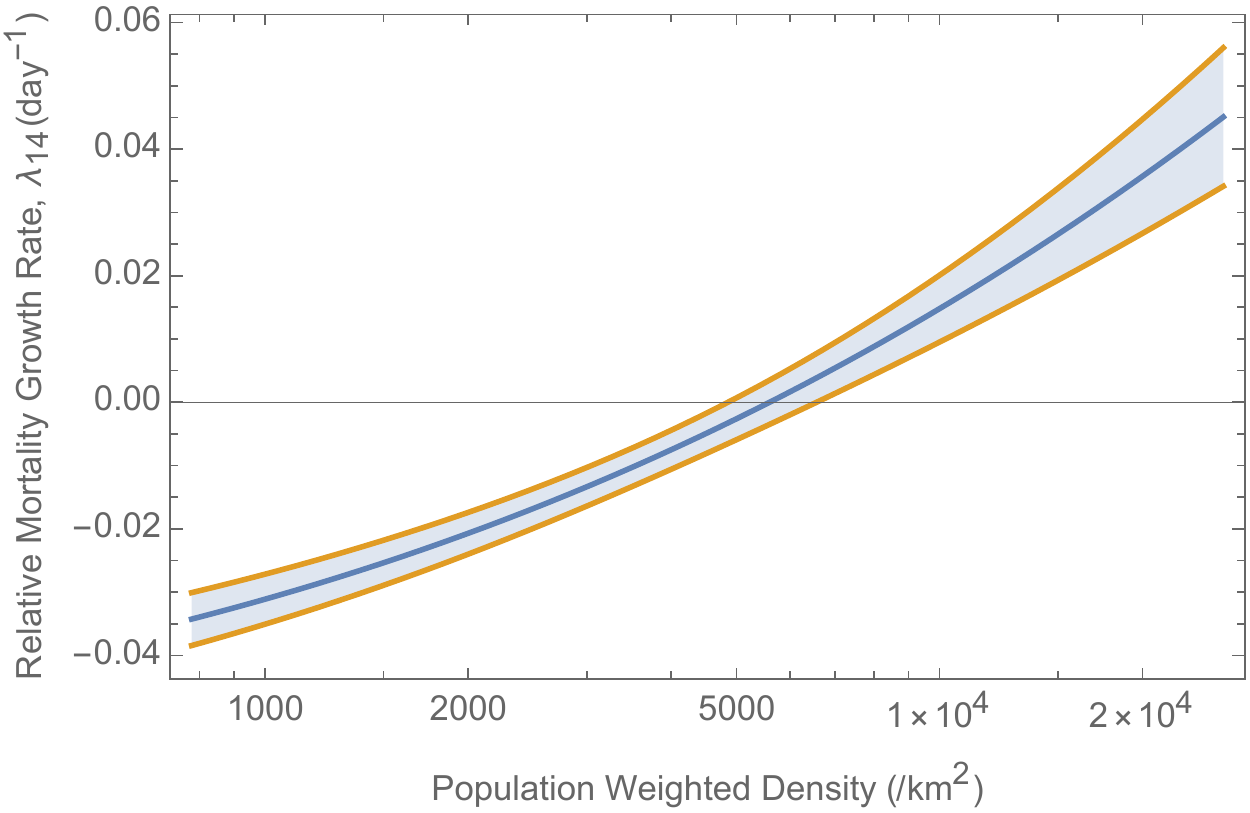}
    \includegraphics[width=0.45\linewidth]{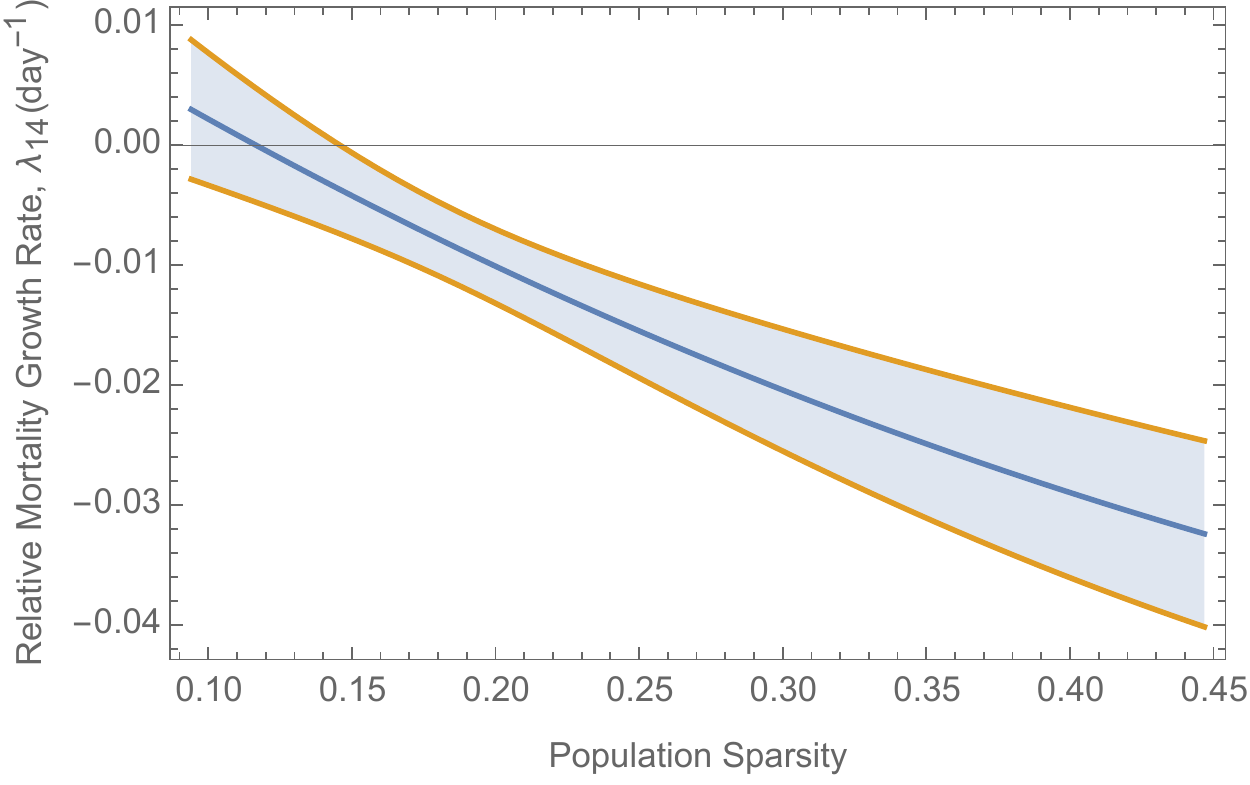}
    \includegraphics[width=0.45\linewidth]{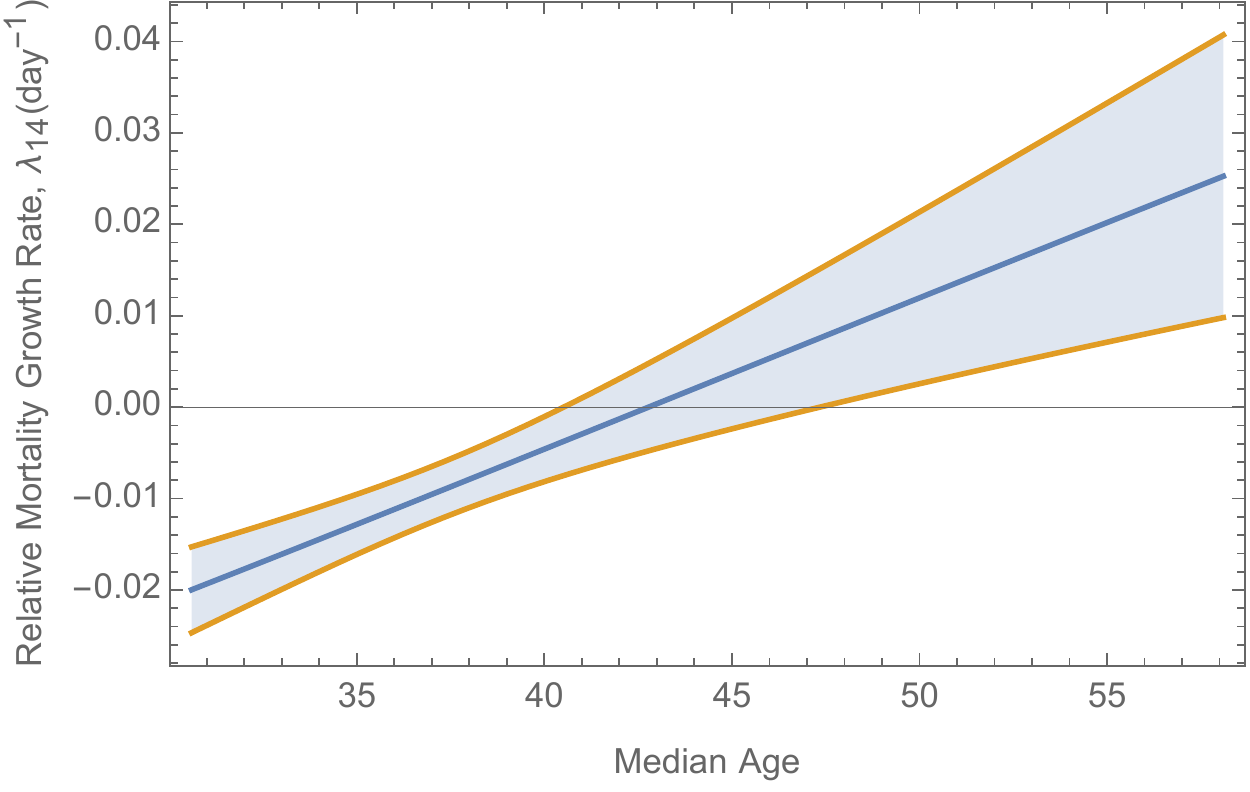}
    \includegraphics[width=0.45\linewidth]{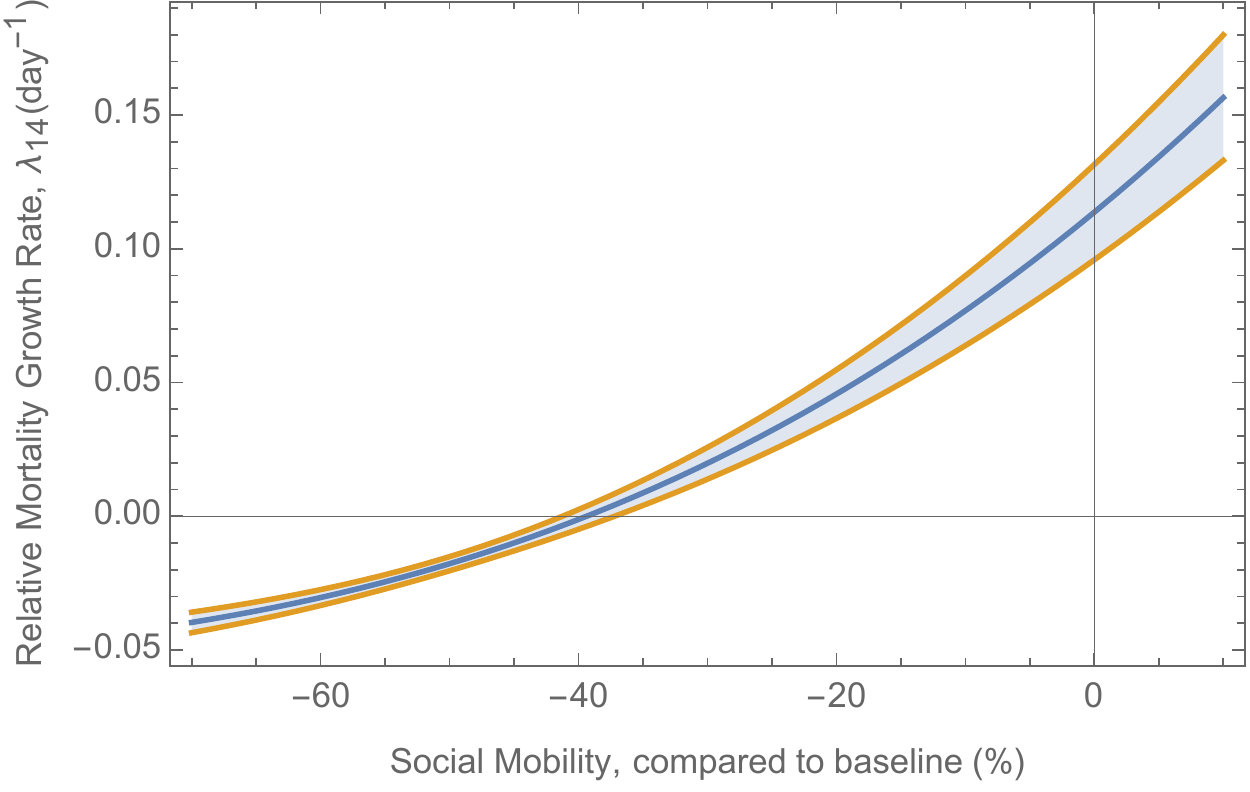}
     \includegraphics[width=0.45\linewidth]{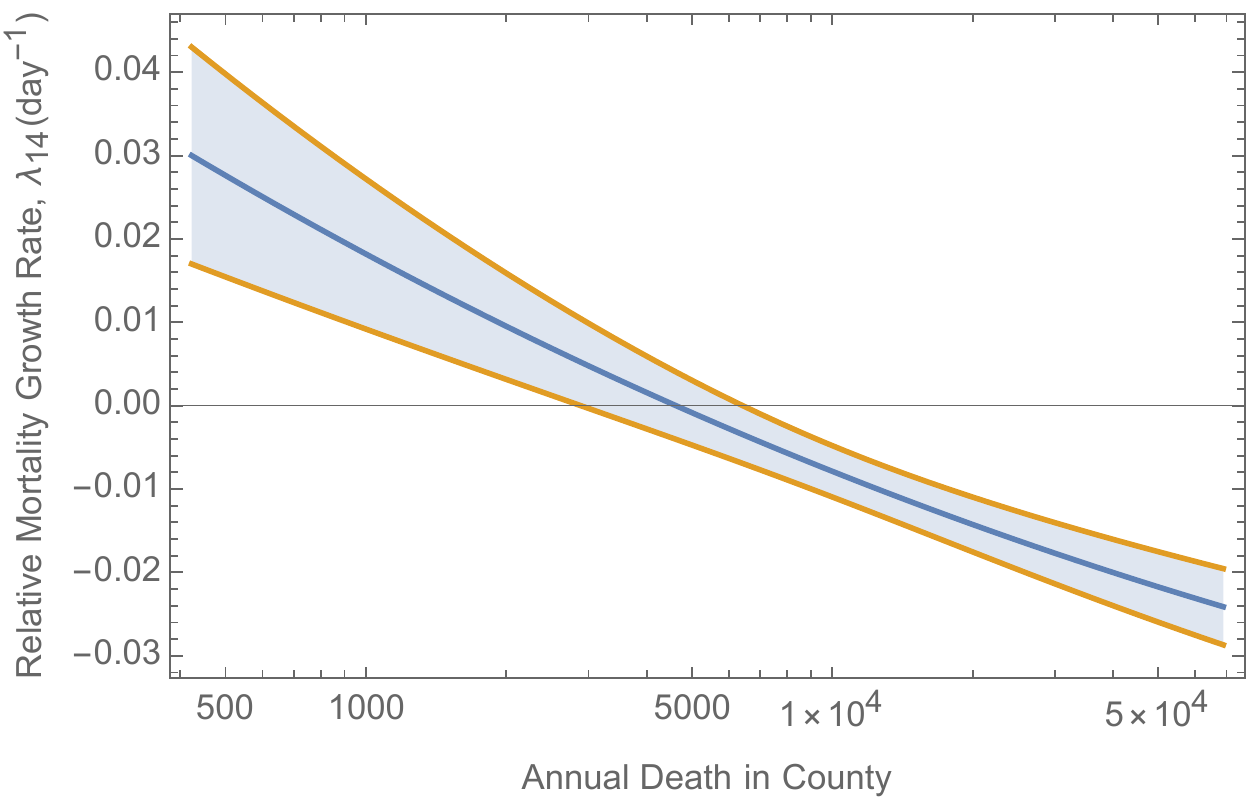}
     \includegraphics[width=0.45\linewidth]{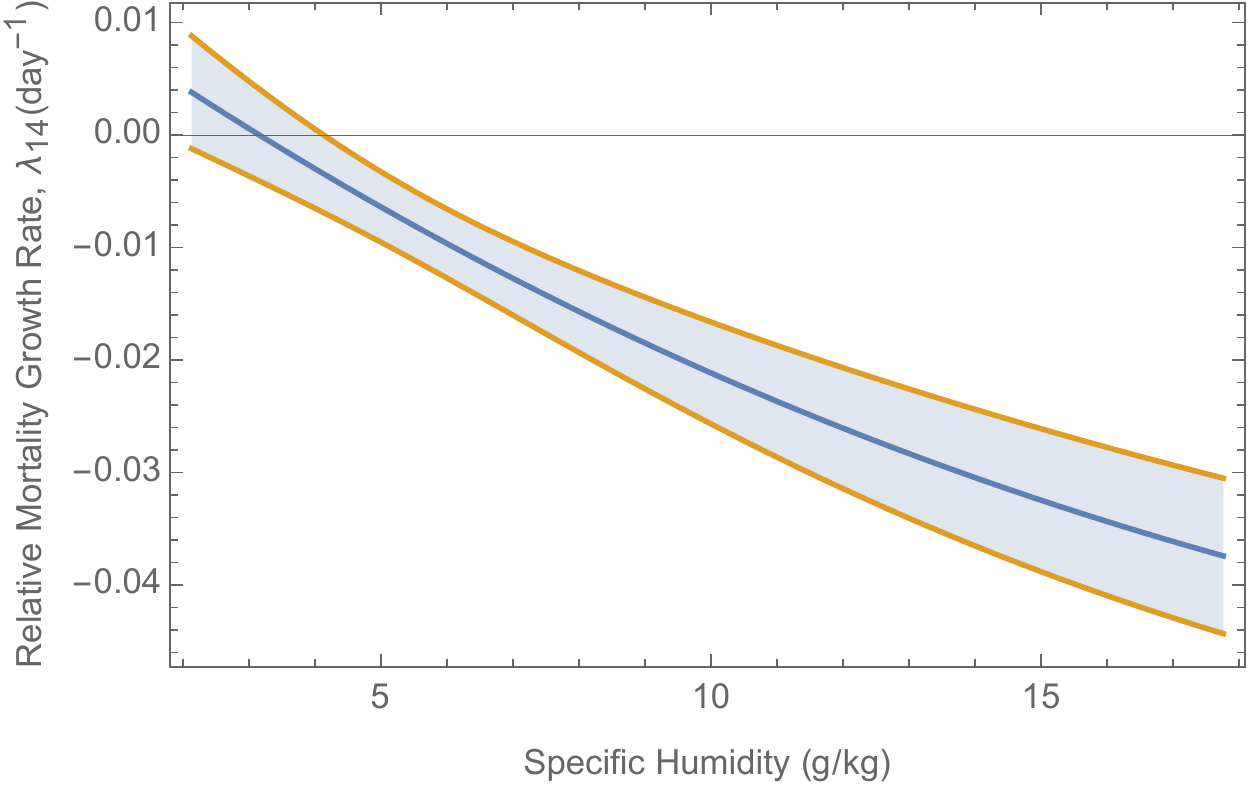}
    \includegraphics[width=0.45\linewidth]{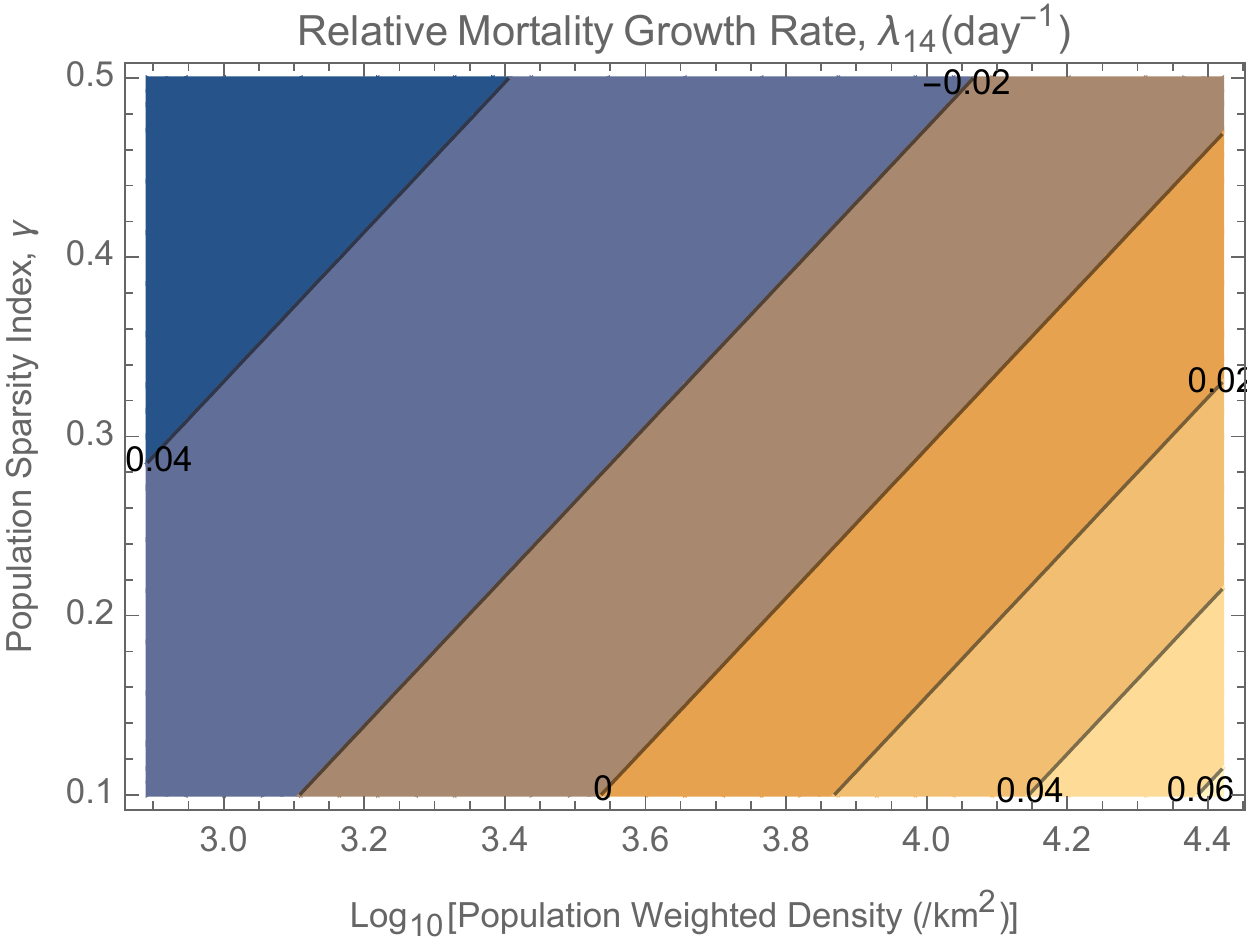}
 \includegraphics[width=0.45\linewidth]{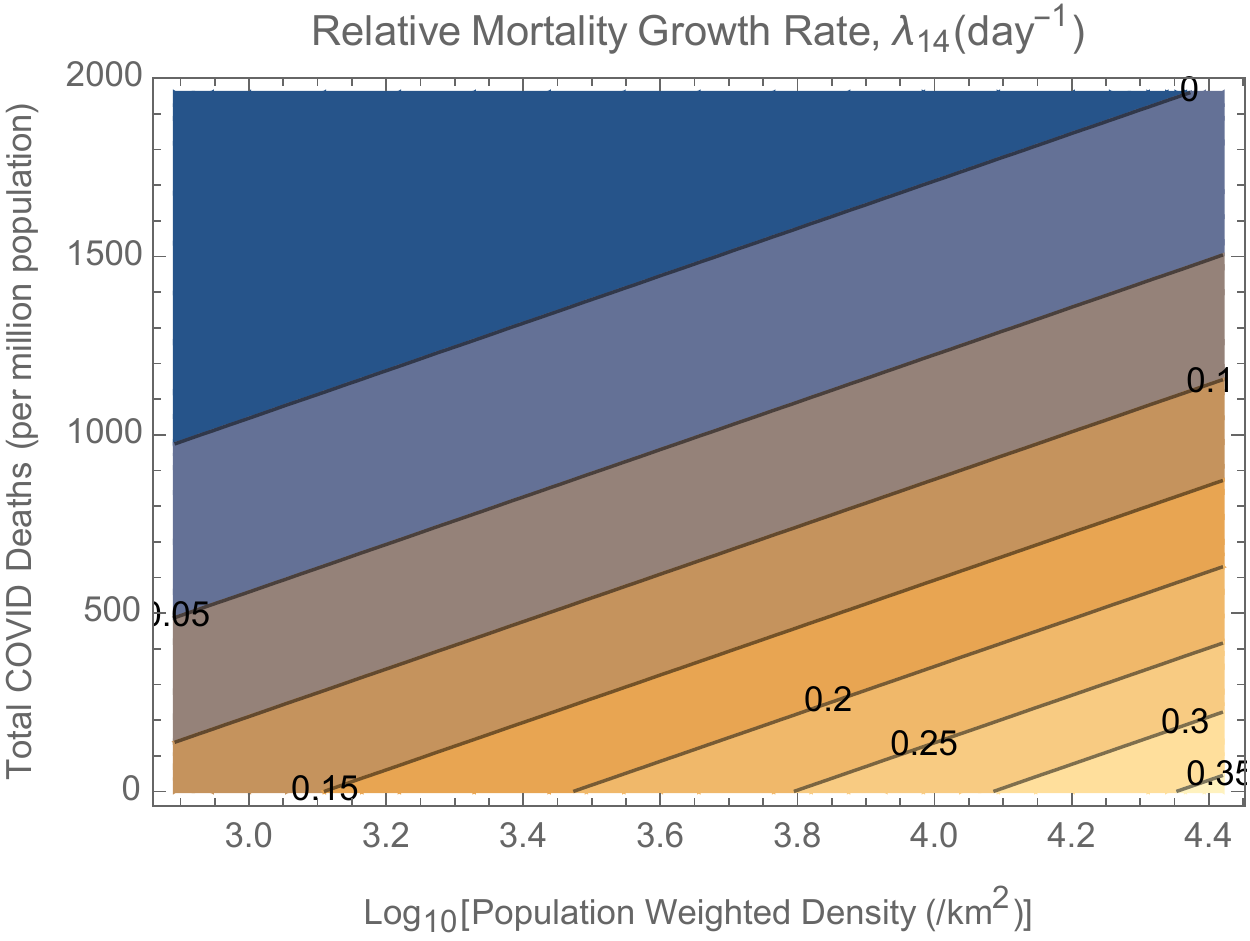}
    \caption{Dependence and 68\% confidence bands of the mortality growth rate --- as specified by the nonlinear model (Eqn.\ \ref{eq:nonlinmodel}) --- on various parameters for an ``average county.'' All parameters not being varied are fixed at their population-weighted mean values (as of 8th June, 2020): log$_{10}$[PWD / km$^{-2}$)] = 3.58, population sparsity = 0.188, COVID death fraction = $5.1 \times 10^{-4}$ (510 deaths/million population), Median Age = 37.5 yr, log(Annual Death) = 4.04, social mobility $\bar{\cal M}=$ -44\% , and specific humidity  $\bar{\cal H}$=5.7 g/kg.   }
    \label{fig:I7vEverything}
\end{figure*}

\begin{figure*}
     \includegraphics[width=0.49\linewidth]{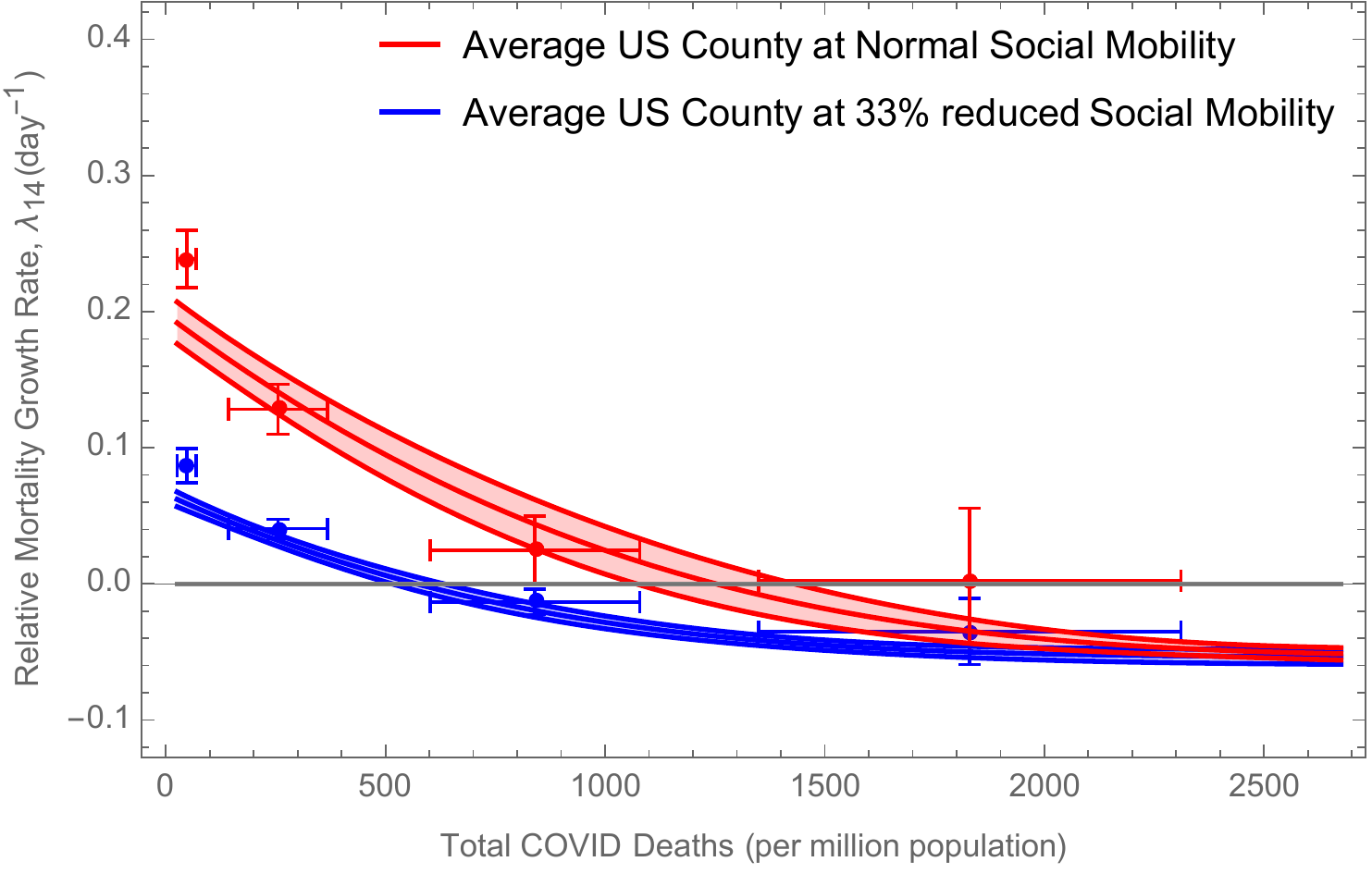}
    \includegraphics[width=0.49\linewidth]{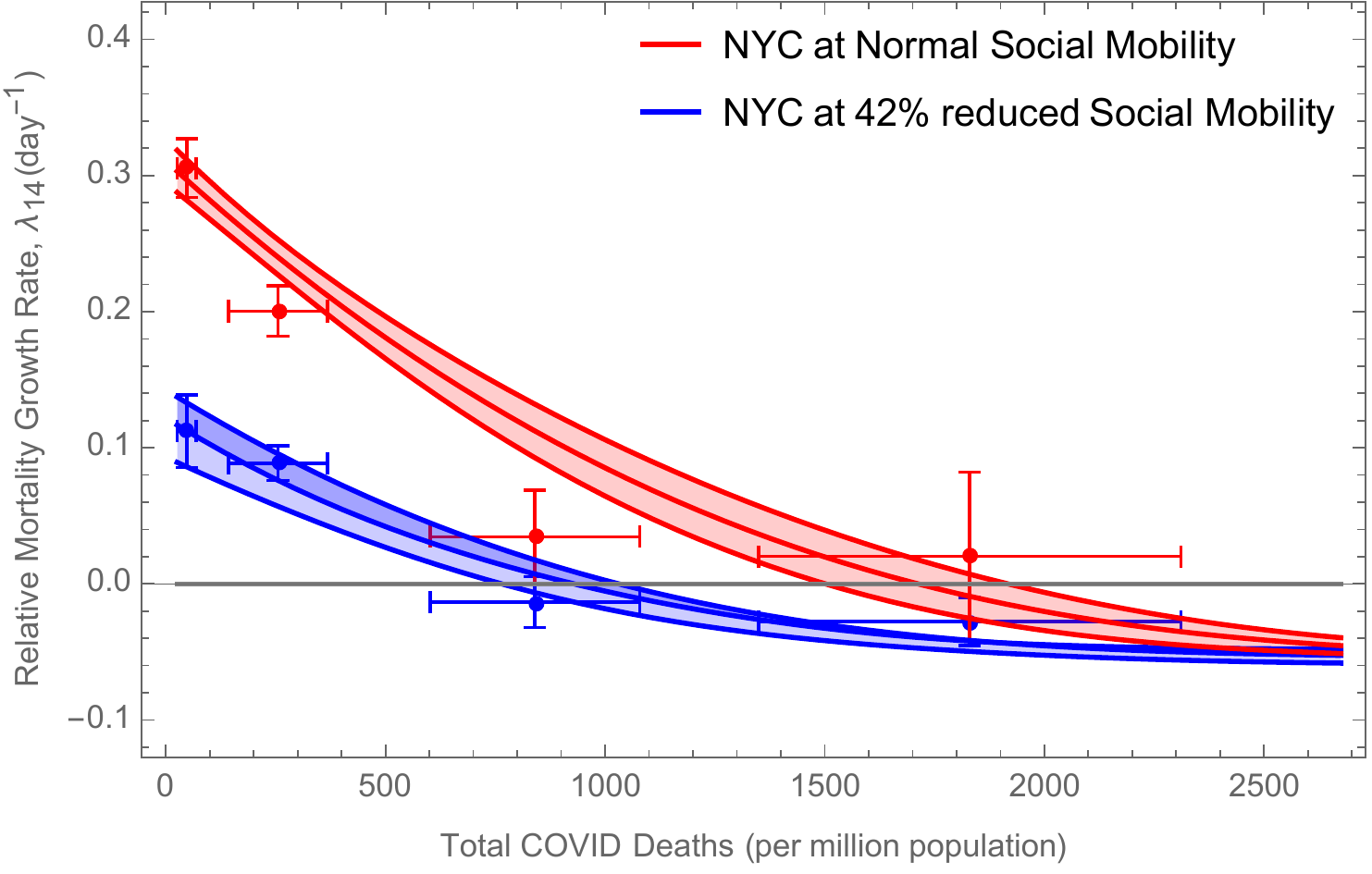}
   \includegraphics[width=0.49\linewidth]{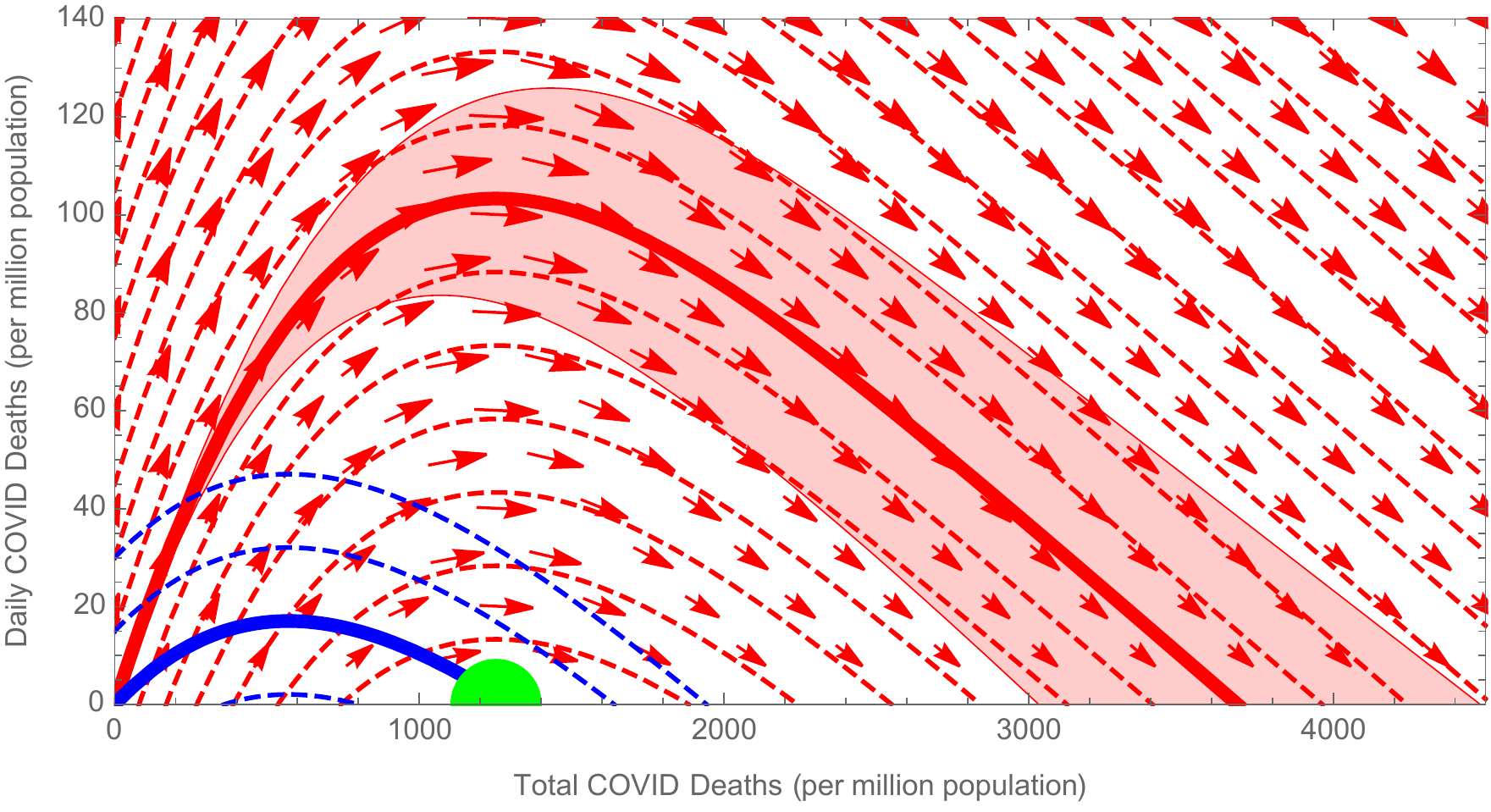}
    \includegraphics[width=0.49\linewidth]{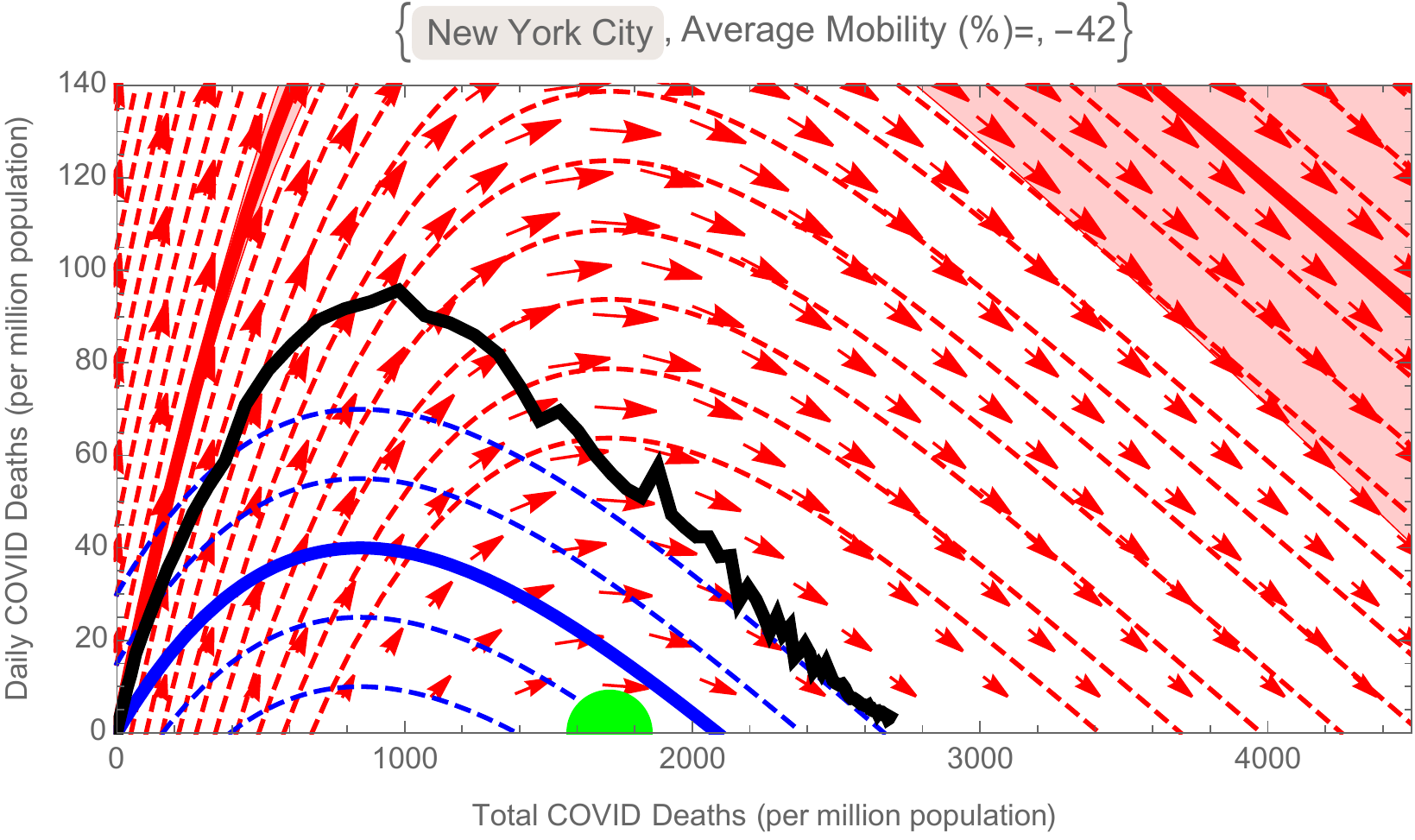}
    \caption{Nonlinear model prediction of the exponential growth rate, $\lambda_{14}$, vs.\ cumulative COVID-19 mortality (top panels), assuming baseline social mobility, $\bar{\cal M}=0$, in the ``average US county'' (see caption of Figure \ref{fig:I7vEverything}) on the left, and New York City, on the right. The curves show 68\% predictions for the nonlinear model (Table \ref{tab:nonlinear}), while the points with errorbars are linear fits to all the data within bins of death fraction.  The threshold for ``herd immunity'' ($\lambda_{14} = 0$) is reached at a mortality of approximately 1300 (1700) per million for an average county (NYC), but this would be higher in counties with more unfavorable values of the drivers. The eventual mortality burden of the average county will be determined by its path through a ``phase space'' of Daily vs.\ Total Mortality (bottom panel). An epidemic without intervention (red curves, with the particular trajectory starting at zero death shown in bold) will pass the threshold for herd immunity (1300 deaths per million; note that at zero daily deaths this is a fixed point) and continue to three times that value due to ongoing infections. A modest $33\%$ reduction in social mobility (blue curves), however, leads to mortality at ``only'' the herd immunity level (the green disk). The black curve on the bottom right panel shows the 7-day rolling average of reported mortality for NYC, which appears to have ``overshot'' the ``herd immunity threshold''.  }
    \label{fig:Herd}
\end{figure*}

\begin{figure*}
    \centering
        \includegraphics[width=0.4\linewidth]{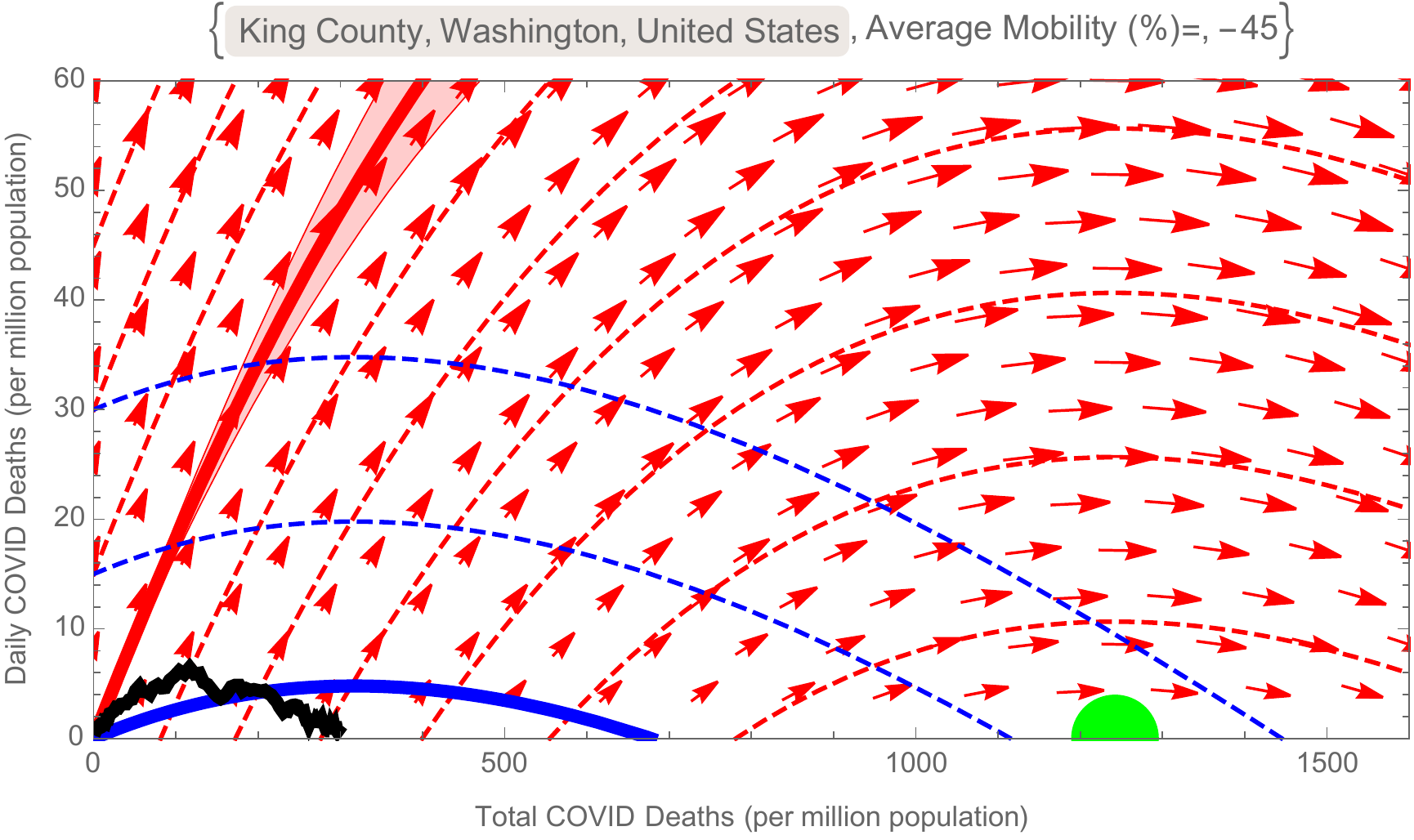}
    \includegraphics[width=0.4\linewidth]{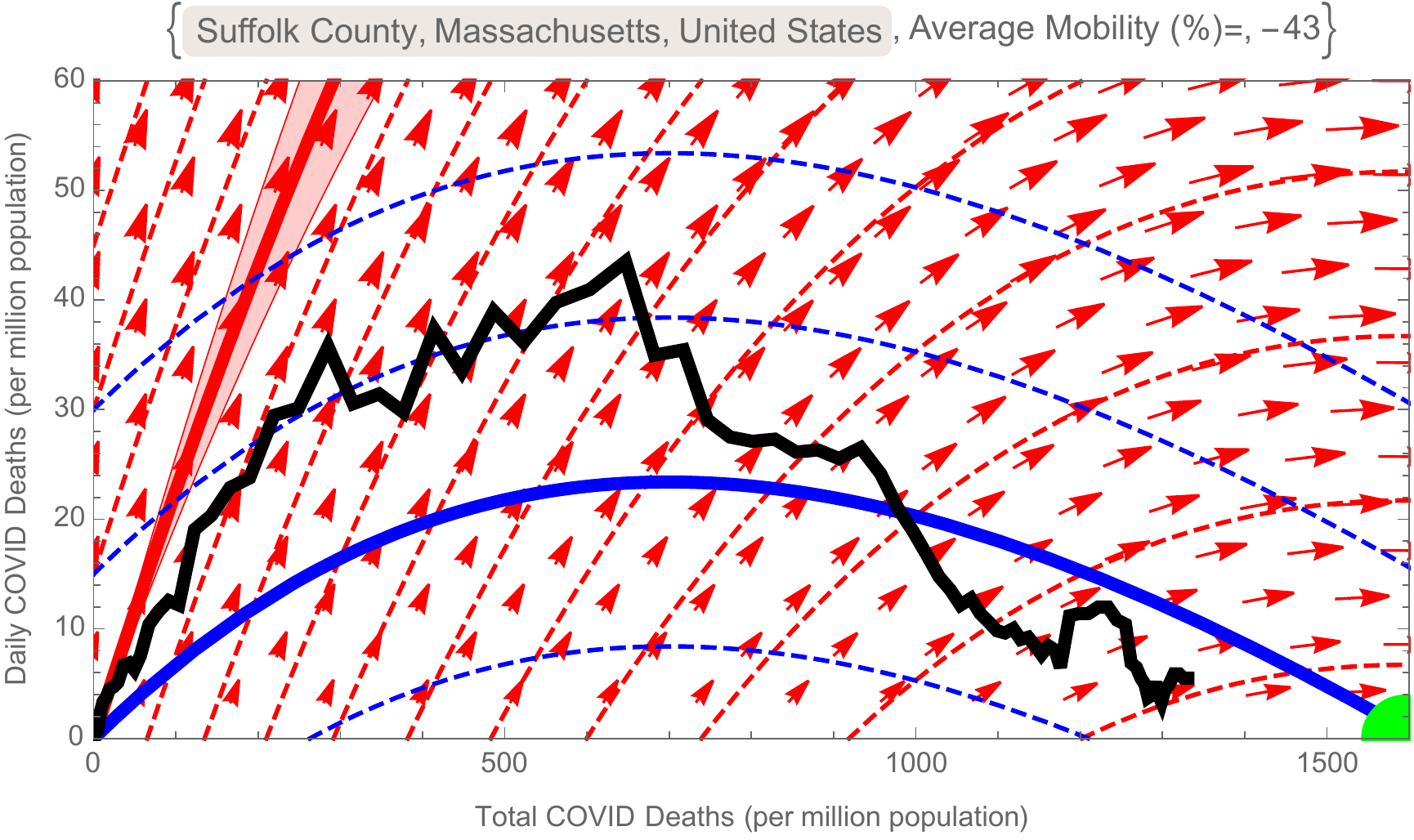}
        \includegraphics[width=0.4\linewidth]{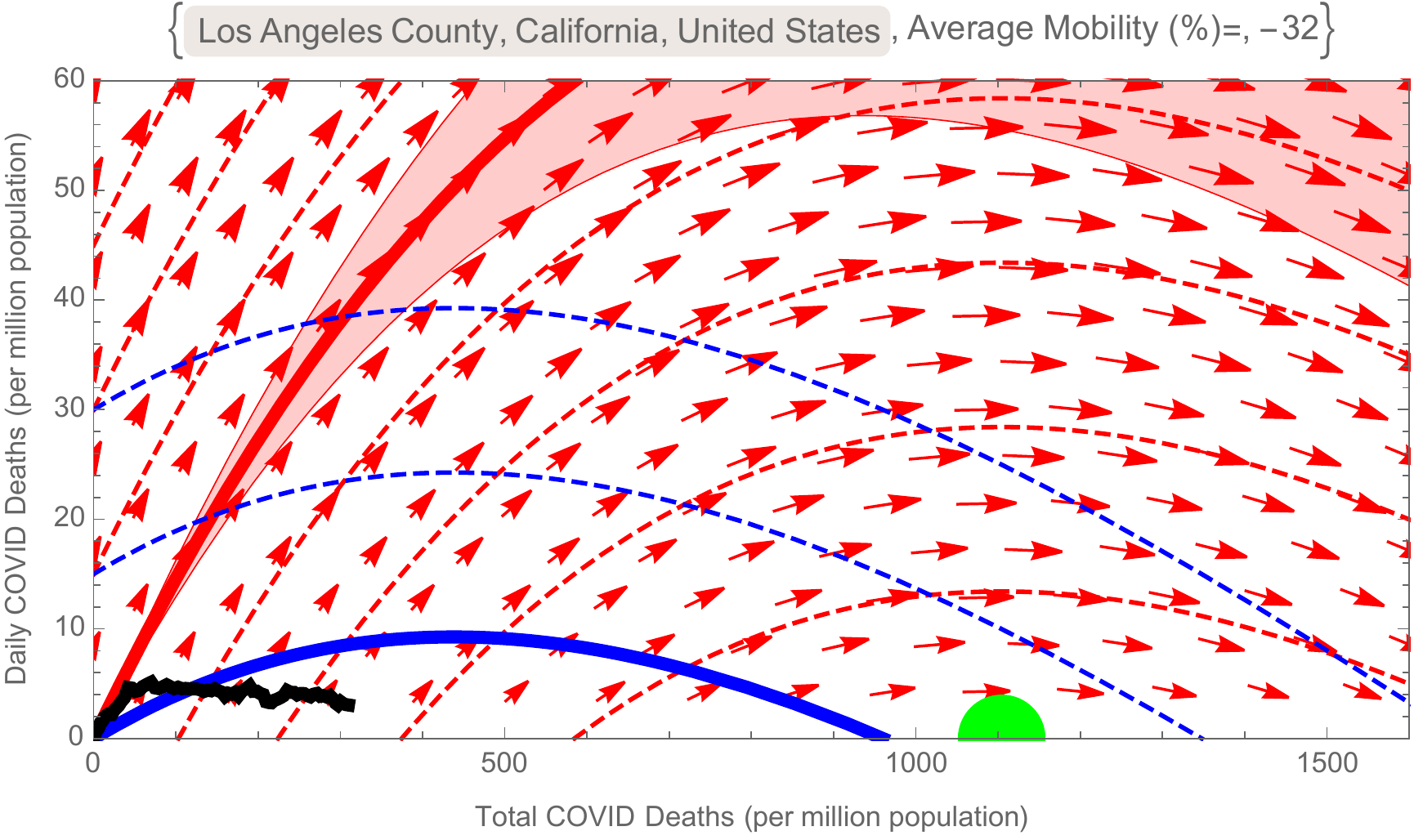}
        \includegraphics[width=0.4\linewidth]{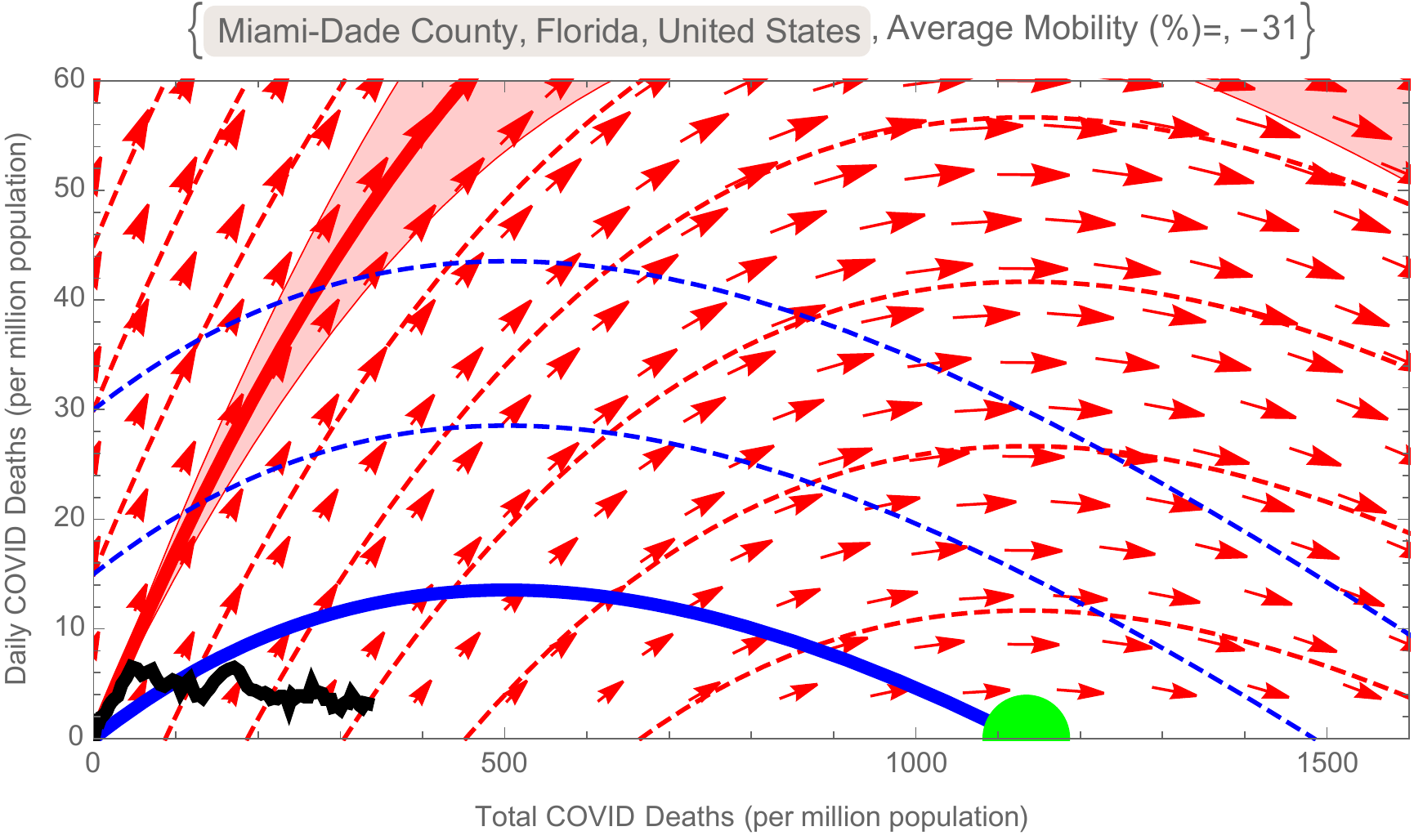}
    \caption{\label{fig:Portraits} Epidemic Phase Portraits for the same four counties as in Figure (\ref{fig:Mortality_prediction}), similar to the Phase portrait in Figure (\ref{fig:Herd}). The blue curves are for the county's average Social Mobility during Feb. 15 through June 12, 2020, while red curves/arrows are at normal (pre-covid) social mobility. The thick black curve is the 7-day rolling average of the official reported mortality, while the green disk shows the threshold for ``herd immunity''.}
\end{figure*}

\begin{figure*}
    \centering
    \includegraphics[width=\linewidth]{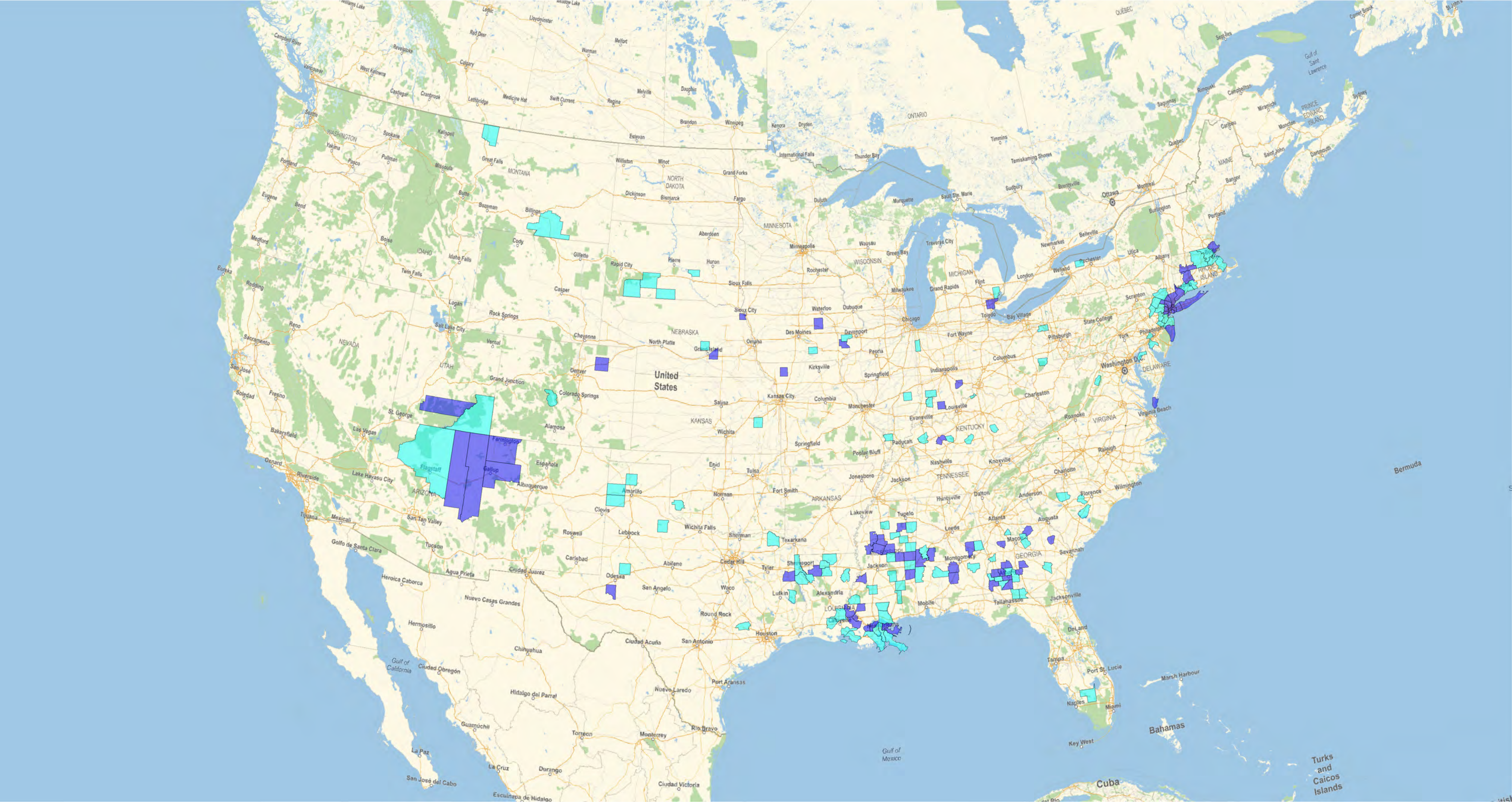}
    \includegraphics[width=\linewidth]{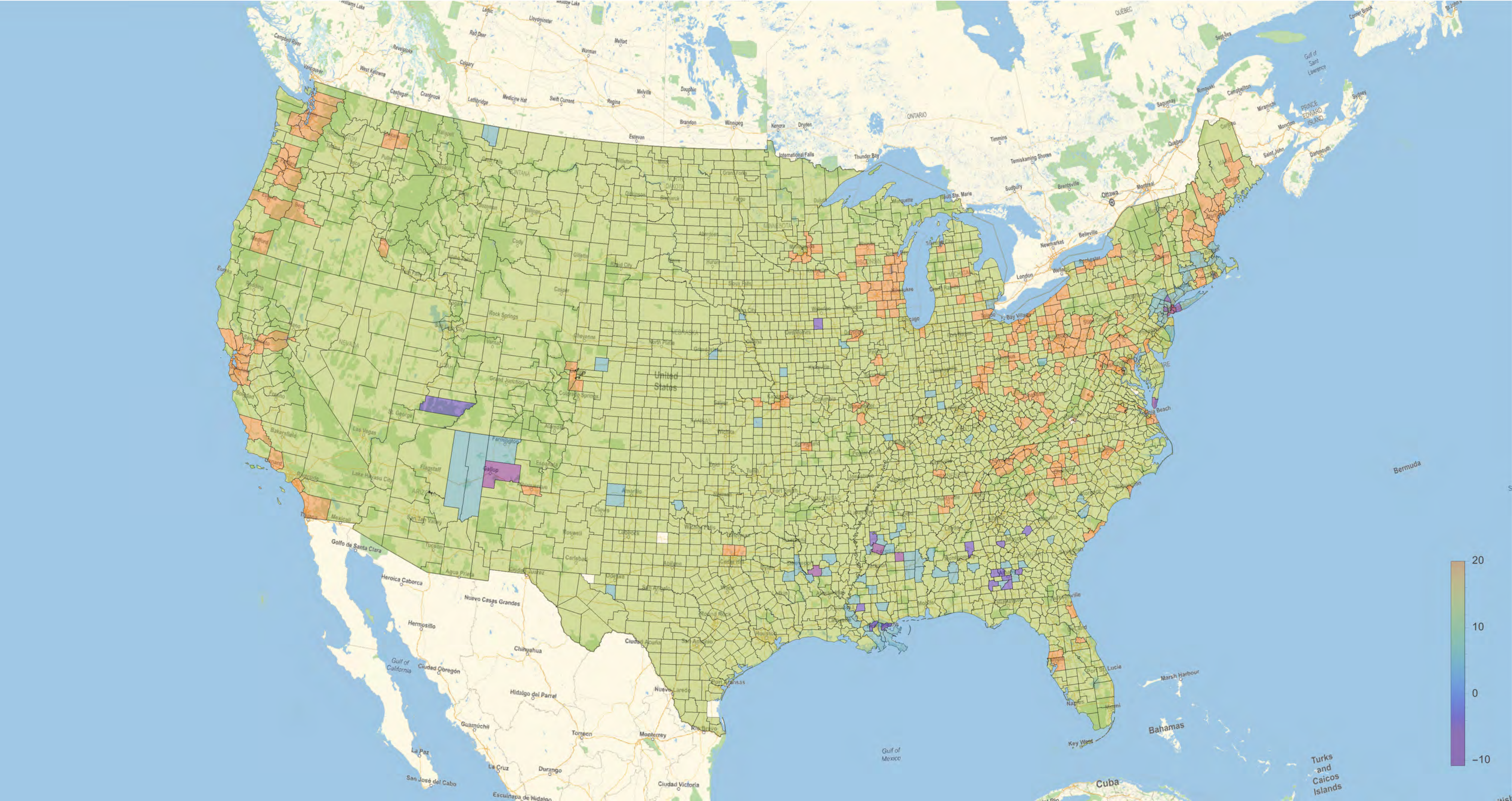}
    \caption{Top: United States counties that have passed (blue), or are within (cyan), the threshold for ``herd immunity'' at the 1-$\sigma$ level, as predicted by the nonlinear model. Bottom: Predicted confidence in the growth of COVID-19 outbreak (defined as predicted daily growth rate divided by its uncertainty), for all counties should they return today to their baseline (pre-COVID) social mobility. Counties that have approached the threshold of herd immunity have lower growth rates due to the depletion of susceptible individuals.}
    \label{fig:USMap}
\end{figure*}

%
%
%
%

Possibly the most pressing question for the management of COVID-19 in a particular community is the combination of circumstances at which the virus fails to propagate, i.e., at which the growth rate, estimated here by $\lambda_{14}$, becomes negative (or, equivalently, the reproduction number $R_t$ falls below one).  In the absence of mobility restrictions this is informally called the threshold for ``herd immunity,'' which is usually achieved by mass vaccination \citep[e.g.,][]{Herd1,Herd2}. Without a vaccine, however, ongoing infections and death will deplete the susceptible population and thus decrease transmission. Varying the parameters of the nonlinear model individually about their Spring 2020 population-weighted mean values (Figure \ref{fig:I7vEverything}) suggests that this threshold will be very much dependent on the specific demographics, geography, and weather in the community, but it also shows that reductions in social mobility can significantly reduce transmission prior the onset of herd immunity. 

To determine the threshold for herd immunity in the absence or presence of social mobility restrictions, we considered the ``average US county'' (i.e., a region with population-weighted average characteristics), and examined the dependence of the growth rate on the cumulative mortality. We found that in the absence of social distancing, a COVID-19 mortality rate of 0.13\% (or 1300 per million population) would bring the growth rate to zero. However, changing the population density of this average county shows that the threshold can vary widely (Figure \ref{fig:I7vEverything}). 

\begin{figure}
     \centering
   \includegraphics[width=\linewidth]{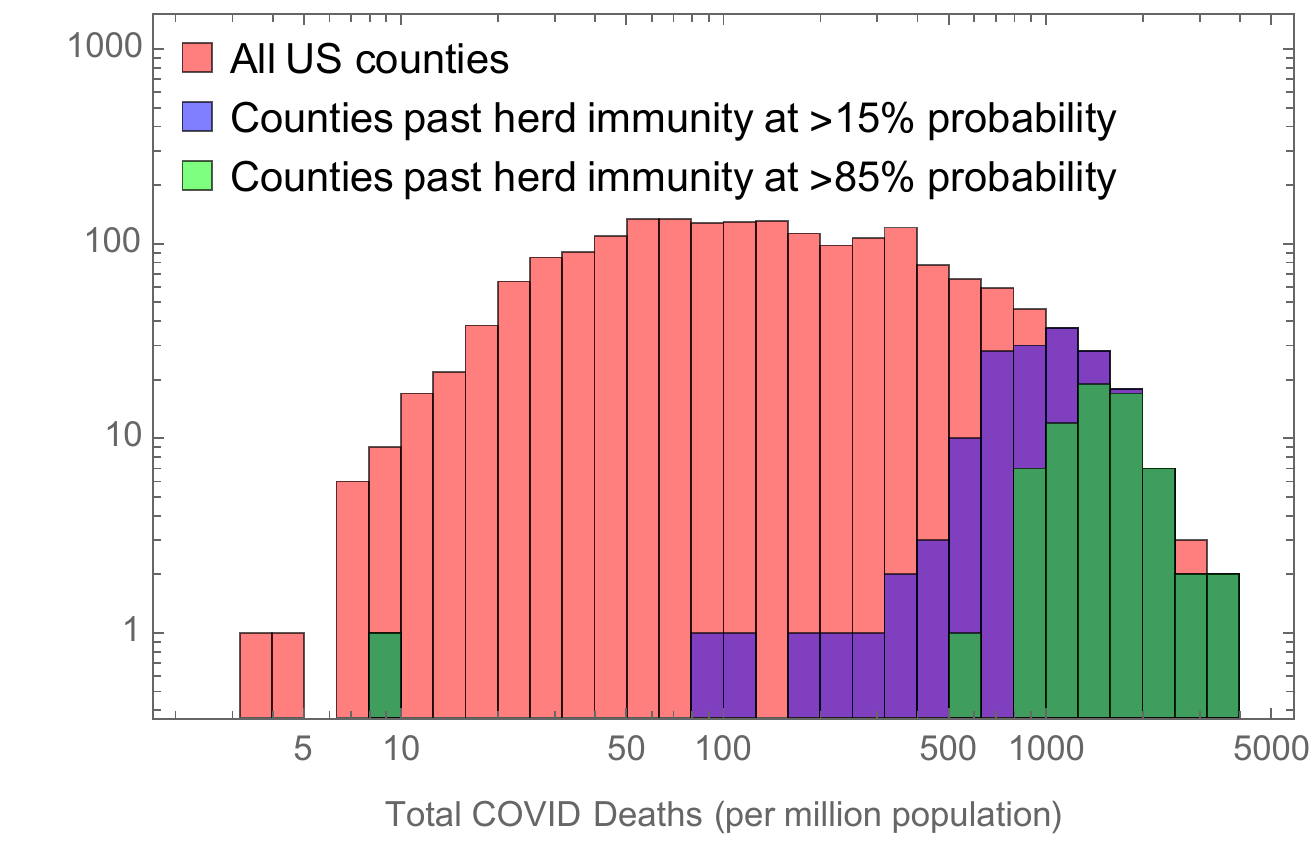}     \caption{Histogram of reported COVID-19 deaths per million for all US counties, showing the proportion that have passed ``herd immunity'' threshold, according to fit of the nonlinear model.}
     \label{fig:herd_counties_death}
\end{figure}
Examination of specific counties showed that the mortality level corresponding to herd immunity varies from 10 to 2500 per million people (Figure \ref{fig:herd_counties_death}).  At the current levels of reported COVID-19 mortality, we found that, as of June 22nd, 2020, only $128\pm 55$ out of 3142 counties (inhabiting $9.4 \pm 2.1$\% of US population) have surpassed this threshold at 68\% confidence level (Figure \ref{fig:USMap}). Notably, New York City, with the highest reported per capita mortality (2700 per million) has achieved mobility-independent herd immunity at the 10$\sigma$ confidence level, according to the model (Figure \ref{fig:Herd}). A few other large-population counties in New England, New Jersey, Michigan, Louisiana, Georgia and Mississippi that have been hard hit by the pandemic also appear to be at or close to the herd immunity threshold. This is not the case for most of the United States, however (Figure \ref{fig:USMap}). Nationwide, we predict that COVID-19 herd immunity would only occur after a death toll of  $340,000 \pm 61,000$, or $1058 \pm 190$ per million of population. 


We found that the approach to the herd immunity threshold is not direct, and that social mobility restrictions and other non-pharmaceutical interventions must be applied carefully to avoid excess mortality beyond the threshold.  In the absence of social distancing interventions, a typical epidemic will ``overshoot'' the herd immunity limit \citep[e.g.,][]{handel2007best,fung2012minimize} by up to 300\%, due to ongoing infections (Figure \ref{fig:Herd}).  At the other extreme, a very strict ``shelter in place'' order would simply delay the onset of the epidemic; but if lifted (see Figures \ref{fig:Herd} and \ref{fig:Portraits}), the epidemic would again overshoot the herd immunity threshold. A modest level of social distancing, however --- e.g.,  a 33\% mobility reduction for the average US county --- could lead to fatalities ``only'' at the level of herd immunity. Naturally, communities with higher population density or other risk factors (see Figure \ref{fig:I7vEverything}), would require more extreme measures to achieve the same.

Avoiding the level of mortality required for herd immunity will require long-lasting and effective non-pharmaceutical options, until a vaccine is available. The universal use of face masks has been suggested for reducing the transmission of SARS-CoV-2, with a recent meta-analysis \citep{chu2020physical} suggesting that masks can suppress the rate of infection by a factor of 0.07--0.34 (95\% CI), or equivalently $\Delta \ln$(transmission) $= -1.9 \pm 0.4$ (at 1$\sigma$). Using our model's dependence of the infection rate constant on mobility, this would correspond to an equivalent social mobility reduction of $\Delta \bar{\cal M}_{\rm mask} \simeq   -24\% \pm 9\%$. Warmer, more humid weather has also considered a factor that could slow the epidemic (\cite[e.g.,][]{wang2020high,2020arXiv200312417N,Xu2020.05.05.20092627}).  Annual changes in specific humidity are $\Delta \bar{\cal H} \simeq \unit{6}{\gram/\kilo\gram}$  (Figure \ref{fig:mobility}b in Supplementary Material), which can be translated in our model to an effective mobility decrease of $\Delta \bar{\cal M}_{\rm summer} \simeq -12\% \pm 5\%$. 
Combining these two effects could, in this simple analysis, yield a modestly effective defense for the summer months: $\Delta \bar{\cal M}_{\rm mask+ summer} \simeq   -37\% \pm 10\%$. Therefore, this could be a reasonable strategy for most communities to manage the COVID-19 epidemic at the aforementioned -33\% level of mobility needed to arrive at herd immunity with the least excess death. More stringent measures would be required to keep mortality below that level. Of course, this general prescription would need to be fine-tuned for the specific conditions of each community.

\section*{Discussion and Conclusions}

By simultaneously considering the time series of mortality incidence in every US county, and controlling for the time-varying effects of local social distancing interventions, we demonstrated for the first time a dependence of the epidemic growth of COVID-19 on population density, as well as other climate, demographic, and population factors.  We further constructed a realistic, but simple, first-principles model of infection transmission that allowed us to extend our heuristic linear model of the dataset into a predictive nonlinear model, which provided a better fit to the data (with the same number of parameters), and which also accurately predicted late-time data after training on only an earlier portion of the data set. This suggests that the model is well-calibrated to predict future incidence of COVID-19, given realistic predictions/assumptions of future intervention and climate factors. We summarized some of these predictions in the final section of Results, notably that only a small fraction of US counties (with less than ten percent of the population) seem to have reached the level of herd immunity, and that relaxation of mobility restrictions without counter-measures (e.g., universal mask usage) will likely lead to increased daily mortality rates, beyond that seen in the Spring of 2020.

In any epidemiological model, the infection rate of a disease is assumed proportional to population density \citep{dejong_diekmann_heesterbaak1995}, but, to our knowledge, its explicit effect in a real-world respiratory virus epidemic has not been demonstrated.  The universal reach of the COVID-19 pandemic, and the diversity of communities affected have provided an opportunity to verify this dependence. Indeed, as we show here, it must be accounted for to see the effects of weaker drivers, such as weather and demographics.  A recent study of COVID-19 in the United States, working with a similar dataset, saw no significant effect due to population density \citep{hamidi2020}, but our analysis differs in a number of important ways.  First, we have taken a dynamic approach, evaluating the time-dependence of the growth rate of mortality incidence, rather than a single static measure for each county, which allowed us to account for the changing effects of weather, mobility, and the density of susceptible individuals.  Second, we have included an explicit and real-time measurement of social mobility, i.e., cell phone mobility data provided by Google \citep{google2020}, allowing us to control for the dominant effect of intervention.  Finally, and perhaps most importantly, we calculate for each county an estimate of the ``lived'' population density, called the population-weighted population density (PWD) \citep{craig1985}, which is more meaningful than the standard population per political area.  As with any population-scale measure, this serves as a proxy ---  here, for estimating the average rate of encounters between infectious and susceptible people --- but we believe that PWD is a better proxy than standard population density, and it is becoming more prevalent, e.g., in census work \citep{dorling_atkins1995,wilson2012}.

We also found a significant dependence of the mortality growth rate on specific humidity (although since temperature and humidity were highly correlated, a replacement with temperature was approximately equivalent), indicating that the disease spread more rapidly in drier (cooler) regions.  There is a large body of research on the effects of temperature and humidity on the transmission of other respiratory viruses \citep{moriyama2020, kudo2019}, specifically influenza \citep{barreca_shimshack2012}.   Influenza was found to transmit more efficiently between guinea pigs in low relative-humidity and temperature conditions \citep{lowen2007}, although re-analysis of this work pointed to absolute humidity (e.g., specific humidity) as the ultimate controller of transmission \citep{shaman2009} . Although the mechanistic origin of humidity's role has not been completely clarified, theory and experiments have suggested a snowballing effect on small respiratory droplets that cause them to drop more quickly in high-humidity conditions \citep{tellier2009, noti2013, marr2018}, along with a role for evaporation and the environmental stability of virus particles \citep{morawska2005, marr2018}.  It has also been shown that the onset of the influenza season \citep{shaman2010, shaman2011} --- which generally occurs between late-Fall and early-Spring, but is usually quite sharply peaked for a given strain (H1N1, H3N2, or Influenza B) --- and its mortality \citep{barreca_shimshack2012} are linked to drops in absolute humidity.  It is thought that humidity or temperature could be the annual periodic driver in the resonance effect causing these acute seasonal outbreaks of influenza \citep{dushoff2004, tamerius2011}, although other influences, such as school openings/closings have also been implicated \citep{earn2012}. While little is yet known about the transmission of SARS-CoV-2 specifically, other coronaviruses are known to be seasonal \citep{moriyama2020, neher2020}, and there have been some preliminary reports of a dependence on weather factors \citep{xu2020, schell2020}.  We believe that our results represent the most definitive evidence yet for the role of weather, but emphasize that it is a weak, secondary driver, especially in the early stages of this pandemic where the susceptible fraction of the population remains large \citep{baker2020}.  Indeed, the current early-summer rebound of COVID-19 in the relatively dry and hot regions of the Southwest suggests that the disease spread will not soon be controlled by seasonality.

We developed a new model of infection in the framework of a renewal equation  (see, e.g., \cite{champredon2018} and references therein), which we could formally solve for the exponential growth rate.  The incubation period in the model was determined by a random walk through the stages of infection, yielding a non-exponential distribution of the generation interval, thus imposing more realistic delays to infectiousness than, e.g., the standard SEIR model. In this formulation, we did not make the standard compartmental model assumption that the infection of an individual induces an autonomous, sequential passage from exposure, to infectiousness, to recovery or death; indeed, the model does not explicitly account for recovered or dead individuals. This freedom allows for, e.g., a back passage from infectious to noninfectious (via the underlying random walk) and a variable rate of recovery or death.  We assumed only that the exponential growth in mortality incidence matched (with delay) that of the infected incidence --- the primary dynamical quantity in the renewal approach --- and we let the cumulative dead count predict susceptible density --- the second dynamical variable in the renewal approach ---  under the assumption that deaths arise from a distinct subset of the population, with lower mobility behavior than those that drive infection (see Supplementary Material).   Therefore, we fitted the model to the (rolling two-week estimates of the) COVID-19 mortality incidence growth rate values, $\lambda_{14}$, for all counties and all times, and used the per capita mortality averaged over that period, $f_D$,  to determine susceptible density.  Regression to this nonlinear model was much improved over linear regression, and, once calibrated on an early portion of the county mortality incidence time series, the model accurately predicted the remaining incidence. 

Because we accounted for the precise effects of social mobility in fitting our model to the actual epidemic growth and decline, we were then able to, on a county-by-county basis, ``turn off'' mobility restrictions and estimate the level of cumulative mortality at which SARS-CoV-2 would fail to spread even without social distancing measures, i.e., we estimated the threshold for ``herd immunity.'' Meeting this threshold prior to the distribution of a vaccine should not be a goal of any community, because it implies substantial mortality, but the threshold is a useful benchmark to evaluate the potential for local outbreaks following the first wave of COVID-19 in Spring 2020.  We found that a few counties in the United States have indeed reached herd immunity in this estimation --- i.e., their predicted mortality growth rate, assuming baseline mobility, was negative --- including counties in the immediate vicinity of New York City, Detroit, New Orleans, and Albany, Georgia.   A number of other counties were found to be at or close to the threshold, including much of the greater New York City and Boston areas, and the Four Corners, Navajo Nation, region in the Southwest.  All other regions were found to be far from the threshold for herd immunity, and therefore are susceptible to ongoing or restarted outbreaks. These determinations should be taken with caution, however. In this analysis, we estimated that the remaining fraction of susceptible individuals in the counties at or near the herd immunity threshold was in the range of 0.001\% to 5\% (see Supplementary Materials).  This is in strong tension with initial seroprevalence studies \citep{rosenberg2020,havers2020} which placed the fraction of immune individuals in New York City at 7\% in late March and 20\% in late April, implying that perhaps 75\% of that population remains susceptible today. We hypothesize that the pool of susceptible individuals driving the epidemic in our model is a subset of the total population --- likely those with the highest mobility and geographic reach --- while a different subset, with very low baseline mobility, contributes most of the mortality  (see Supplementary Material).  Thus, the near total depletion of the susceptible pool we see associated with herd immunity corresponds to the highly-mobile subset, while the low-mobility subset could remain largely susceptible.   One could explicitly consider such factors of population heterogeneity in a model --- e.g., implementing a saturation of infectivity as a proxy for a clustering effect \citep{capasso1978,mollison1985,deboer2007,farrell2019} --- but we found (in results not shown) that the introduction of additional of parameters left portions of the model unidentifiable. Despite these cautions, it is interesting to note that the epidemic curves (mortality incidence over time) for those counties that we have predicted an approach to herd immunity are qualitatively different than those we have not. Specifically, the exponential rise in these counties is followed by a peak and a sharp decline --- rather than the flattening seen in most regions --- which is a typical feature of epidemic resolution by susceptible depletion. 

At the time of this writing, in early Summer 2020, confirmed cases are again rising sharply in many locations across the United States --- particularly in areas of the South and West that were spared significant mortality in the Spring wave. The horizon for an effective and fully-deployed vaccine still appears to be at least a year away.  Initial studies of neutralizing antibodies in recovered COVID-19 patients, however, suggest a waning immune response after only 2--3 months, with 40\% of those that were asymptomatic becoming seronegative in that time period \citep{long2020}. Although the antiviral remdesivir \citep{beigel2020, grein2020, wang2020} and the steroid Dexamethasone \citep{horby2020} have shown some promise in treating COVID-19 patients, the action of remdesivir is quite weak, and high-dose steroids can only be utilized for the most critical cases. Therefore, the management of this pandemic will likely require non-pharmaceutical intervention --- including universal social distancing and mask-wearing, along with targeted closures of businesses and community gathering places --- for years in the future.  The analysis and prescriptive guidance we have presented here should help to target these approaches to local communities, based on their particular demographic, geographic, and climate characteristics, and can be facilitated through our \href{https://wolfr.am/COVID19Dash}{online simulator dashboard}.  Finally, although we have focused our analysis on the United States, due to the convenience of a diverse and voluminous data set, the method and results should be applicable to any community worldwide, and we intend to extend our analysis in forthcoming work.

\section*{Acknowledgement}

We are indebted to helpful comments and discussions by our colleagues, in particular Bruce Bassett, Ghazal Geshnizjani, David Spergel, and Lee Smolin.  NA is partially supported by Perimeter Institute for Theoretical Physics. Research at Perimeter Institute is supported in part by the Government of Canada through the Department of Innovation, Science and Economic Development Canada and by the Province of Ontario through the Ministry of Colleges and Universities.








\printbibliography[title={Bibliography}] 
\newpage

\section*{Supplementary Material}

\subsection*{Data \& Methods}

\subsubsection*{Datasets, Resources, Definitions}

\begin{figure}
    \centering
    \includegraphics[width=\linewidth]{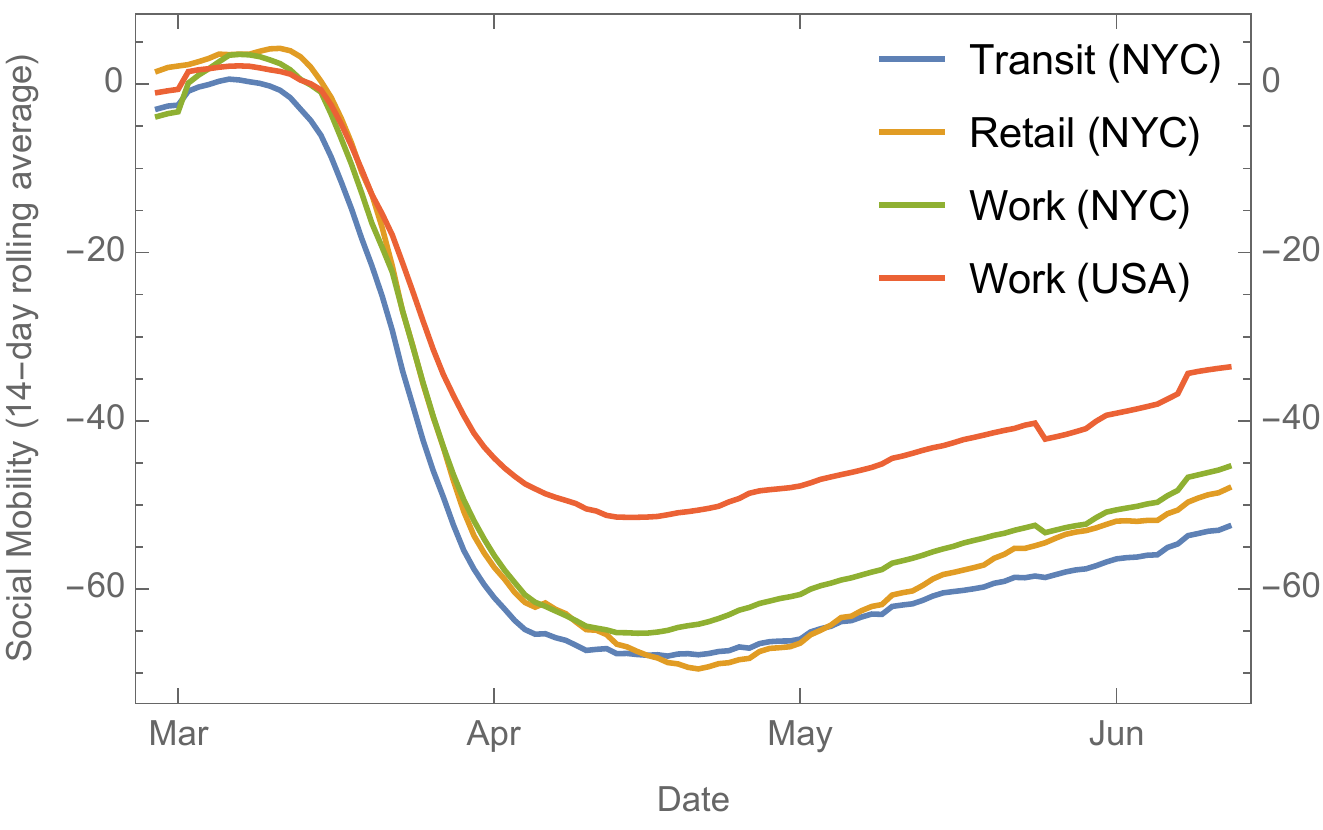}
    \includegraphics[width=\linewidth]{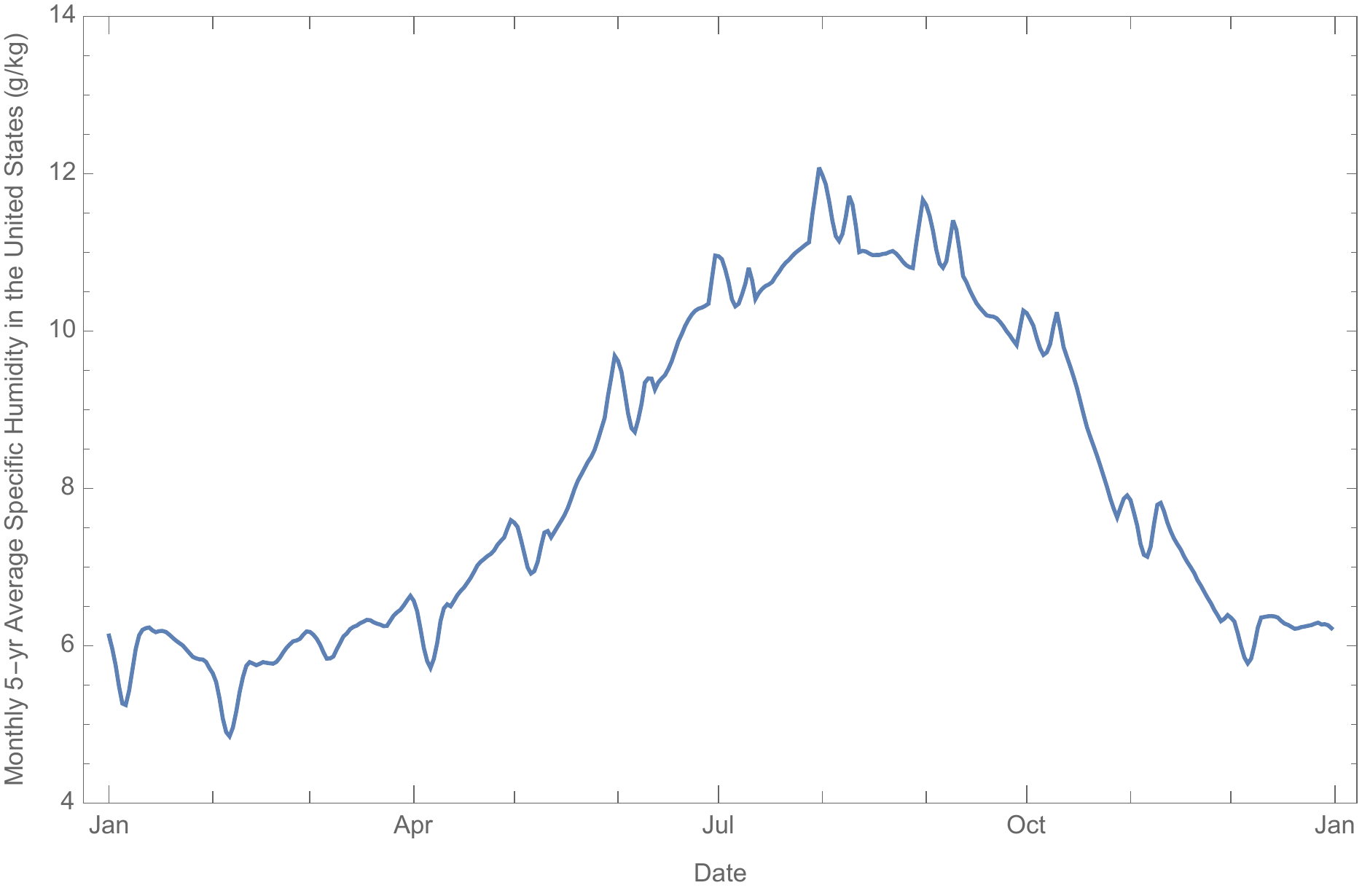}
    \caption{(Top)(a)The 14-day rolling average of (population-weighted) social mobility \citep{google2020} for NYC, as well as all US counties considered here. For our analysis, we only use ``work places'' as an indicator, as others do not appear to show any {\it independent} correlation with $\lambda_{14}(t)$. 
    (Bottom)(b)
    28-day moving average of historical annual specific humidity in the United States (weighted-averaged by population). }
    \label{fig:mobility}
\end{figure}

All data for cases and mortality, demographics, mobility, and weather were incorporated into the publicly-available Wolfram COVID-19 Dataset and the Wolfram|Alpha Knowledgebase  \citep{wolfram2020}. COVID-19 confirmed case and mortality data were obtained from both the NYTimes and Johns Hopkins University Github repositories \citep{JHU2020, NYTimes2020}; the former was used for the analysis initial case data in metropolitan regions, while the average of the two data sets was used for all other analyses. In each case, daily confirmed counts were utilized. Demographic data by county, including people per household, estimated 2019 population, annual births, and annual deaths were obtained from the US Census 2019 {\em County Population Estimates} data set \citep{census2019a}. Median ages were determined from the US Census 2018 {\em County Characteristics Resident Population Estimates} data set \citep{census2018}.  For the Median Age, Wolfram|Alpha has curated the raw data from \textit{United States Census Bureau, American Community Survey 5-Year Estimates: B01002, the Median Age By Sex, American FactFinder}; for the People per Household and Annual Death, the source of curated data is \textit{United States Census Bureau, State \& County QuickFact}s. County outline polygons were obtained from the US Census 2019 {\em TIGER/Line shapefiles} database \citep{census2019b}. Local weather data (Figure \ref{fig:mobility} was obtained from the NOAO {\em Global Surface Summary of the Day} (GSOD) database \citep{noao2020}. The nearest WBAN station with daily dew point and pressure values (for calculation of specific humidity), and daily average temperature was chosen for each county or metropolitan region. Weather data was averaged over a two-week period for $\lambda_{14}$, and over a window equal to the growth period for metropolitan regions.  Google's {\em COVID-19 Community Mobility Reports} dataset \citep{google2020}, specifically ``Workplace mobility,'' was used to estimate the human social mobility in each county (Figure \ref{fig:mobility}).

Population-weighted population density (or, population weighted density, PWD) \citep{craig1985,wilson2012, dorling_atkins1995}, was calculated using the Global Human Settlement Population raster dataset  \citep{GHS2019}, which contains $\unit{250}{\meter}$-resolution population values worldwide, taken from census data. The value of PWD for a county --- or for a set of counties, in the metropolitan region analysis --- was calculated as the population-weighted average of density over all $(\unit{250}{\meter})^2$-area pixels contained within the region, i.e.,
	\begin{equation}
	{\rm PWD} = \sum_j \frac{\left(p_j/a_j\right) p_j}{\sum_i p_i}\,,
	\label{eq:PWPD}
	\end{equation}
where $p_j$ is the value (i.e., the population) of the $j$th pixel, $a_j = \unit{0.0625~}{\kilo\meter}^2$ is the area of each pixel (the GHS-POP image uses the equal-area Molleweide projection), and $\sum_i p_i$ is the total population of the region. This measure has also been called the {\em lived population density} because it is the population density experienced by the average person.

\begin{figure}
    \centering
    \includegraphics[width=\linewidth]{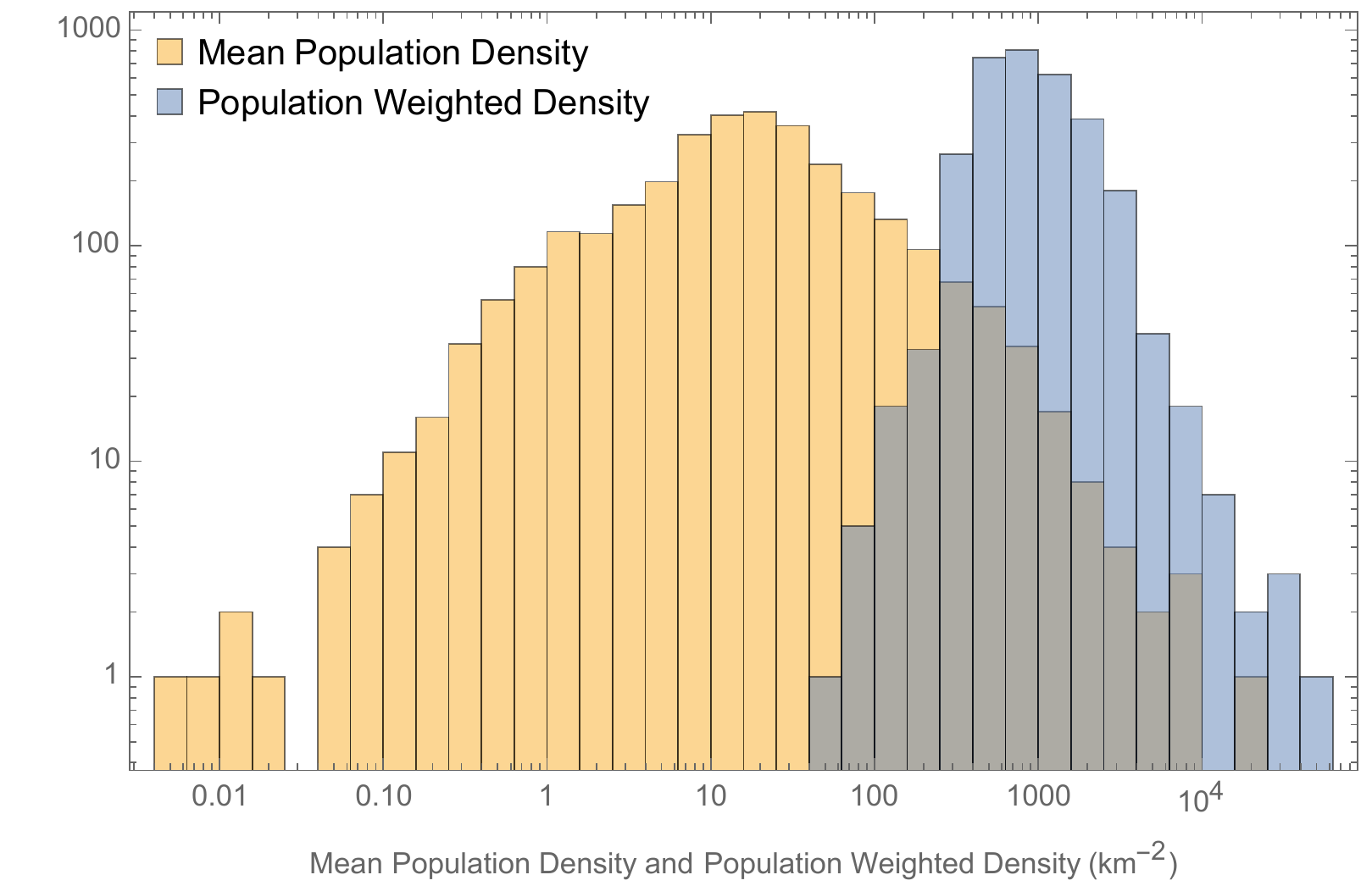}
        \includegraphics[width=\linewidth]{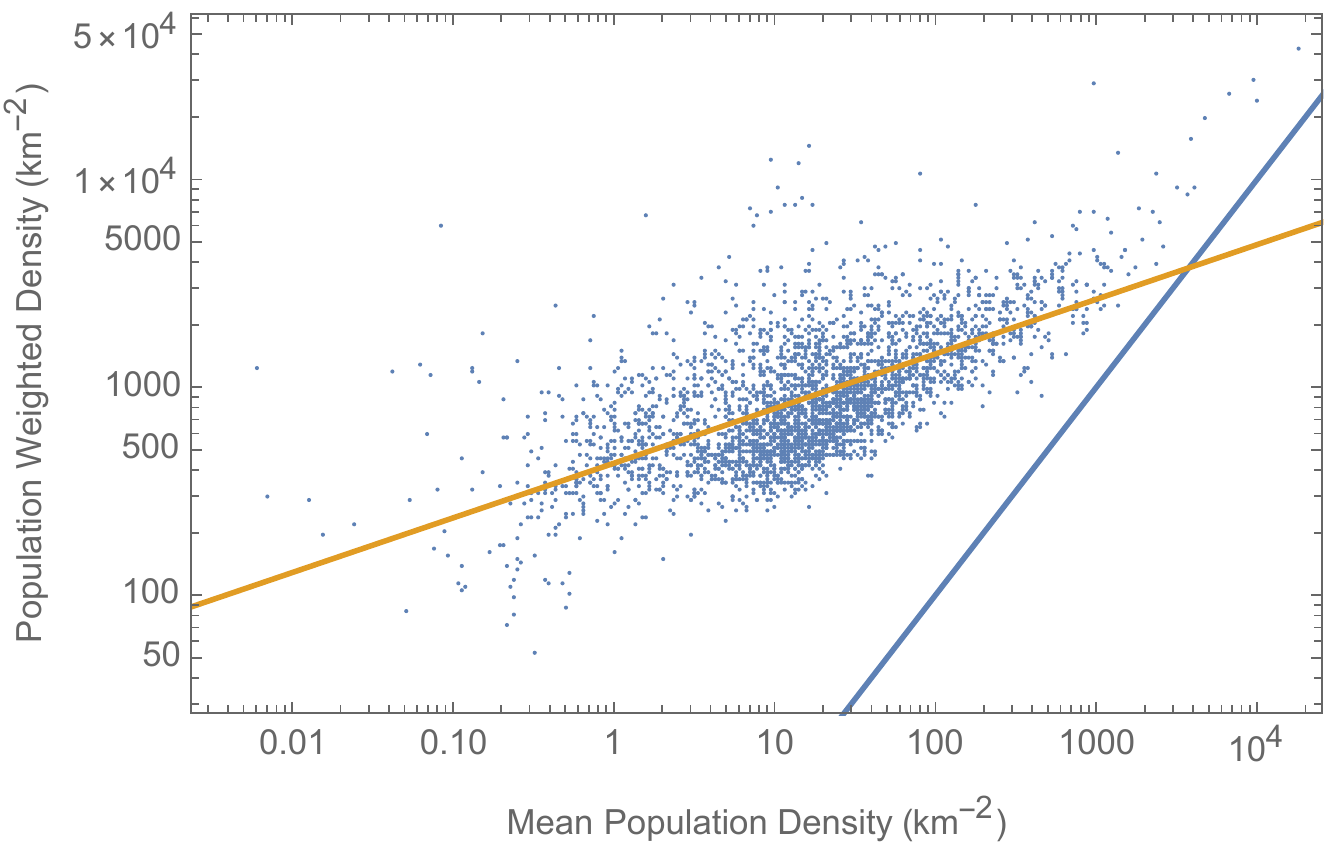}
    \caption{Comparison of the distribution of Population Weighted Density with the Crude population density of US counties: (a)(Top): Histograms, (b)(Bottom): Relative distributions: The blue line shows the one-to-one correspondence, while the orange line is the best-fit power-law ${\rm PWD}_{\rm 250}({\rm km}^{-2}) \simeq 430 \times \left[D({\rm km}^{-2})\right]^{1/4}$. }
    \label{fig:PWPD}
\end{figure}

\begin{figure}
    \centering
    \includegraphics[width=\linewidth]{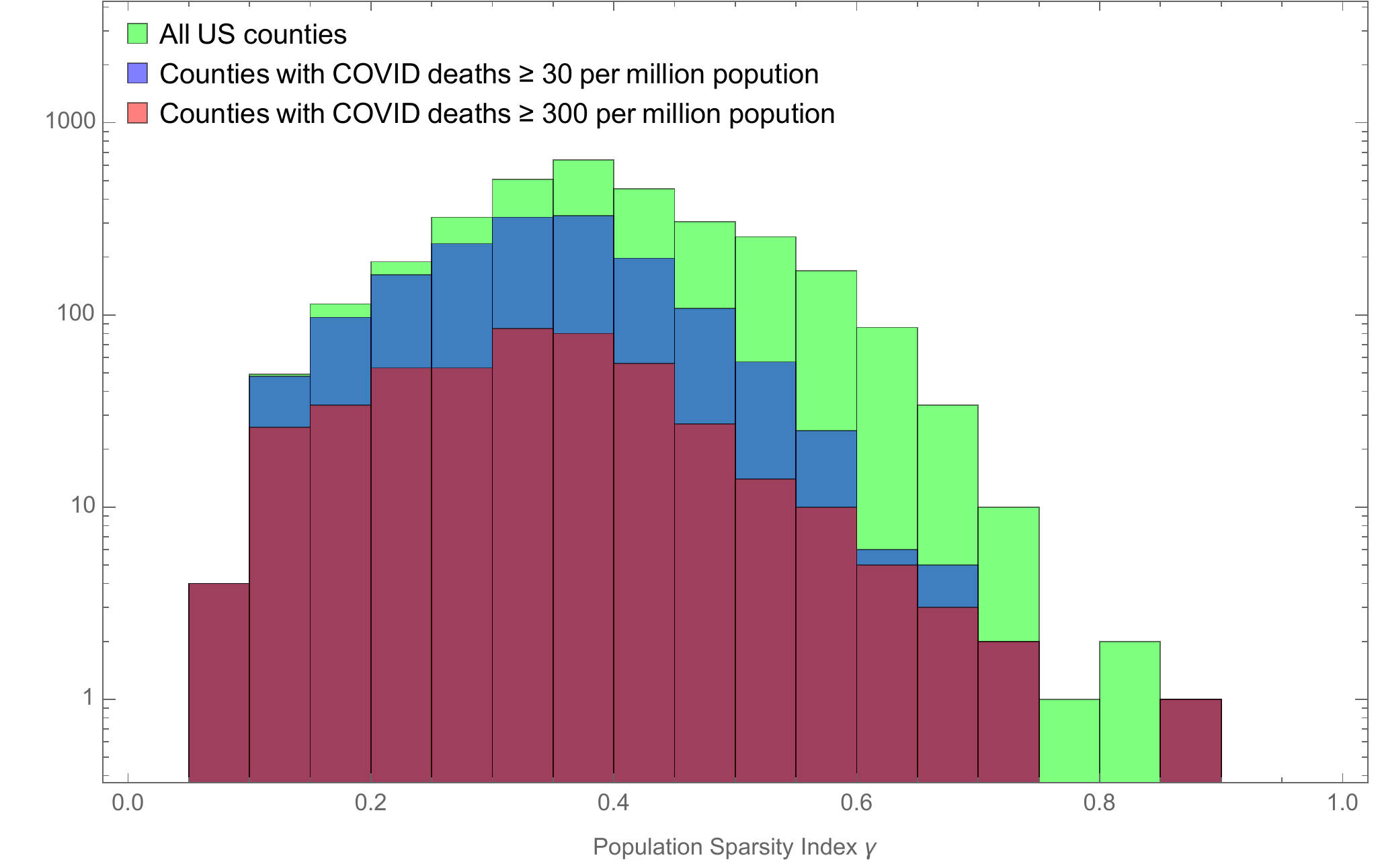}
    \includegraphics[width=\linewidth]{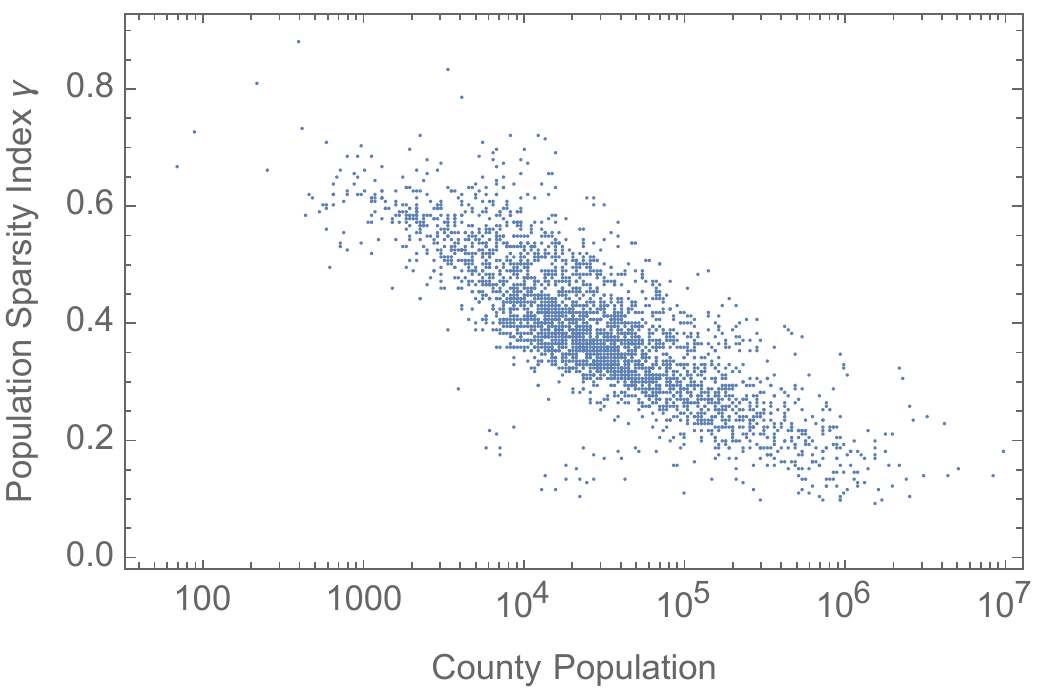}
    \includegraphics[width=\linewidth]{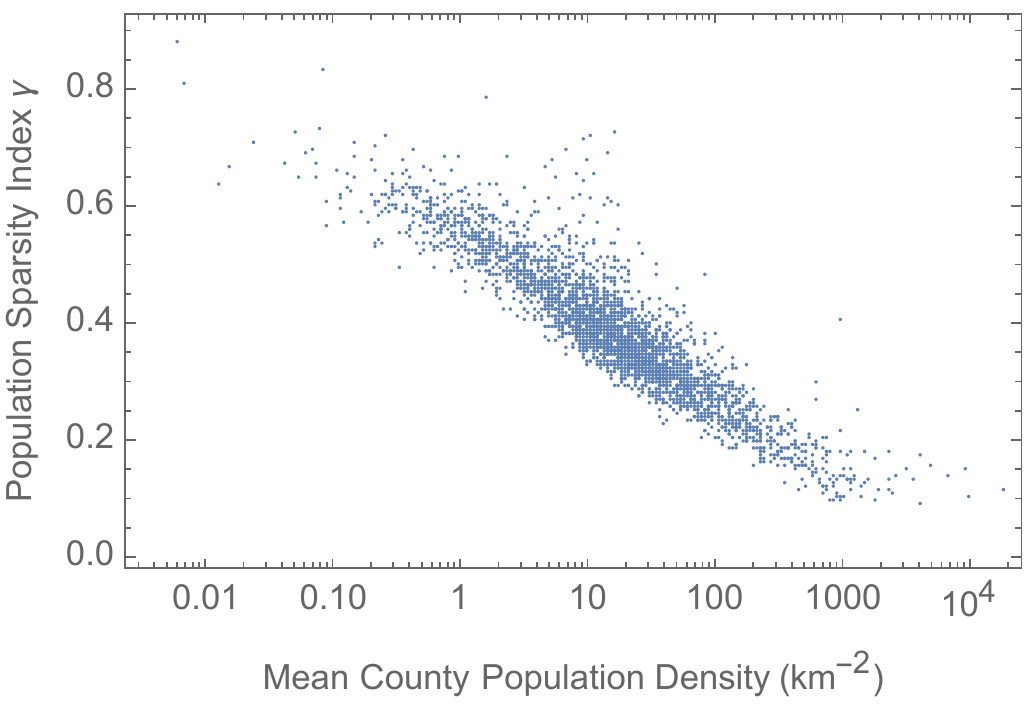}
    \caption{(a)(Top)Distribution of Population Sparsity Index $\gamma$ (b)(Middle) its correlation with total population of the county (c)(Bottom) and its average population density}
    \label{fig:gamma}
\end{figure}

In high density counties, the population weighted density PWD is close to the mean density of the county $D={\rm Pop}/{\rm Area}$, suggesting a uniform distribution of population (see Figure \ref{fig:PWPD}). However, in lower density counties, the mean density is much lower than the population weighted density, due to heterogeneous dense pockets of population amidst vast empty spaces outlined by political boundaries. To represent the degree to which the population density changes across the region (county or metropolitan region) we define the {\em population sparsity index}, $\gamma$. Assuming that the population-weighted population density declines approximately as a power law with ``pixel'' area, ${\rm PWD}_{\sqrt{\rm Area}} \sim {\rm Area}^{-\gamma}$, we define:
	\begin{equation}
	\gamma = \frac{\log\left({\rm PWD}_{\unit{250}{\meter}}\right)- \log(D)}{\log\left[{\rm Area}\right] -\log\left[\left(\unit{250}{\meter}\right)^2\right]}\,.
	\label{eq:gamma}
	\end{equation}
In other words we estimate the assumed power-law decline using two data points. The distribution of $\gamma$ and its correlation with county population and population density are shown in Figure (\ref{fig:gamma}). We can see that $\gamma$ ranges from $0.09$ (i.e., very uniform) for the most populous/dense counties to $0.88$ (i.e. very sparse) for least populated/dense counties. For reference, the value of $\gamma$ for New York City is $0.14$. 

\subsubsection*{Initial growth of confirmed cases for metropolitan regions}

For each of the top 100 metropolitan regions \citep{census2019c}, a logarithmic-scale population heat map, windowed from the full GHS-POP raster image, was used to select a minimal connected set of US counties enclosing the region of population enhancement.  In this process, overlap and merger reduced the total number of metropolitan regions under consideration to 89.  

As is discussed in the main text, nearly every metropolitan region saw, in mid-March 2020, an exponential increase of daily confirmed cases, followed by a flat/plateau period of nearly constant daily confirmed cases. In a few cases, the second phase --- primarily caused by the country-wide lockdown ---  lasted only days or weeks (possibly signaling a depletion of the susceptible population, see discussion in main text), but for most metropolitan regions the plateau persisted for months (indeed, persists or is again increasing at the time of this writing). Thus, the initial value of the exponential growth rate, $\lambda$, of daily confirmed cases could be reliably and automatically estimated by fitting the case numbers to a logistic function
	\begin{equation}
	f_{\rm logistic} (t) = \frac{f_{\rm max}}{1 + \exp\left[ - \lambda \left(t - t_0\right)\right]}
	\end{equation}
where $t_0$ represents the transition time from exponential growth to a constant, $f_{\rm max}$ is the plateau value in case numbers, and $f_{\rm logistic}(t) \propto \exp[\lambda t]$ for $t\ll t_0$. Fits were performed on the logarithm of the case numbers, yielding the maximum likelihood estimation of parameters under the assumption lognormally-distributed errors (an analysis of the fit residuals, not shown, confirmed this assumption: case number fluctuations exhibit a variance far in excess of Poisson noise, but are well modeled by a log-normal probability density function with constant width), and associated estimates of the variance in each parameter. To avoid polluting the exponential growth phase with singular early cases, a ``detection limit'' of 3 was imposed, and all daily case values less than or equal to that limit were ignored in fitting. The only manual intervention required for this fit was the specification of the upper limit of its range, i.e., the {\em end} of the plateau region, for each data set. 

To analyze the effect of demographic, population, mobility, and weather variables on this initial growth rate, we perform a weighted linear regression to the lambda values (and their standard errors) of the 89 metropolitan regions. 

To choose representative cities for the visual examples in Figures \ref{fig:metroregions}A and \ref{fig:metroregions}C, we performed an additional logistic fit to the mortality incidence data of each region and retained for Figure \ref{fig:metroregions}C only those that had (1) less than 15\% error in both growth rates, and (2) $|\lambda_{\rm case} - \lambda_{\rm death}|  < \unit{0.15}{d^{-1}}$. This was done in an effort to specifically comment on or highlight only those cities for which the growth rate was accurately determined, and was well correlated with the more reliable measure, mortality growth, that we used for the remainder of the analysis.

\subsubsection*{Linear Dynamical Model of Mortality Data} 

A standard weighted least squares analysis was performed on the measured exponential growth rate, $\lambda_{14}$, as a function of demographic, mobility, population and weather variables, with weights equal to inverse root of the estimated variance.

\subsubsection*{\label{sec:Model} Nonlinear Model}

We construct a model where, in the standard analogy to chemical reaction kinetics, the incidence of infections per unit area at time, $i(t)$, is proportional to the product of the density of susceptible individuals, $S(t)$, and the density of infected individuals, $I(t)$. But, we allow for the rate constant for infection\footnote{In a physical picture of collisions, the rate constant of infection is $\beta(C) = \langle \sigma v\rangle_{\rm eff}(C)$, i.e., the scattering cross section of an encounter between a susceptible individual and an infectious individual in stage $C$, $\sigma$, multiplied by their relative velocity, $v$.} in the encounter, $\beta$,  to depend on the infected individual's ``stage'' of infection, $C$, with $C=0$ immediately following infection. The incidence then has the form:
	\begin{equation}
	i(t) = -\dot{S}(t) = S(t) \int_0^{\infty} \, \beta(C) \, \mathcal{I}(C,t) \, {\rm d}C\, ,
	\label{eq:Sdot}
	\end{equation}
where $\mathcal{I}(C,t)$ is the density of infected per stage at time $t$, and the first equality expresses that we neglect changes to the susceptible population by all means other than infection.  The density of infected individuals is found by integration over the stages of infection, 
	\begin{equation}
	I(t) = \int_0^{\infty} \, \mathcal{I}(C, t) \, {\rm d}C \, .
	\end{equation}
If the rate constant were taken to be independent of stage, i.e., $\beta(C)= \bar{\beta}$, we would obtain the familiar expression $\dot{S}(t) = - \bar{\beta} S(t) I(t)$. We will assume spatial homogeneity and that the total density of individuals is constant and equal to $S(0)$ for a particular region, but, that the density could vary when comparing different regions.

We assume that an infected individual's evolution through the stages of infection, $C$, follows a Gaussian random walk in time, but modulated by an exponential rate, $d$, of death or recovery. Therefore we have
	\begin{equation}
	\mathcal{I}(C,t) = \int_0^{\infty} \, i(t - a) \, f_{\rm rw}\left(C; a\right) \, {\rm e}^{-d \, a} \, {\rm d} a
	\end{equation}
where $a$ is the ``age'' of an infection (time since infection), and the probability density function for the stage at a given age is given by
	\begin{equation}
	f_{\rm rw}\left(C; a \right) =  \sqrt{\frac{2 \tau}{\pi a}} \, 
	\exp \left[ -\frac{\tau C^2}{2 a}\right] \,,
	\end{equation}
where $\tau$ is the characteristic time scale of the random walk\footnote{More precisely: $C$ is the absolute value of the position of a 1D random walker, taking one step every $\Delta t$, with step size drawn from a normal distribution with mean zero and variance $\Delta t/\tau$. The variance of the walker position at time $t$ is then $t/\tau$}. Integrating the expression for $\mathcal{I}(C,t)$ over all stages and taking the derivative with respect to time yields the familiar expression $\dot{I}(t) = i(t) - d I(t)$, showing that the model reduces to the SIR model if a stage-independent rate constant, $\bar{\beta}$, is assumed.

As we show here, using the random walk to specify the dependence of infection stage on time allows for both a non-exponential distribution of delays to infectiousness (which is more realistic than that of the simplest model with incubation, the SEIR model) and a formal solution for the exponential growth rate.   Inserting the expression for $\mathcal{I}(t,C)$ into the incidence equation yields
	\begin{equation}
	i(t) = S(t) \int_0^{\infty}  i(t-a)\left[ {\rm e}^{-d \,a}   \int_0^{\infty}  \beta(C)  \, f_{\rm rw}\left(C; a\right)  {\rm d}C \right]{\rm d}a 
	\end{equation}
which is in the form of a {\em renewal equation} \citep{heesterbeek_dietz1996, champredon2015, champredon2018}, with the bracketed expression being the expected infectivity of an individual with infection age $a$. To obtain the simplest nontrivial incubation period, we assume that $\beta(C) = \bar{\beta} \, \Theta(C-1)$ --- where $\Theta(x)$ is the Heaviside step function ---  meaning that an infected individual is only infectious once they reach stage $C=1$, and the infection rate constant is otherwise unchanging. This implies that the incidence is
	\begin{equation}
	i(t) =  S(t) \int_0^{\infty}  i(t-a) \, \bar{\beta}\, \mathcal{F}(a) \,{\rm d}a \, .
	\end{equation}
where 
	\begin{equation}
	\begin{split}
	 \mathcal{F}(a) &=  {\rm e}^{-d \,a}   \int_1^{\infty}   \, f_{\rm rw}\left(C; a\right)  {\rm d}C  \\
	 & =  {\rm e}^{-d \,a} \left[1 -  {\rm erf}\left(\sqrt{\frac{\tau}{2 a}}\right)\right]
	 \end{split}
	 \end{equation}
is the probability that an individual infected $a$ time units ago is infectious. 

If we now assume that the density of susceptibles is constant $S(t) = \bar{S}$ over some interval of time, and that the incidence grows (or decays) exponentially in that interval, $i(t) = A {\rm e}^{\lambda t}$, we find
	\begin{equation}
	1 = \bar{\beta} \bar{S} \int_0^{\infty} \exp[-a \lambda] \, \mathcal{F}(a) \, {\rm d} a
	\end{equation}
which, assuming $\lambda + d > 0$ (i.e., the exponential growth rate cannot go below $-d$),  can be integrated to obtain
	\begin{equation}
	\left( \lambda + d \right) \, \exp\left[ \sqrt{2 \left(\lambda + d\right) \tau}\right] = \bar{\beta} \bar{S}\,.
	\end{equation}
This expression for $(\lambda + d)$ has a formal solution in terms of the {\em Lambert W-function}, with simple asymptotic forms:
	\begin{equation}
	\begin{split}
	\lambda + d &= \frac{2}{\tau} \left[ {\rm W}\left(\sqrt{\frac{\bar{\beta} \bar{S} \tau}{2}}\right)\right]^2  \\
	&\approx \left\{ 
	\begin{array}{lc}
	\frac{1}{2 \tau} \left\{ \ln \left[ \frac{2 \bar{\beta} \bar{S} \tau}{\ln^2\left(\bar{\beta} \bar{S} \tau / 2\right)}\right]  \right\}^2 & \left(\bar{S} \bar{\beta} \tau\gg 1\right)\\
	\\
	\bar{\beta} \bar{S} & \left(\bar{S} \bar{\beta} \tau\ll 1\right)
	\end{array}
	\right.
	\end{split}
	\label{eq:nmodel}
	\end{equation}
For the early stages of the epidemic, when we can assume that the population of susceptibles is approximately constant and large, we see that the growth rate depends approximately linearly on the square of the logarithm of the density.  In later stages, when either the base contact rate declines due to social distancing interventions, or the population of susceptibles decreases, we see that the exponential rate takes the value $\lambda \approx - d$.

In practice, we utilize the exact Lambert W-function expression as our ``nonlinear model'' for fitting $\lambda_{14}$, where we parameterize $\beta$ and $\tau$ by the demographic, population, and weather variables (see main text).  To estimate the susceptible density, $\bar{S}$, in this procedure we must use the reported mortality statistics.  Thus far we have not specified the dynamics of death. We now make the assumption that the probability of death increases proportionally to the number of exposures an individual experiences.  As we prove in a separate section, below, this implies that the susceptible density is related to the fraction of dead in the community, $f_D = D_{\rm tot}/N$ (where $D_{\rm tot}$ is the cumulative mortality and $N$ is the total population), by $S(t) = S(0) \exp\left[-C_D \, f_D\right]$.

The {\em basic reproduction number}, $R_0$, and the distribution of {\em generation intervals}, $g(t_g)$, are defined \citep{champredon2015,nishiura2010} through the function $\mathcal{F}(a)$:
	\begin{equation}
	g(t_g) = \frac{\bar{\beta}S(0)\, \mathcal{F}(t_g)}{R_0} \quad {\rm with} \quad R_0 \overset{\mathrm{def}}{=} \int_0^{\infty} \bar{\beta} S(0) \, \mathcal{F}(t_g) \, {\rm d} t_g\,.
	\end{equation} 
The generation interval (or, generation time), $t_g$, is the time between infections of an infector-infectee pair, and is often estimated from clinical data by the {\em serial interval}, which is the time between the start of symptoms \citep{britton2019}, and the basic reproduction number is the average number of infectees produced by a single infected individual, assuming a completely susceptible population. These quantities can be calculated exactly for our model, as
	\begin{equation}
	R_0 = \frac{\bar{\beta}S(0)}{d} \, {\rm e}^{-\sqrt{2 d \tau}}
	\end{equation}
and 
	\begin{equation}
	g(t_g) = d \, {\rm e}^{\sqrt{2 d \tau} - d t_g} \int_1^{\infty} \sqrt{\frac{2 \tau}{\pi t_g}} \exp\left[ - \frac{\tau C^2}{2 t_g}\right] {\rm d}C\,, \label{eq:generation}
	\end{equation}
where the expected value and variance of the generation interval are then:
	\begin{equation} \label{eq:genEandVar}
	E\left[t_g\right] = \frac{1}{d} + \sqrt{\frac{\tau}{2 d}} \quad {\rm and} \quad {\rm Var}\left[t_g\right] = \frac{1}{d^2} + \sqrt{\frac{\tau}{8 d^3}}\,.
	\end{equation}

\subsection*{Extended Results and Analysis}

\subsubsection*{Relation between the remaining susceptible density, $S(t)$ and the death fraction, $f_D(t)$}

In epidemic models the infection of susceptible individuals is typically determined by
    \begin{equation}\label{eq:infectionS}
    \dot{S} = - \beta \, S\, I_*
    \end{equation}
where $I_*$ is the density of infectious (contagious) individuals, and for our model, $\beta S I_*$ is the right-hand side of Eqn.\ \ref{eq:Sdot}. This can be solved, formally, as:
    \begin{equation} \label{eq:solveS}
    S(t) = S(0) \exp\left[- \beta \int^tI_*(s) \, {\rm d}s\right]
    \end{equation}
Alternatively, the susceptible density can be expressed in terms of the cumulative number of infected individuals, $I_{\rm tot}$, i.e.,
    \begin{equation}
    S(t) = S(0) \left[ 1 - f_I(t) \right]
    \end{equation}
where $f_I(t) = I_{\rm tot}/N$, with $N$ the total population. When fitting the exponential growth rate of mortality, $\lambda_{14}(t)$ to our nonlinear model, Eqn.\ \eqref{eq:nmodel} (see main text), we must estimate the value of the susceptible density driving growth at that time. Without any reliable information about the true infected or infectious populations, we must do so using the mortality statistics.  We show here how the previous two equations can be used, along with reasonable assumptions about distinct sub-populations driving infection and death, to determine a relationship between the reported cumulative mortality (per capita) and the remaining susceptible density.

Our basic assumption is that there are two different categories of susceptible individuals underlying the dynamics of the epidemic: (A) highly mobile individuals with a large geographic reach that frequently interact with other individuals (in particular, infectious individuals) and thus drive the dynamics of infection (these could be termed ``super-spreaders'' \citep{liu2020secondary}); and (B) essentially non-mobile individuals that have quite rare contacts with infectious individuals, but have a much higher probability of death once infected, and therefore make up the majority of the mortality burden.  The dynamics of each susceptible population is governed by an equation of the form in Eqn.\ \eqref{eq:infectionS}, with a common density of infectious individuals, $I_*$, but with different rate constants, $\beta_A \gg \beta_B$.  From Eqn.\ \eqref{eq:solveS}, we see that the susceptible densities of the two populations are then related, at any time, by:
    \begin{equation}
    \frac{S_A(t)}{S_A(0)} = \left[\frac{S_B(t)}{S_B(0)}\right]^{\beta_A/\beta_B}\,.
    \end{equation}
Expressing the non-mobile population in terms of the cumulative fraction infected, we have
    \begin{equation}
    \frac{S_A(t)}{S_A(0)} = \left[ 1 - f_I^{(B)}(t) \right]^{\beta_A/\beta_B}\,,
    \end{equation}
and, assuming that the infection fatality rate (IFR) is a constant factor, $f_D(t) = {\rm IFR} \times f_I(t-\Delta t)$, where $\Delta t$ is the delay from infection to death, we can write:
    \begin{equation}
    \frac{S_A(t)}{S_A(0)} = \left[ 1 - \frac{f_D^{(B)}(t+\Delta t)}{\rm IFR} \right]^{\beta_A/\beta_B}\,.
    \end{equation}
Finally, having assumed that the ratio of infection rates is large, we can approximate this as:
    \begin{equation}
    \frac{S_A(t-\Delta t)}{S_A(0)} \approx \exp\left[- \frac{\beta_A}{\beta_B\, {\rm IFR}} \,f_D^{(B)}(t)\right]
    \end{equation}

The ``A'' category of individuals, as defined above, are exactly those individuals driving the infection in our model (and, presumably, in the real world), and, therefore, the susceptible density $S_A$ is exactly that which must be estimated for Eqn.\ \eqref{eq:nmodel}. On the other side, with people aged 65 and over accounting for $\sim$80\% of COVID-19 deaths, and with approximately $\sim$45\% of deaths linked to nursing homes, the mortality statistics are clearly tracing individuals similar to category ``B.'' Therefore, we use this relationship,
    \begin{equation*}
    S(t-\Delta t) = S(0) \, \exp\left[- C_D \, f_D(t)\right]\,,
    \end{equation*}
to estimate the susceptible density in terms of the reported per capita mortality, where we assume $S(0)$ is proportional to the population weighted density (PWD).

\begin{figure}
    \centering
    \includegraphics[width=\linewidth]{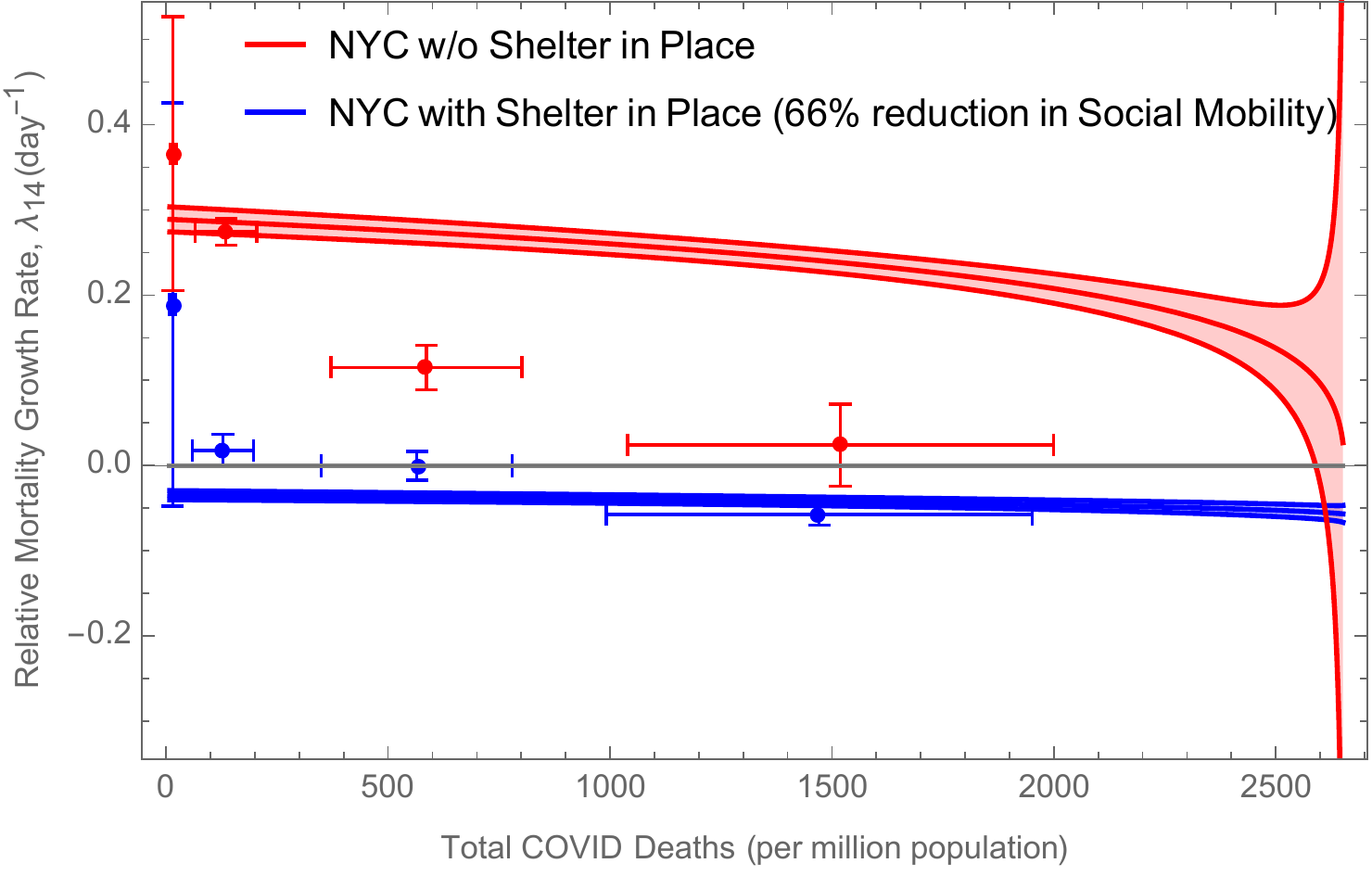}
    \includegraphics[width=\linewidth]{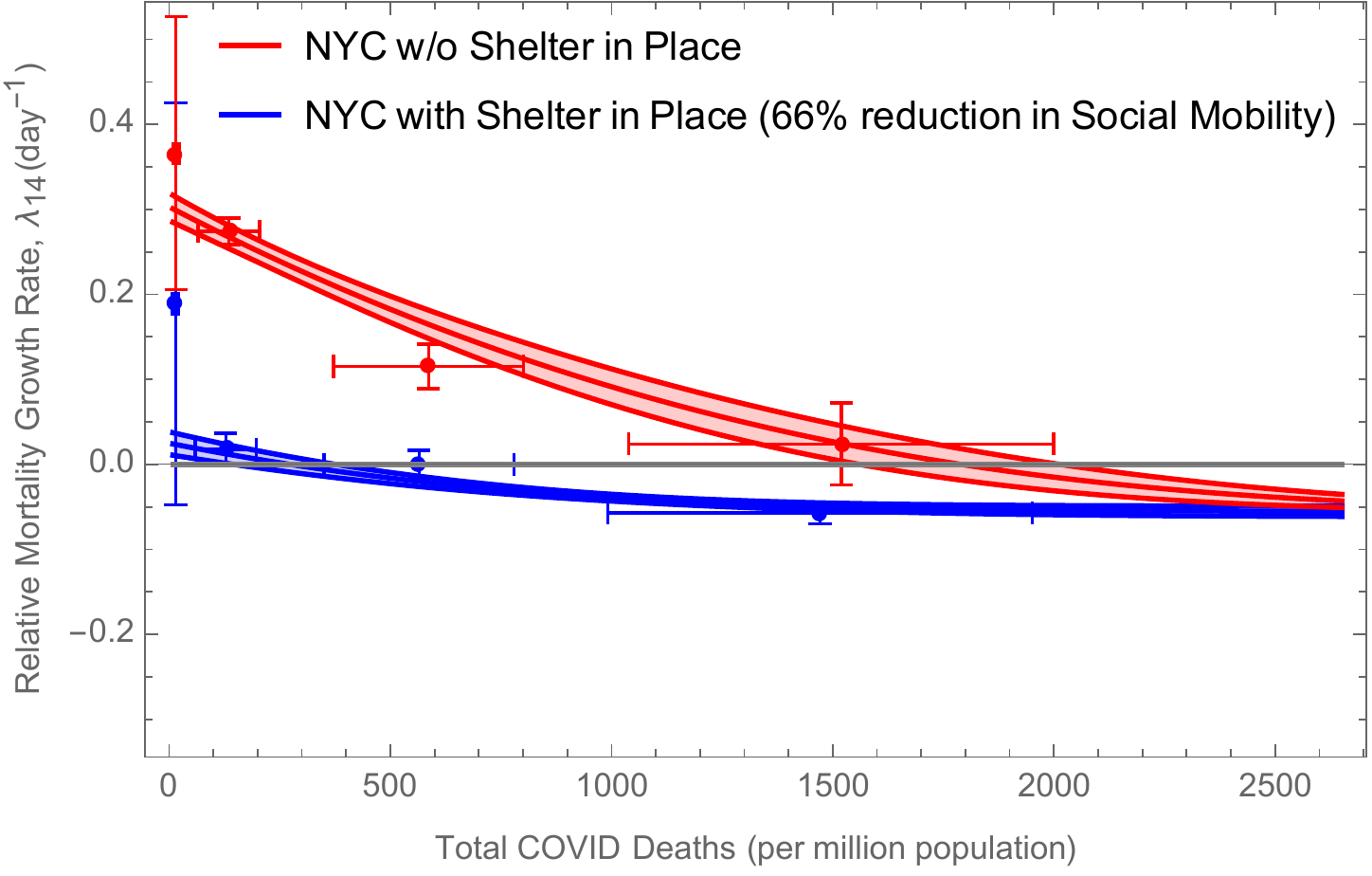}
    \caption{Comparison of hypotheses regarding the relationship between the susceptible fraction, $S(t)/S(0)$, and the dead fraction, $f_D(t)$. A model in which deaths are suffered by a largely immobile population, while the infection is driven by a mobile category of individual (bottom) was preferred to one with a single homogenous population (top) at the $9\sigma$ confidence level. Similar to Figure \ref{fig:Herd}, the points show the best-fit linear model prediction for NYC, fitted independently in different bins of mortality to population ration, while the lines show best-fit $\pm$ 1$\sigma$ nonlinear models. Note that while we show the nonlinear model predictions for a county similar to NYC, we use all US data to find the best fits.}
    \label{fig:Herd_Lin_Exp}
\end{figure}
We also considered the standard approach, in which the population is a single homogeneous group.  In that case, the susceptible density could be estimated as
    \begin{equation}
    S(t) = S(0) \left[1 - f_I(t)\right] = S(0) \left[ 1 - \frac{f_D(t+\Delta t)}{\rm IFR}\right]\,.
    \end{equation}
In testing both models, we found that the two-component population scenario was preferred by the data at the $\sim 10\sigma$ confidence level, with the homogeneous population model failing to capture the observed dependence of the growth rate on the per capita mortality (Figure \ref{fig:Herd_Lin_Exp}).

The broader implications of our assumption of two populations is that the required proportion of individuals with immunity for ``herd immunity'' to take effect, is lower than  population with homogeneous mobility characteristics, i.e., the epidemic will slow as a significant proportion of the ``super-spreader'' category have been infected (category A, above), regardless of the level of infection and immunity in the rest of the population.  Indeed, the effect of population heterogeneity on lowering the ``herd immunity'' threshold for COVID-19 was recently noted  \citep{britton2020mathematical}, and will be  important in interpreting the results of randomized serology tests across the entire population \citep{Havers2020.06.25.20140384}.       





\subsection*{Incubation Time}

\begin{figure}
    \includegraphics[width=1.6\linewidth]{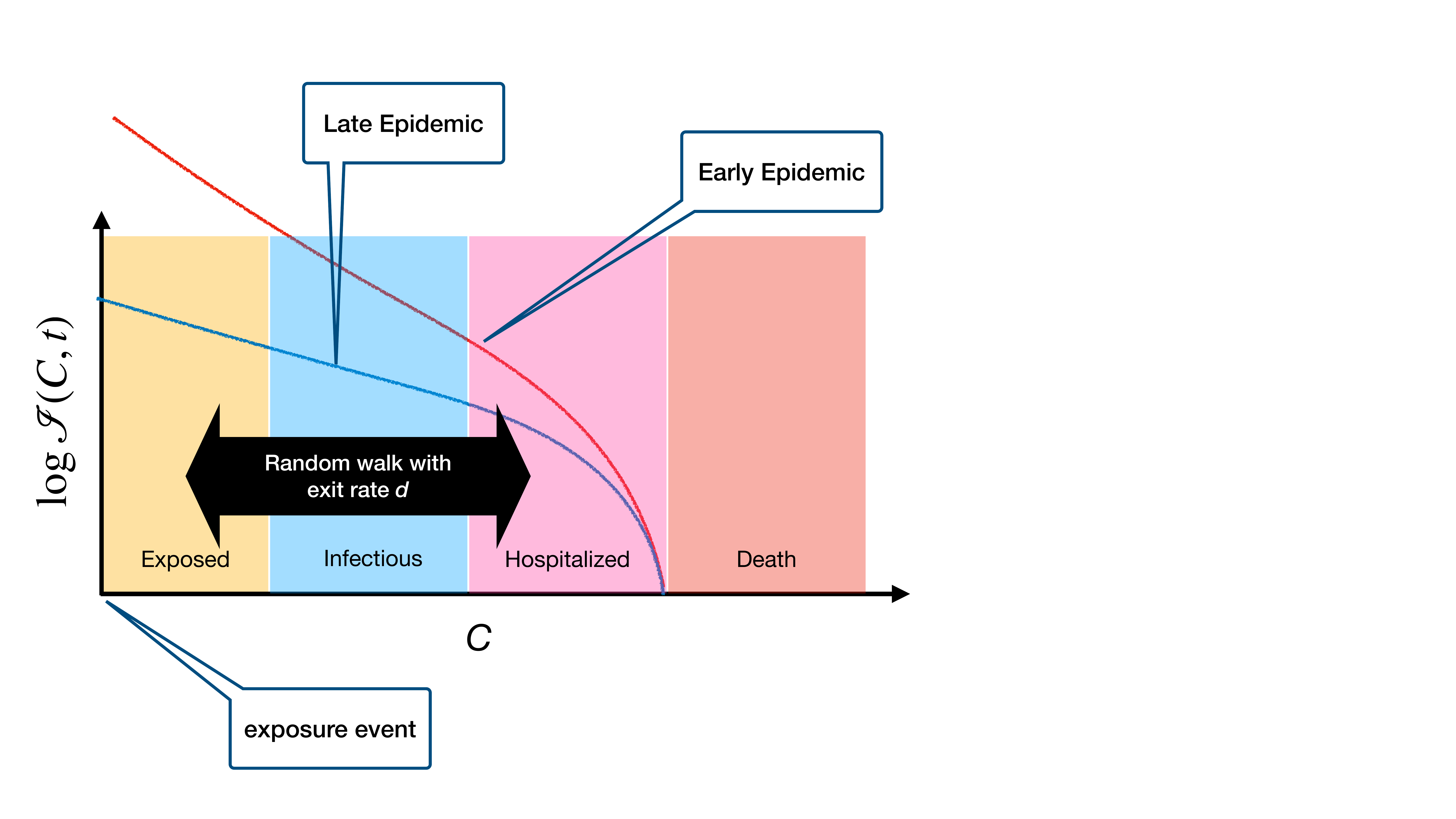}
    \caption{The stages of COVID-19 infection, as a continuous variable $C$. We model the disease as a random walk, starting at $C=0$, and a uniform exit rate (due to either quarantine or recovery). The curves show the resulting distribution in $C$, during the growing (early) and decaying (late) epidemic. Note that the {\it Hospitalization} and {\it Death} are not directly modelled in the Nonlinear model, as they should have a small effect on the spread of the epidemic.  }
    \label{fig:Stages}
\end{figure}

\begin{figure}
    \centering
    \includegraphics[width=\linewidth]{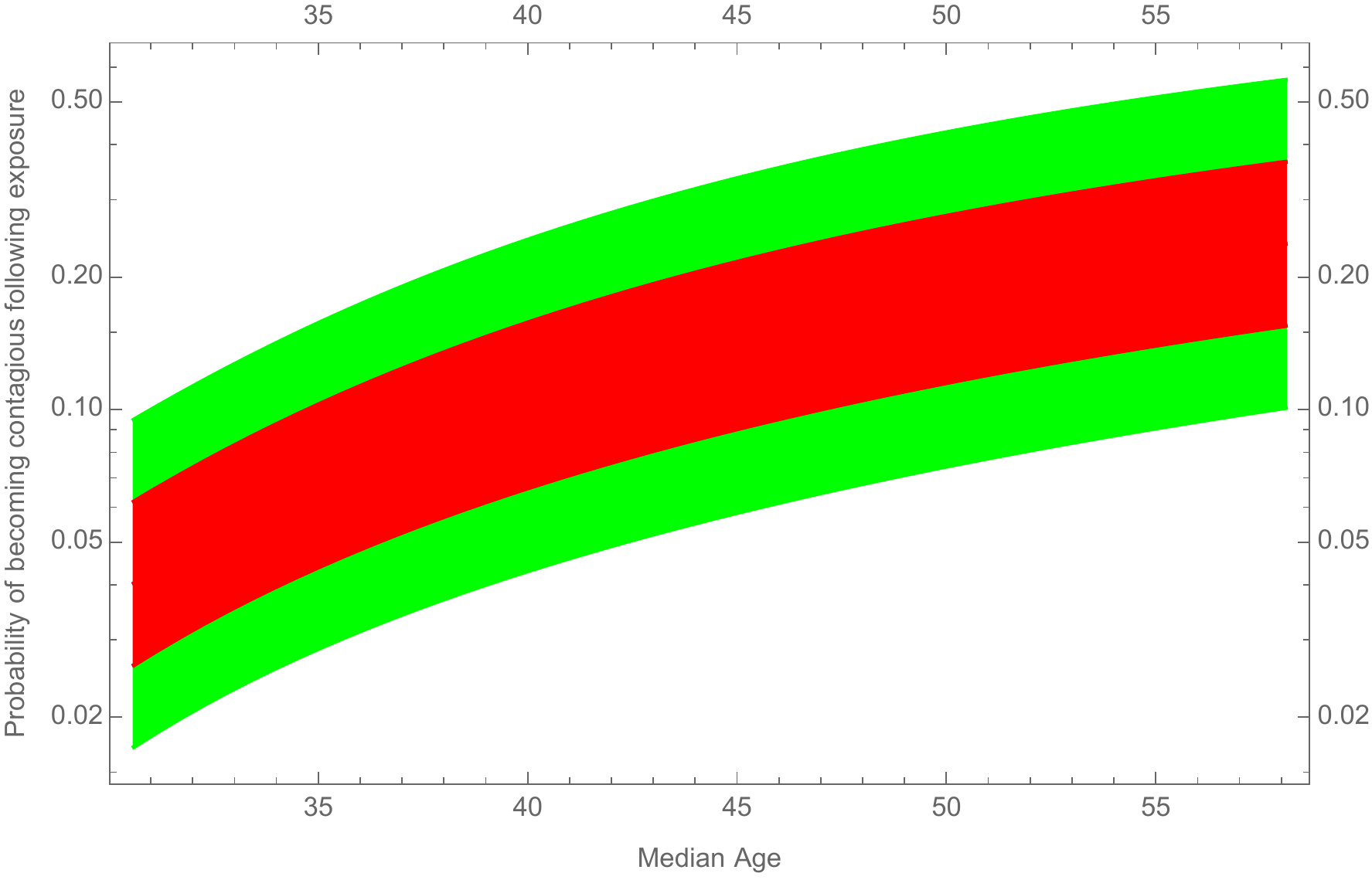}
    \includegraphics[width=\linewidth]{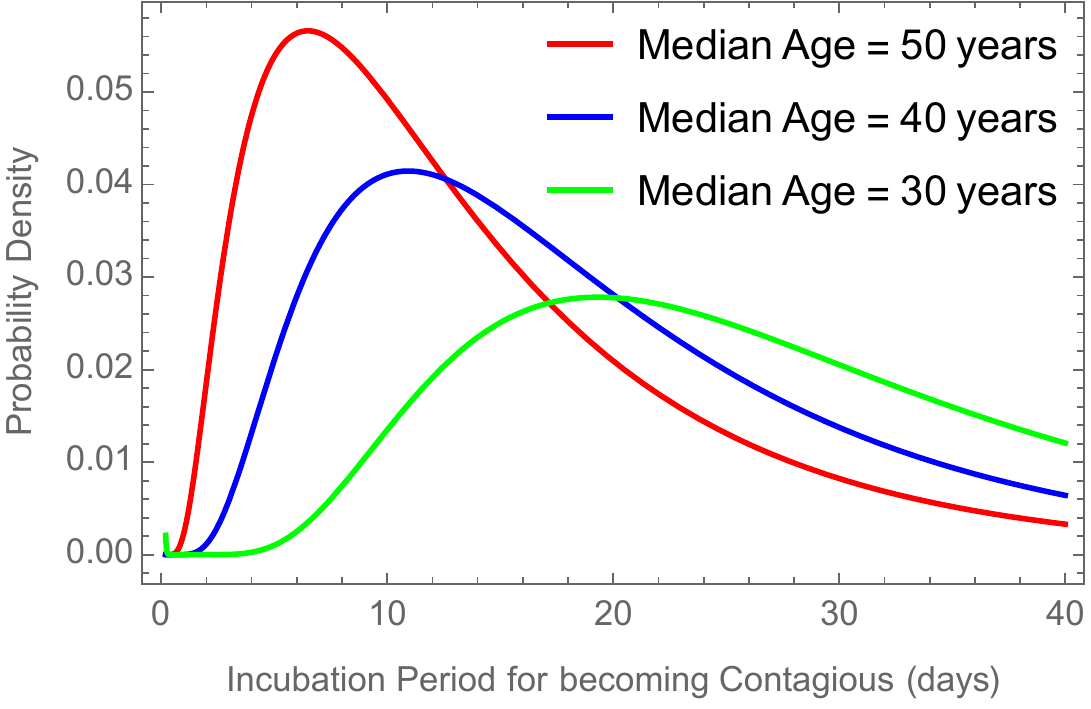}
     \includegraphics[width=1.08\linewidth]{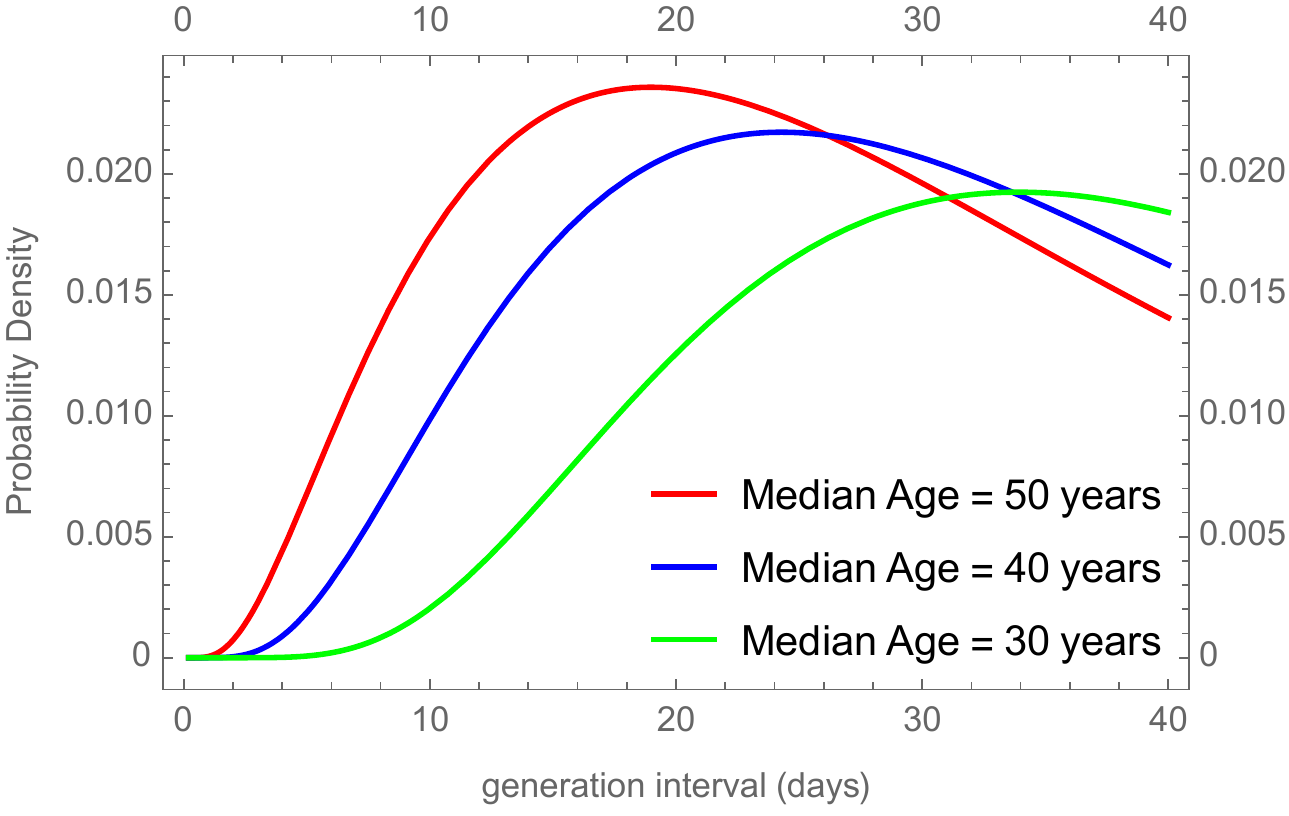}
    \caption{(Top)(a) The probability that an individual exposed to the virus will ever become infectious. (Middle)(b) The probability distribution for incubation period for onset of virus shedding. (Bottom)(c) Distribution of the generation Interval, $g(a)$, i.e., the time from an individual's infection to them infecting another. }
    \label{fig:incubation}
\end{figure}

The nonlinear epidemic model described above posits that the incubation of SARS-COV2 virus within an infected individual can be modelled by a stochastic random walk starting at zero, with excursions beyond $\pm1$ corresponding to episode(s) of infectiousness. This makes our model distinct from the standard SE$^m$I$^n$R compartmental models (see, e.g., \citep{champredon2018}), where the progress of the disease is only in one direction --- $E^1 \to E^2 \to \ldots \to I^1 \to I^2 \to \ldots$ --- while in our model (Figure \ref{fig:Stages}), the individual can jump back and forth between different stages (with the obvious exception of Death), with a constant exit rate of $d$ for quarantine, recovery, or death. This can be described using a (leaky) diffusion equation:
\begin{equation}
    \frac{\partial \mathcal{I}}{\partial t} = \frac{1}{2\tau} \frac{\partial^2 \mathcal{I}}{\partial C^2} - d \mathcal{I}.\label{eq:diffusion}
\end{equation}
Based on this picture, and the best-fit parameters to the US county mortality data  (Table \ref{tab:nonlinear}), we can infer the probabilities associated with the different stages of the disease. For example, by looking at the steady-state solutions of Equation (\ref{eq:diffusion}), we can compute the probability that an exposed individual (who starts at $C=0$) will ever become infectious (i.e., make it beyond $C=C_{\rm inf} =1$):
\begin{equation}
    P_{\rm inf} = \exp[-\sqrt{2 d \tau} C_{\rm inf}],~~~{\rm where}~~~ C_{\rm inf}=1,
\end{equation}
This is plotted as a function of the median age of the community in Figure (\ref{fig:incubation}a). For example, for the median age of all US counties, $A= 37.4$-yr, we get:
\begin{equation}
    P_{\rm inf}(37.4~{\rm years}) = 0.08^{+0.04}_{-0.03} ~(68\%~ {\rm C.L.}),
\end{equation}
i.e., less than 12\% of exposed individuals will ever be able to infect others, although this fraction increases in older communities. 

Next, we can compute the distribution of times for the onset of infectiousness, i.e., the incubation period. This can be done by using a first crossing probability of a random walk, which we did by solving Equation (\ref{eq:diffusion}) using a discrete Fourier series in the $(0,+1)$ interval. The resulting probability density is shown in Figure (\ref{fig:incubation}b), again showing a shorter incubation period in older communities. 

Finally, we can compute the probability density function for the generation interval, $g(t_g)$, defined in Equation (\ref{eq:generation})  (Figure \ref{fig:incubation}c). This shows a similar qualitative dependence on age as the incubation period, but the median incubation period is, as expected, shorter than the generation interval for each age group. Using Eqn.\ \eqref{eq:genEandVar} and our parameterization of $\tau$, we find a mean generation interval of
    \begin{equation}
    E[t_g](37.4~{\rm years}) = 43\pm23~{\rm d} \,.
    \end{equation}
This estimate is much longer than those found by tracking the serial interval (time from between the start of symptoms for an infector-infectee pair) in COVID-19 patients \citep{ganyani2020, nishiura2020}, which are on the order of 5--10~d. It is possible that the long tail of these distributions, generated by the slow asymptotic exponential decay at rate $d\approx\unit{0.06}{d^{-1}}$, raises the mean generation interval, while a clinical study, is necessarily biased toward shorter serial intervals.

\subsection*{Error Diagnostics and Forecasting COVID-19 Mortality}

\begin{figure}
    \centering
    \includegraphics[width=\linewidth]{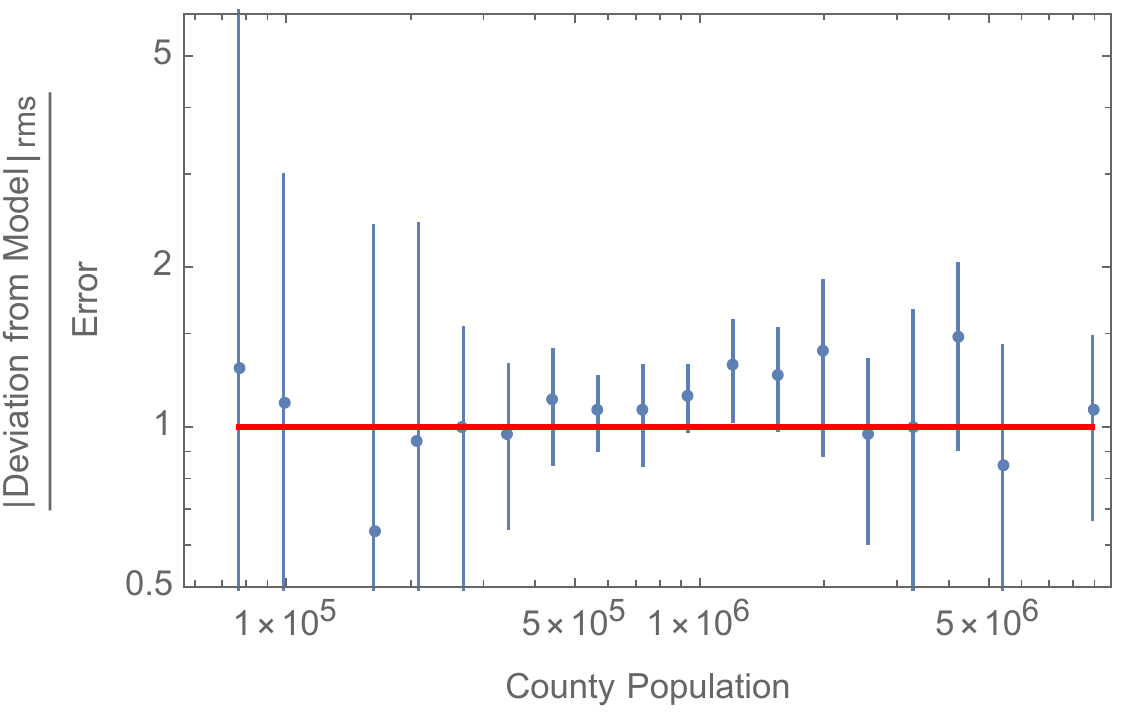}
    \caption{Root Mean Square of the Ratio of the Simplified Nonlinear model residual to measurement errors (i.e. the $\sqrt{\chi^2_{\rm red}}$), as a function of the population of the county. We see that the residuals are consistent with measurement errors.  }
    \label{fig:chi2}
\end{figure}

One of the most pressing questions in any exercise in physical modelling is whether we have a good understanding of the uncertainty in the predictions of the model. While we have an estimate of the measurement uncertainties for the mortality growth rates, $\lambda_{14}$, which we discussed in the main text, we also should characterize whether the deviation of the best-fit model from the measurements are consistent with statistical errors. To evaluate this, we can look at the average of the ratio of the variance of the model residuals to that of the measurement errors, otherwise known as reduced $\chi^2$, or $\chi^2_{\rm red}$. This is shown in Figure (\ref{fig:chi2}), demonstrating that we see no systematic error in model that is significantly bigger than statistical errors, across counties with different populations. 

As another consistency check, Table (\ref{tab:big_v_small}) examines whether the parameters of the model change significantly from urban counties with large, uniform populations, to rural counties with a small and more sparse population (Figures \ref{fig:PWPD}-\ref{fig:gamma}). From counties with enough COVID mortality data, roughly those with population $\gtrsim 10^6$ inhabit half of the total population, which we chose as our threshold, separating large from small counties. We notice no statistically significant difference, and Table (\ref{tab:big_v_small}) even suggests that Fisher errors quoted here might be overestimating the true errors. This comparison brings further confidence in the universality of the nonlinear model across geography and demography.    

\begin{table*}

\begin{tabular}{l|l|l|l }
\hline
Parameter & Small Counties& Large Counties & Difference/Error \\
\hline
$	\tau_0( {\rm day})$	&	126.	$\pm$	58.9	&	219.	$\pm$	124.	&	-0.37	\\
$d^{-1}({\rm day})$	&	18.4	$\pm$	3.92	&	18.1	$\pm$	2.32	&	0.0104	\\
$\ln\left[\beta_0\tau_0^{-2}({\rm m}^2/{
\rm day}^3) \right]$&	-2.23	$\pm$	1.57	&	-0.519	$\pm$	2.1	&	-0.747	\\
$C_D$	&	2743.	$\pm$	845.	&	4425.	$\pm$	958.	&	-0.323	\\
$C_A$	&	-4.73	$\pm$	2.31	&	-2.87	$\pm$	1.75	&	-0.338	\\
$100 C_{\cal M}$	&	0.0527	$\pm$	0.0162	&	0.0732	$\pm$	0.0233	&	-0.227	\\
$C_\gamma$	&	-1.97	$\pm$	2.13	&	-8.37	$\pm$	4.48	&	0.744	\\
$C_T$	&	-0.0415	$\pm$	0.0251	&	-0.0768	$\pm$	0.0342	&	0.405	\\
$C_{A_D}$	&	-1.36	$\pm$	0.632	&	-0.588	$\pm$	0.641	&	-0.521	\\

\end{tabular}
\caption{Comparison of nonlinear model parameters between small (population < 1 million) with large (population > 1 million) counties. We see no significant statistical difference, as demonstrated by the values in the last column that remain below 1. 
Note that, in contrast to Table (\ref{tab:nonlinear}), we use temperature rather than specific humidity ($C_{T}$ rather than $C_{\cal H}$), as the latter was not available for some small counties. Nevertheless, the parameters remain also statistically consistent with Table (\ref{tab:nonlinear}).  }
\label{tab:big_v_small}
\end{table*}

On average, we find that (either county-weighted or population-weighted) $\chi_{\rm red}^2 \simeq 1.28$,  suggesting that the model errors are only 13\% bigger than statistical errors.  

\begin{figure}
    \centering
    \includegraphics[width=\linewidth]{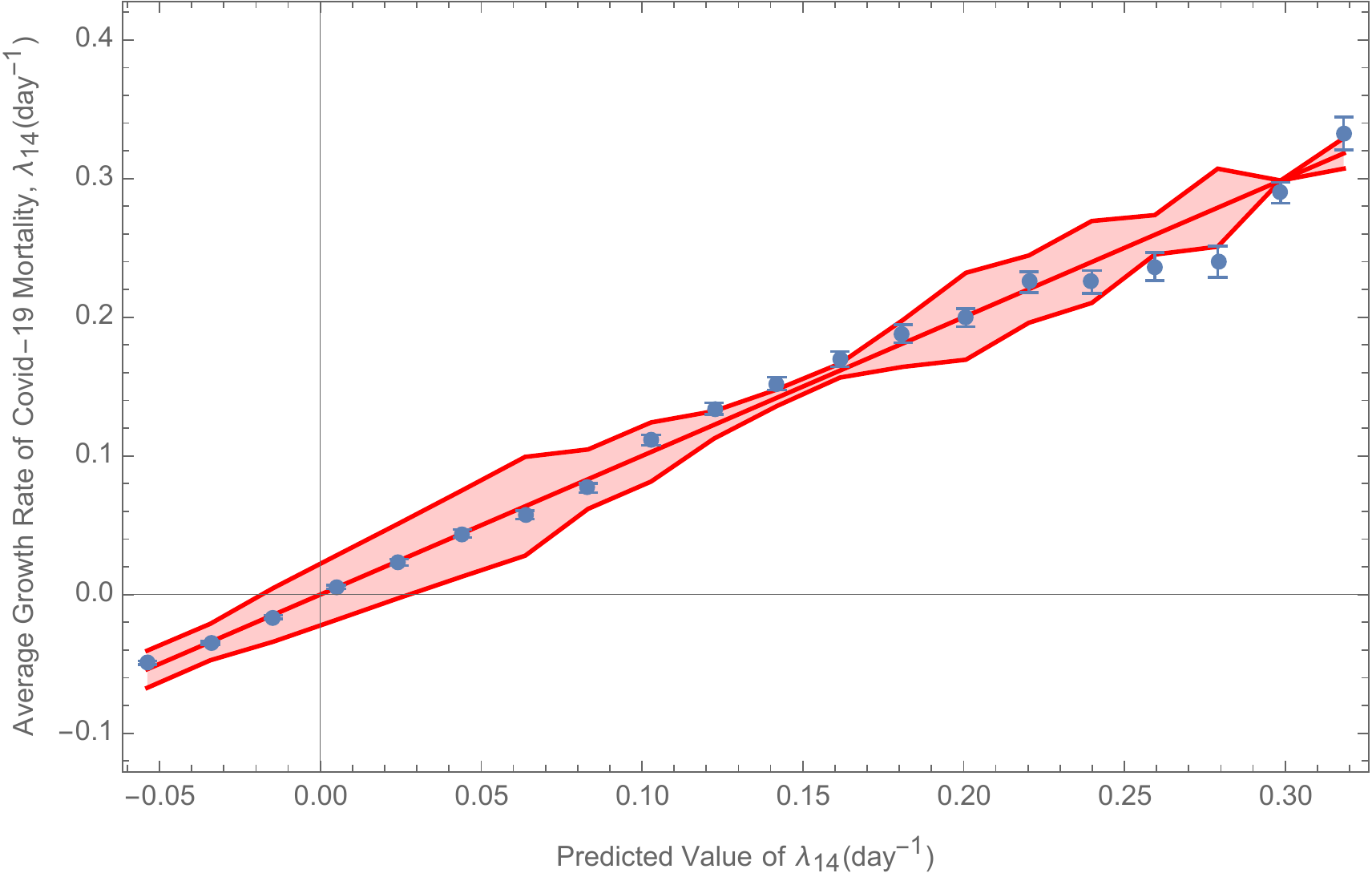}
    \caption{Observed versus Best-Fit model prediction for bins of $\lambda_{14}$. The points show the mean of measured $I_7$ within each predicted bin, as well as the error on the mean. The red region shows the mean excess model error, on top of measurement uncertainties.  }
    \label{fig:I7_Observed_v_Predict}
\end{figure}

We further compare the model-prediction vs measured mortality growth rate in Figure (\ref{fig:I7_Observed_v_Predict}) for all our available data. We find that the 1-$\sigma$ error in the model prediction (in excess measurement errors) is on average $\pm \sigma_{14} = \pm 0.0180$, i.e. 1.8\% error in the daily mortality growth rate. This is shown in Figure (\ref{fig:I7_Observed_v_Predict}) as the red region, which compares the model prediction with the observed mortality growth rates. We can also see that there appears to be no significant systematic deviation from the predictions, at least for $\lambda_{14} < 0.23$/day.  

Given an understanding of the physical model and its uncertainties, we can provide realistic simulations to forecast the future of mortality in any community, similar to those provided in the main text (Figure \ref{fig:Mortality_prediction}), which can be made on-demand using our online dashboard: \href{https://wolfr.am/COVID19Dash}{https://wolfr.am/COVID19Dash}.

In order to perform these simulations, we follow these simple steps. To predict the daily mortality on day $t+1$, $D(t+1)$, we use the prior 13 days of $D(t)$, as well as the total mortality up to that point:
\begin{enumerate}
    \item Use Equation (\ref{eq:nonlinmodel}), plugging in prior total mortality, county information, weather, mobility and parameters in Table \ref{tab:nonlinear} to find $\lambda_{14}$. Every simulation uses a random realization of model parameters (from their posterior fits), which remain fixed through that simulation.
    \item Add the random model uncertainty to $\lambda_{14}$ using:
    \begin{equation}
        \lambda_{14}(t+1) \to \lambda_{14}(t+1)+ \frac{\sigma_{14}}{\sqrt{14}}\sum_{t'=t-12}^{t+1} g_{t'},
    \end{equation}
    where $g_{t'}$s are random independent numbers drawn from a unit-variance normal distribution. This captures the model uncertainty mentioned above, while ensuring that it remains correlated across the 14 days that are used to define $\lambda_{14}$.
    \item Having fixed the logarithmic slope for daily mortality $\lambda_{14}$, find the best-fit intercept and its standard error for $\ln[D(t')+1/2]$ for the preceding 13 days, i.e. $t-12 \leq t'\leq t$, which can then be used to find a random realization for $\ln[D(t+1)+1/2]$
    \item Advance to next day and return to step 1.
\end{enumerate}

\end{document}